\documentclass[10pt]{amsart}

\RequirePackage{amsmath,amsthm,amssymb,graphicx,mathrsfs}

\usepackage[colorlinks=true,allcolors={blue}]{hyperref}
\usepackage[margin=0pt]{subfig}

\usepackage[style=ieee-alphabetic,backend=biber,useprefix=false,url=false]{biblatex}
\usepackage{fancyhdr}
\numberwithin{equation}{section}

\RequirePackage{amsfonts} \RequirePackage{mathrsfs} 

\providecommand{\CC}{\mathbb{C}}
\providecommand{\HH}{\mathbb{H}}
\providecommand{\NN}{\mathbb{N}}

\providecommand{\RR}{\mathbb{R}}
\providecommand{\ZZ}{\mathbb{Z}}

\providecommand{\Aa}{\mathcal{A}}
\providecommand{\Dd}{\mathcal{D}}
\providecommand{\Ff}{\mathcal{F}}
\providecommand{\Hh}{\mathcal{H}}
\providecommand{\Mm}{\mathcal{M}}
\providecommand{\Rr}{\mathcal{R}}
\providecommand{\Ss}{\mathcal{S}}

\RequirePackage{mathtools}

\providecommand{\Mellin}{\mathcal{M}}
\providecommand{\testfn}{\psi}
\newcommand{\measure}[1]{\left|#1\right|_\dimension}
\DeclareMathOperator{\Res}{Res}

\newcommand{\e}{\varepsilon}
\providecommand{\ph}{\varphi}

\newcommand{\dimension}{N}
\newcommand{\RRN}{\RR^\dimension}
\newcommand{\RRNplus}{\RR^{\dimension+1}}

\providecommand{\simdim}{\dim_S}
\providecommand{\lowersimdim}{\underline{\dim}_S}

\providecommand{\xilf}{\xi_{L,f}}
\providecommand{\xilfphi}{\xi_{\Phi,f}}
\providecommand{\xilfphia}{\xi_{L_\Phi^\alpha,f}}

\newcommand{\Region}{\Omega}
\newcommand{\Regionbd}{\partial\Region}
\newcommand{\Regionplus}{Y}

\newcommand{\heatContent}{E_\Region}
\newcommand{\heatzeta}{\widehat{\zeta}_{\Region}}
\newcommand{\partialheatzeta}{\widehat{\xi}_\Region}
\newcommand{\genheatzeta}{\widehat{\zeta}_{\Region,F}}

\newcommand{\sigmaRem}{\sigma_0}
\newcommand{\rempow}{{(\dimension-\sigma_0)}}
\newcommand{\remset}{{\Omega_R}}

\newcommand{\decompRem}{R_\Region}

\newcommand{\heatop}{\Hh}

\newcommand{\heatmean}{\mathscr{M}}
\newcommand{\hyperspace}{\text{Tem}_{+}}
\newcommand{\hypospace}{\text{Tem}_{-}}
\newcommand{\wlower}{w_{\Regionplus,f}^-}
\newcommand{\wupper}{w_{\Regionplus,f}^+}
\newcommand{\wsoln}{w_{\Regionplus,f}}
\newcommand{\parameasure}{\omega_{\Regionplus,p}}

\newcommand{\tubezeta}{\widetilde\zeta}

\newcommand{\Knr}{{K_{n,r}}}
\newcommand{\Phinr}{{\Phi_{n,r}}}

\newcommand{\nth}{^{\text{th}}}

\DeclarePairedDelimiter{\bracket}{\langle}{\rangle}
\newcommand{\enclose}[1]{\left(#1\right)}
\providecommand{\set}[1]{\left\{ #1 \right\}}
\newcommand{\suchthat}{\,\middle|\,} \providecommand{\norm}[1]{\left\lVert#1\right\rVert}

\RequirePackage{dsfont} \newcommand{\1}{\mathds{1}}

 \usepackage[breakable,skins]{tcolorbox}
\tcbuselibrary{skins}

\newcounter{theo}[section]
\numberwithin{theo}{section}

\newtcolorbox{mybox}[1]{
    colback=blue!5,
    colbacktitle=blue!20,
    coltitle=black,
    colframe=blue,
    fonttitle=\bfseries,
    title={#1},
    enhanced,
    attach boxed title to top text left={xshift=0mm,yshift=-2mm},
    boxed title style={arc=0mm},
    arc=0mm,
    breakable
}

\newcommand{\envSetup}[2]{
    \newenvironment{#1}[1][]
    {\refstepcounter{theo}\begin{mybox}{\strut #2~\thetheo\ifstrempty{##1}{}{:~##1}}\relax }
    {\end{mybox}}
}

\envSetup{Theorem}{Theorem}
\envSetup{Definition}{Definition}
\envSetup{Lemma}{Lemma}
\envSetup{Proposition}{Proposition}
\envSetup{Corollary}{Corollary}
\envSetup{Assumption}{Assumption}
\envSetup{myAbstract}{Abstract}

\title{On Complex Dimensions and Heat Content of Self-Similar Fractals}
\author{William E. Hoffer}

\author{Michel L. Lapidus}

\begin{document}

\begin{abstract}
    Complex fractal dimensions, defined as poles of appropriate fractal zeta functions, describe the geometric oscillations in fractal sets. In this work, we show that the same possible complex dimensions in the geometric setting also govern the asymptotics of the heat content on self-similar fractals. 

    We consider the Dirichlet problem for the heat equation on bounded open regions whose boundaries are self-similar fractals. The class of self-similar domains we consider allow for non-disjoint overlap of the self-similar copies, provided some control over the separation. The possible complex dimensions, determined strictly by the similitudes that define the self-similar domain, control the scaling exponents of the asymptotic expansion for the heat content. 

    We illustrate our method in the case of generalized von Koch snowflakes and in particular extend known results for these fractals with arithmetic scaling ratios to the generic (in the topological sense), non-arithmetic setting.
\end{abstract}

\maketitle


\section{Introduction}
\label{sec:intro}

In this work, we study the \textit{heat content} of regions with \textit{self-similar, fractal} boundaries. Given a region $\Region\subset\RR^\dimension$, here an open bounded set, its boundary $\Regionbd$ is said to be self-similar if it the the invariant set of an \textit{iterated function system} consisting only of similitudes of Euclidean space. On such regions, we consider the Dirichlet problem for the heat equation in the sense of Perron-Wiener-Brelot with boundary conditions chosen as a model for heat flow into the region. The heat content is the integral of this solution over the region as a function of the time parameter and we will concern ourselves with the asymptotic expansion of this quantity.

The main results of this work are to provide explicit formulae for the heat content of regions with self-similar fractal boundary in terms of the poles of associated scaling zeta functions and the residues of a heat zeta function, up to an order determined by estimates of the remainder in an initial decomposition. These formulae take the form of a sum over the possible complex dimensions of the boundary with coefficients determined by an associated heat zeta function. We prove these results in the context of solutions to scaling functional equations, for which we state and prove sufficient conditions for requisite asymptotic estimates needed to explicitly compute inverse Mellin transforms in terms of limits of sums of residues of the integrand at poles of a meromorphic function. 

\subsection{Main Results}
\label{ssc:mainResults}

Let $\Region\subset\RR^\dimension$ be a bounded open set and suppose that its boundary $\Regionbd$ is the invariant set of a self-similar system $\Phi$ (Definition~\ref{def:selfSimSys}) for which $(\Regionbd,\Region)$ is an osculant fractal drum (Definition~\ref{def:oscRFD}). Let $\heatContent$ denote the total heat content (Definition~\ref{def:heatContent}) of $\Region$ with respect to Problem~\ref{prob:specificHeatProblem} and suppose that it decomposes according to self-similar copies up to the error term
\[ \decompRem(t) := \heatContent(t) - \sum_{\ph\in\Phi} E_{\ph[\Region]}(t),  \]
with estimate $|\decompRem(t)|=O(t^{(\dimension-\sigmaRem)/2})$ as $t\to0^+$, where $\sigmaRem\in\RR$ is the smallest such parameter. 

We state and prove two admissibility conditions for this remainder (Theorem~\ref{thm:latticeCaseAdmissibility} and Theorem~\ref{thm:lowerDimAdmissibility}) depending on $\Phi$ and $\sigmaRem$ under which the following results hold. In the pointwise setting, the $k\nth$ antiderivative of the heat content, $E_\Region^{[k]}(t)$, for $k\geq2$ is shown to satisfy
\[ 
    \heatContent^{[k]}(t) = \sum_{\omega\in\Dd_\Phi(\HH_{\sigmaRem})} 
        \Res\Bigg(\cfrac{t^{(\dimension-s)/2+k}}{((\dimension-s)/2+1)_k}\heatzeta(s/2;\delta);\omega\Bigg)
        + \Rr^k(t),
\]
for all $t\in(0,\delta)$ for fixed $\delta>0$ and where the remainder term satisfies the estimate $\Rr^k(t)=O(t^{\rempow/2-\e+k})$ as $t\to0^+$ for any $\e>0$. The function $\heatzeta$, which we call a heat zeta function (Definition~\ref{def:heatZetaFn}), is defined in analogy to a tube zeta function and is given in terms of a scaling zeta function associated to $\Phi$, viz. Theorem~\ref{thm:heatZetaFormula}. Further, this identity is valid for any $k\in\ZZ$ when interpreted as a distributional identity, and in particular the heat content itself is given by
\[
    \heatContent(t) = \sum_{\omega\in\Dd_\Phi(\HH_{\sigmaRem})} 
                \Res\Bigg(t^{(\dimension-s)/2}\heatzeta(s/2;\delta);\omega\Bigg)
                + \Rr^{[0]}(t).
\]
The pointwise results is the content of Theorem~\ref{thm:heatFormulaPtw} and the distributional result (without specialization to the case of $k=0$) is the content of Theorem~\ref{thm:heatFormulaDist}.

In these formulae, the summations are indexed by the possible complex dimensions (Definition~\ref{def:cDims}) of $\Regionbd$ relative to $\Region$ in an open right half-plane determined by the order of the estimate on $\decompRem$, $\sigmaRem$. This set, $\Dd_\Phi(\HH_{\sigmaRem})$, is the set of singularities (here, the poles) of the scaling zeta function 
\[
    \zeta_\Phi(s) = \cfrac{1}{1-\sum_{\ph\in\Phi} \lambda_\ph^s},
\]
where $\lambda_\ph$ is the scaling ratio of the similitude $\ph\in\Phi$. For an osculant fractal drum (with an estimate of the volume in a residual set comparable to that of $\decompRem$), the set of complex dimensions of $\Regionbd$ relative to $\Region$ is exactly a subset of the poles of this scaling zeta function \cite[Theorem 5.5]{Hof25}. These poles are the complex solutions $\omega$ of the Moran equation 
\[
    1 = \sum_{\ph\in\Phi}\lambda_\ph^\omega.
\]

These results are obtained through the study of solutions to scaling functional equations, which heat contents are shown to satisfy by virtue of their scaling properties (viz. Proposition~\ref{prop:heatContentScaling} and Corollary~\ref{cor:heatScalingLaw}) when there exists a decomposition induced by a self-similar system for the heat content with suitably estimated remainder (cf. Definition~\ref{def:heatDecompRem}). In fact, they are stated and proved in the context of general solutions to scaling functional equations (viz. Theorem~\ref{thm:pointwiseFormula} and Theorem~\ref{thm:distFormula}) with admissible remainders such that the Mellin transform of the remainder term and the scaling zeta function $\zeta_\Phi$ associated to the SFE are \textit{jointly languid}, a notion we introduce in Section~\ref{ssc:languidity}. The first admissibility criterion occurs when $\sigmaRem$, from the remainder estimate, is strictly smaller than a lower bound for the poles of $\zeta_\Phi$, here called the lower similarity dimension of $\Phi$ (Definition~\ref{def:lowerSimDim}). The second criterion is when the scaling ratios of $\Phi$ lie in the lattice case (Definition~\ref{def:latticeDichotomy}).

Lastly, we illustrate our results in Section~\ref{ssc:heatAppGKF} in the case of generalized $(n,r)$-von Koch fractals (GKFs), such as the examples in Figure~\ref{fig:threeGKFs}. The admissibility criteria are shown to be satisfied when either $n\geq5$ or in the lattice case (see Definition~\ref{def:latticeDichotomy}). In particular, the ordinary von Koch snowflake (see Figure~\ref{fig:threeGKFs}) falls in the lattice case. Further, in the lattice case, when the poles of $\zeta_\Phi$ (and thus the complex dimensions of the GKF) are simple, the heat content is given by a distributional expansion of the form 
\[ 
    \heatContent(t) = \sum_{\omega\in\Dd_\Phinr(\HH_{0})} r_\omega\,t^{(2-\omega)/2} + \Rr(t),
\]
where $r_\omega$ is a constant determined by the residue of the heat zeta function (Corollary~\ref{cor:distHeatFormulaGKFLattice}). The general versions of these results are given in Theorem~\ref{thm:pointwiseHeatFormulaGKF} and Theorem~\ref{thm:distHeatFormulaGKF}. The estimates for the decomposition remainder were established in \cite{vdB00_generalGKF} and our results here extend the explicit formulae in \cite{vdB00_generalGKF} to the nonlattice case (see Definition~\ref{def:latticeDichotomy}) when $n\geq5$. Further, we establish a precise connection between the spectral (here, regarding the heat content) complex dimensions and the underlying geometric complex dimensions.

\subsection{Organization of the Work}

This work is organized as follows. Section~\ref{sec:preliminaries} contains the relevant background information for stating and proving the main results, including an overview of: Perron-Wiener-Brelot solutions to the heat equation (the details of which are relegated to Appendix~\ref{app:PWBsolutions}), the geometry of self-similar fractals and their complex dimensions, and the growth estimates known as languidity. An additional appendix regarding Mellin transforms and some relevant properties for the proofs of our results are contained in Appendix~\ref{app:Mellin}. Section~\ref{sec:SFE} contains the discussion of scaling functional equations as well as the theorems and proofs of our main results regarding their solutions in the general setting. The applications of these results to the heat content of regions with self-similar fractal boundary are contained in Section~\ref{sec:heatAnalysis}, including the illustration thereof in the case of generalized von Koch fractal domains.

\section{Preliminaries}
\label{sec:preliminaries}

\subsection{The Heat Equation on Bounded Open Sets}
\label{ssc:heatPrelim}

\subsubsection{Perron-Wiener-Brelot Solutions}

Let $Y\subset\RR^{\dimension+1}$ be an arbitrary bounded open set. To consider the heat equation on an arbitrary open set, care must be taken to interpret the prescribed boundary conditions. In this work, we study a solution in the sense of Perron-Wiener-Brelot (PWB) per Definition~\ref{def:PWBsolution}, the details of which we include in Appendix~\ref{app:PWBsolutions}. In brief, a PWB solution adapts Perron's famous method to solving the Laplace equation \cite{Per23} in the context of the heat equation. It is named additionally for the contributions of Wiener \cite{Wie24_Potential,Wie24_Average} and Brelot \cite{Bre55,Bre67}. For a more thorough background of the contributors to heat potential theory, see Section~\ref{ssc:PWBhistory}. 

The advantage of the PWB approach, over a weak/variational formulation, is that it yields a solution which is pointwise-defined everywhere and continuously differentiable in the interior. Convergence to prescribed boundary values is a pointwise limit, though only on a subset of the boundary called the regular essential boundary (see Section~\ref{ssc:PWBboundary}).

We restrict our attention to a cylindrical open set, the Cartesian product of a fixed spatial region $\Omega\subset\RR^\dimension$ (without loss of generality a connected open set) and $(0,\infty)$. The Dirichlet problem for the heat equation is then given by
\begin{equation}
    \label{prob:generalHeatProblem}
    \left\{
    \begin{aligned}
        \partial_t u - C\Delta u &= 0   &&\text{in }\Region\times(0,\infty), \\
            u &= f                      &&\text{on }\Region\times\set{0}, \\
            u &= g                      &&\text{on }\partial\Region\times (0,\infty).
    \end{aligned}
    \right.
\end{equation}
Here, $C$ is a positive constant called the diffusivity constant. In order for a solution $u$ to exist in the PWB sense, the prescribed boundary values specified by the functions $f:\Region\to\RR$ and $g:\Regionbd\to\RR$ must be resolutive (per Definition~\ref{def:resolutiveFn}). Letting $F:\partial(\Region\times(0,\infty))\to\RR$ represent the shared boundary conditions, that is with $F|_{t=0}\equiv f$ and $F|_{x\in\partial\Omega}\equiv g$, an equivalent characterization is that $F$ is integrable with respect to each element of a family of parabolic measures relative to $\Omega\times(0,\infty)$ (per Definition~\ref{def:paraMeasure}).

In what follows, we will specialize to the following boundary conditions. Initially, we suppose that the PWB solution to the heat equation inside of $\Region$ is zero. On the boundary of $\Region$ for all time, we suppose that the boundary $\partial\Region$ is held at a constant temperature, which we suppose to be equal to one. Explicitly,
\begin{equation}
    \label{prob:specificHeatProblem}
    \left\{
    \begin{aligned}
        \partial_t u - C\Delta u &= 0   &&\text{in }\Region\times[0,\infty), \\
            u &= 0                      &&\text{on }\Region\times\set{0},    \\
            u &= 1                      &&\text{on }\partial\Region\times (0,\infty).
    \end{aligned}
    \right.
\end{equation}
These boundary conditions model the flow of heat into $\Region$ from its boundary, supposing that the edges are held at a constant temperature. These boundary conditions define a resolutive boundary function for any bounded open set, so that a PWB solution exists, and furthermore, they are scale invariant. 

To see the former, let $\Regionplus=\Region\times(0,\Region)$ and define the function 
\[ F(x,t) = \1_{\partial\Region}(x), \]
which is a measurable function for each of the parabolic measures $\parameasure$ (Definition~\ref{def:paraMeasure}), $p\in\Regionplus$, since each measure is a Borel measure and $\partial\Region$ is closed (whence its indicator function is measurable). Further, $F$ is integrable since the measure of the (essential) boundary (see Section~\ref{ssc:PWBboundary}) of $\Regionplus$ is equal to one, i.e. $\parameasure(\partial_e\Regionplus)=1$, since they are probability measures. Thus $F$ is parabolically integrable and Problem~\ref{prob:specificHeatProblem} admits a PWB solution. 

We also have that the solution to Problem~\ref{prob:specificHeatProblem} is unique due to the results of Widder \cite{Wid44} since the solution is nonnegative (which can be seen from the minimum principle for the heat equation). 

The scale invariance of these boundary conditions will be essential to establishing the relationship between Problem~\ref{prob:specificHeatProblem} and the analogous problem on the image $\ph[\Region]$ under a similitude $\ph$ of Euclidean space. By (parabolically) scale invariant, we mean that $F(x,t)=F(\ph(x),\lambda^2 t)$ for any similitude $\ph$ of $\RR^\dimension$ whose scaling ratio is $\lambda>0$. That $F$ is scale invariant follows immediately from it being piecewise constant (and the fact that $\lambda^2\cdot0=0$). In general, the results of this work may be generalized to parabolically integrable, scale-invariant choices of boundary conditions.

\subsubsection{Heat Content and its Properties}

First, we define our main object of study, the heat content of (a measurable subset of) the spatial domain. 
\begin{Definition}[Heat Content]
    \label{def:heatContent}
    \index{Heat content}
    Let $u_\Region$ be the PWB solution to the Dirichlet Problem~\ref{prob:generalHeatProblem} on an open bounded set $\Region\subset\RR^\dimension$ with resolutive boundary conditions (per Definition~\ref{def:resolutiveFn}). The \textbf{heat content} $\heatContent$ inside any measurable set $F\subseteq\Region$ is defined as 
\begin{equation}
        \heatContent(t;F) := \int_F u_\Region(x,t) \,dx.
    \end{equation}
    When $F=\Region$, we write $\heatContent(t)=E_\Region(t;\Region)$ and call it the \textbf{total heat content}.
\end{Definition}

In the case of Problem~\ref{prob:specificHeatProblem}, we note that the heat content $\heatContent(t;F)$ is uniformly bounded for all $t\in[0,\infty)$ and any measurable set $F$. This follows as a corollary of the maximum principle for the heat equation, the boundedness of the region $\Region$, and elementary properties of the Lebesgue measure. Since the boundary conditions of Problem~\ref{prob:specificHeatProblem} are bounded by one, we have the following explicit estimate. 
\begin{Proposition}[Heat Content Boundedness]
    \label{prop:heatContentBounded}
    Let $u_\Region$ be the PWB solution to Problem~\ref{prob:specificHeatProblem}. Then for any $t\in[0,\infty)$ and any measurable set $F\subseteq\Region$, we have $|\heatContent(t;F)| \leq \measure{\Region}$.
\end{Proposition}

Next, we deduce a scaling property for the heat content. This will require imposing (parabolically) scale-invariant boundary conditions, the simplest of which are constant boundary conditions.
\begin{Proposition}[Heat Content Scaling Property]
    \label{prop:heatContentScaling}
    \index{Scale invariance}
    \index{Scale invariance!Parabolic scale invariance}
    \index{Scaling law!of heat content}
    Let $u_\Region$ and $u_{\ph[\Region]}$ be PWB solutions to Problem~\ref{prob:generalHeatProblem} on a bounded open set $\Region$ and $\ph[\Region]$ in $\RR^\dimension$, respectively, where $\ph$ is a similitude of $\RR^\dimension$ with scaling ratio $\lambda>0$. Suppose that $f$ and $g$ are chosen so that $u_\Region$ is unique, e.g. if $f$ and $g$ are both nonnegative (in which case $u_\Region\geq 0$, which is sufficient by \cite{Wid44}).
    \medskip
    
    Suppose that the boundary functions $f$ and $g$ have the following scale invariance properties with respect to $\ph$: for any $x\in\Region$, $f$ is \textbf{scale invariant} (i.e. $f(\ph(x),0)=f(x,0)$) and $g$ is \textbf{parabolic scale invariant} (i.e. $g(\ph(x),\lambda^2t)=g(x,t)$). Then 
    \[ E_{\ph[\Region]}(t) = \lambda^\dimension E_\Region(t/\lambda^2). \]
\end{Proposition}
\begin{proof}

    First, we have that for any $x\in \Region$ and for all $t>0$,
    \begin{equation}
        \label{eqn:heatScaling}
        u_{\ph[X]}(\ph(x),\lambda^2 t)=u_\Region(x,t).
    \end{equation}
    To see this, we note that the function $v(x,t):=u_{\ph[\Region]}(\ph(x),\lambda^2 t)$ satisfies the heat equation for any $x\in\Region$ and for all $t>0$, since $(\partial_t-\Delta)v=\lambda^2(\partial_t-\Delta)u_{\ph[\Region]}=0$ on the points $(\ph(x),t)\in\ph[\Region]\times(0,\infty)$. 
    
    Furthermore, $v$ satisfies the same boundary conditions as the function $u_\Region$. Let us use the notation in the discussion of Problem~\ref{prob:arbitraryHeatProb}: let $u_\Region=u_{\Regionplus,F}$ and $u_{\ph[\Region]}=u_{\ph[\Regionplus],F}$, where $\Regionplus=\Region\times(0,\infty)$ and where $F$ represents respective boundary conditions $f$ and $g$ when appropriately restricted. Let $p_n=(x_n,t_n)\to q=(y,s)$ be a sequence of points in $\Regionplus$ approaching the boundary point $q$, and in the case of a semi-singular boundary point we assume that $t\to s^+$. Let $p'=(\ph(x_n),\lambda^2 t)$ and $q'=(\ph(y),\lambda^2t)$ denote the corresponding points under parabolic transformation. 
    
    By assumption, at any regular boundary point $q$ (respectively $q'$) we have that 
    \begin{align*}
        u_{\Regionplus,F}(p_n)&\to F(q) &\text{ as }p_n\to q\in\partial_e\Regionplus, \\
        u_{\ph[\Regionplus],F}(p'_n)&\to F(q') &\text{ as }p_n\to q\in\partial_e\Regionplus.
    \end{align*}
    In other words, $u_{\Regionplus,F}$ (respectively $u_{\ph[\Regionplus],F}$) converge on the regular essential boundary in the sense of Definition~\ref{def:convergenceOnEssenBd}. We note that if $q$ is a regular boundary point, then so too is $q'$ for its corresponding boundary. This can be seen from the barrier criterion (see for instance \cite[Theorem~8.46]{Wat12}). Namely, if there is a barrier $w$ defined near the point $q$, then the corresponding function $w(\ph(x),\lambda^2t)$ will be a barrier at $q'$. 
    
    By definition, $v(p)=u_{\ph[\Regionplus],F}(p')$, so $v(p)\to F(q')$. Under the parabolic scale invariance assumption, $F(q')=F(q)$. Thus, we conclude by uniqueness that $v(x,t)=u_\Region(x,t)$, which is exactly \eqref{eqn:heatScaling}. 
Using properties of the Lebesgue integral and \eqref{eqn:heatScaling}, it follows that
    \begin{align*}
        E_{\ph[\Region]}(t) &= \int_\Region u_{\ph[\Region]}(\ph(x) ,t)\,d\ph(x) \\ 
        &= \lambda^{\dimension}\int_\Region u_{\ph[\Region]}(\ph(x),\lambda^2(t/\lambda^2))\,dx \\ 
        &= \lambda^{\dimension}\int_\Region u_{\Region}(x,t/\lambda^2)\,dx \\ 
        &= \lambda^{\dimension}E_\Region(t/\lambda^2).
    \end{align*}
\end{proof}

Note that an important part of this scaling property is that the heat content in a region is related to a \textit{rescaled} problem's heat content (cf. $E_{\lambda\Region}(t)$), not a \textit{restricted} heat content (cf. $E_\Region(t;\lambda\Region)$).

\subsection{Geometry of Self-Similar Fractals}
\label{ssc:fractalPrelim}

\subsubsection{Self-Similar Iterated Function Systems}

Let $(X,d)$ be a complete metric space. An \textit{iterated function system (IFS)} on $(X,d)$ is a finite set $\Phi$ of contraction mappings on $X$. Explicitly, this means that for each map $\ph\in\Phi$, there exists a constant $r_\ph\in(0,1)$ such that for all $x,y\in X$,
\[ d(\ph(x),\ph(y)) \leq r_\ph\, d(x,y). \]
Hutchinson showed that there is a unique nonempty set $K=K_\Phi$ which is closed, bounded, and satisfying  
\[ K = \bigcup_{\ph\in\Phi} \ph[K], \]
which is called the attractor or invariant set of $\Phi$; it is compact and obtained as the closure of the set of fixed points of finite compositions of mappings in the IFS \cite{Hut81}. For example, many classical fractals may be obtained through this framework using the space of nonempty, compact subsets of Euclidean space $\RR^\dimension$ equipped with the Hausdorff metric. This includes Cantor sets, fractal snowflakes, Sierpinski carpets, Menger spongers, and more \cite{Fal90,Bar88,LapRad24_IFG}. In what follows, we shall be focused on sets like these in Euclidean space, rather than elements of more general complete metric spaces.

All of these specific examples share an additional property, \textit{self-similarity}. The mappings of an IFS are required to be contractions, but for self-similar (iterated function) systems we shall also require that the mappings are similitudes. 
\begin{Definition}[Self-Similar System]
    \label{def:selfSimSys}
    A \textbf{self-similar system} $\Phi$ on a complete metric space $(X,d)$ is a finite set of (nontrivial) contractive similitudes, 
    $\Phi := \set{\ph_k:X\to X}_{k=1}^m$.     
\end{Definition}
Explicitly, for each of the mappings $\ph_k$, $k=1,...,m$, to be both a contraction and a similitude, we must have that for every $x,y\in X$, there is some $r_k\in(0,1)$, called the \textit{scaling ratio} of $\ph_k$, so that
\[ 
    d(\ph_k(x),\ph_k(y)) = r_k\, d(x,y). 
\]
Note that we have ruled out the presence of a constant mapping (by requiring $r_k> 0$) and have enforced contractivity (by imposing that $r_k<1$). In general, a similitude may be defined simply by the equality without the shrinking distance requirement, allowing any positive scaling ratio. We note that maps defined this way are equivalently compositions of isometries---namely, compositions of translations, rotations, and reflections--- of Euclidean space and uniform scaling maps.

Given a nonempty compact set $X\subset\RR^\dimension$, $X$ is said to be a \textit{self-similar set} if there exists a self-similar system $\Phi$ such that $X$ is the attractor of $\Phi$.

\subsubsection{Lattice/Nonlattice Dichotomy}

In the study of self-similar fractals with multiple scaling ratios, there is a \textit{lattice/nonlattice dichotomy} in behavior depending on whether or not the distinct scaling ratios are \textit{arithmetically related} or not. This dichotomy has also been called the arithmetic/non-arithmetic dichotomy, and has been discussed in the work of Lalley in \cite{Lal88,Lal89,Lal91}, in the work of Strichartz on self-similar measures \cite{Str1,Str2,Str3}, in the work of the second author and collaborators on fractal harps, fractal drums, and the theory of complex dimensions (such as in \cite{LapvFr13_FGCD,LRZ17_FZF,Lap93_Dundee,KL93,LP10,LPW11,HL06,Lap19}), and for the generalized von Koch snowflakes by van den Berg and collaborators in their work on heat content \cite{vdB00_generalGKF,vdB00_squareGKF,vdBGil98,vdBHol99}, among others. See \cite{DGM+17} and the references therein for more information about this dichotomy in fractal geometry.

\begin{Definition}[Lattice/Nonlattice Dichotomy]
    \label{def:latticeDichotomy}
    A set of (distinct) scaling ratios $\set{\lambda_k}_{k=1}^K$ is said to be in the \textbf{lattice case} if the group 
    \[ G = \prod_{k=1}^K \lambda_k^\ZZ \]
    is a discrete subgroup of the positive real line with respect to multiplication. In this case, there exists a generator $\lambda_0$ such that every $\lambda_k=\lambda_0^{m_k}$ for some positive integer $m_k$. In this case, the (distinct) scaling ratios are said to be \textbf{arithmetically related}.
    \medskip
    
    If $G$ is not a discrete subgroup of the positive real line (in which case it is a dense subgroup), then the (distinct) scaling ratios are said to be \textbf{non-arithmetically related}. This is called the \textbf{nonlattice case}.
\end{Definition}
Note that if $\Phi$ is a self-similar system, the set $\set{\lambda_\ph}_{\ph\in\Phi}$ is necessarily the set of distinct scaling ratios of the similitudes in $\Phi$, rather than viewed as a multiset. It has cardinality less than or equal to $|\Phi|$. By a slight abuse of language, to say that the scaling ratios of the mappings in $\Phi$ are arithmetically related is to say that the \textit{set of} scaling ratios, i.e. the distinct scaling ratios, are arithmetically related.

An important feature of this dichotomy is that the \textit{complex dimensions} of self-similar fractals behave very differently in the lattice and nonlattice case. The structure of complex dimensions has important implications for the asymptotics of the quantities on regions with self-similar fractal boundary. For more information about the structure results in the different cases, see \cite[Theorem~2.16]{LapvFr13_FGCD} for the one dimensional setting and \cite[Theorem~3.6]{LapvFr13_FGCD} for a generalized result which may be applied to self-similar sets in higher dimensions.

\subsubsection{Relative Fractal Drums}

A relative fractal drum (RFD), a notion originally introduced in \cite{LRZ17_FZF}, is a geometric object that considers a fractal of interest, here the boundary $\Regionbd$, relative to some set which is close to the fractal in an appropriate sense. For this work, it the relative set simply the interior region $\Region$ on which we study Problem~\ref{prob:generalHeatProblem} and the closeness condition is automatic from the definition of the boundary of a set.

\begin{Definition}[Relative Fractal Drum (RFD)]
    \label{def:RFD}
    Let $X,\Omega\subset\RR^\dimension$. Suppose that $\Omega$ is open, has finite Lebesgue measure, and has the property that there exists $\delta>0$ such that 
    \[ 
        X \subseteq \Omega_\delta := \set{x\in\RR^\dimension \suchthat \exists\,y\in\Omega,\,d(x,y)<\delta}, 
    \]
    a $\delta$-neighborhood of $\Omega$. Then the pair $(X,\Omega)$ is called a \textbf{relative fractal drum (RFD)}. 
\end{Definition}

So, the RFDs of this work are of the form $(\Regionbd,\Region)$, where $\Region\subset\RR^\dimension$ is a bounded open set (hence having finite Lebesgue measure). Note that any $\delta>0$ is sufficient for $\Region_\delta$ to contain $\Regionbd$. Often, one may assume that $\Region$ is connected since we may independently study each of the relative fractal drums $(\partial U,U)$ for each connected component $U$ of $\Region$. (In other words, $\Region$ may be chosen to be a bounded domain of $\RR^\dimension$.) Note that, in general, $\Omega$ has at most countably many connected components.

Since our fractal sets (i.e. the boundaries of the given region) will be defined via iterated function systems, we will impose conditions which control the way in which this relative set interacts with the IFS. The first is a separation condition called the open set condition.\begin{Definition}[Open Set Condition]
    \label{def:OSC}
    An iterated function system $\Phi=\set{\ph_i}_{i=1}^m$ on $\RR^\dimension$ satisfies the \textbf{open set condition (OSC)} if there exists a nonempty, open set $U\subset\RR^\dimension$ (called a feasible open set for $\Phi$) such that:
    \begin{enumerate}
        \item $U\supset \bigcup\limits_{i=1}^m \ph_i[U]$;
        \item For each $i,j\in\set{1,...,m}$ with $i\neq j$, $\ph_i[U]\cap \ph_j[U]=\emptyset$. 
    \end{enumerate}
\end{Definition}

We need slightly more information about such a feasible open set when studying an RFD $(X,\Omega)$. Namely, we need to know that the images of this set remain closest to the corresponding images of the attractor under the mapping. This condition, introduced as the osculating set condition in \cite{Hof25}, ensures that points in the image $\ph[\Omega]$, for $\ph\in\Phi$, remain closest to the corresponding image $\ph[X]$ under the same contraction $\ph$, as opposed to another distinct image.
\begin{Definition}[Osculating Sets and RFDs]
    \label{def:oscRFD}
    Let $\Phi=\set{\ph_i}_{i=1}^m$ be an iterated function system on $\RR^\dimension$, and let $X$ be its attractor. A nonempty, open set $\Omega\subset\RR^\dimension$ is said to be an \textbf{osculating set} for $\Phi$ if the following hold:
    \begin{enumerate}
        \item $\Phi$ satisfies the open set condition with respect to $\Omega$;
        \item For each $i=1,..,m$, if $y\in\ph_i[\Omega]$, then $d(y,X)=d(y,\ph_i[X])$. 
    \end{enumerate}
    A relative fractal drum $(X,\Omega)$ is called an \textbf{osculant} fractal drum if there exists an IFS $\Phi$ for which $X$ is its attractor and $\Omega$ is an osculating set thereof. 
\end{Definition}

\subsubsection{Tube Zeta Functions and Complex Dimensions}

Fractal zeta functions arise out of the use of the Mellin transform to analyze the oscillatory behavior characteristic of fractals in the space of scales. The versions of fractal zeta functions developed for the higher-dimensional theory, distance and tube zeta functions, extend the theory of complex (fractal) dimensions to this setting \cite{LRZ17_FZF}. In this work, we will focus on relative tube zeta functions, considering a boundary relative to its corresponding interior region. 

To start, the \textbf{tube function} of the RFD $(X,\Omega)$ (also called a relative tube function) is the Lebesgue measure of a tubular neighborhood of $X$ intersected with $\Omega$ as a function of the distance parameter defining the neighborhood. Explicitly, the tube function of $X$ relative $\Omega$ is the function 
\[ 
    V_{X,\Omega}(t) := \measure{X_t\cap\Omega},
\]
for all $t>0$. Here, we denote by $\measure{\,\cdot\,}$ the Lebesgue measure in $\RR^\dimension$ and we need only impose that $\Omega$ is measurable, since a tubular neighborhood (as a union of open balls) is necessarily an open set, even if $X$ itself is arbitrary. For the fractal drums we consider, $\Omega$ is open and thus measurable. 

A \textit{truncated Mellin transform} is an ordinary Mellin transform but with restriction of the domain to an interval subset of $(0,\infty)$. The Mellin transform is the multiplicative cousin of the Fourier transform, defined for the positive real numbers with the operation of multiplication and the natural Haar measure on this group. In particular, an upper cutoff improves the convergence properties of the transform at the cost of complicating the self-similarity properties of the transform. In general, a truncated Mellin transform can introduce lower or upper cutoffs.\begin{Definition}[Truncated Mellin Tranform]
    \label{def:truncatedMellinTransform}
    Let $f$ be an integrable function on $(a,b)$, where $0\leq a<b<\infty$. Then the \textbf{truncated Mellin transform} $\Mellin_a^b[f]$ of $f$ is the complex analytic function 
    \[ 
        \Mellin_a^b[f](s) := \int_a^b t^{s-1}f(t)\,dt 
    \]
    for all $s\in\CC$ for which this Lebesgue integral converges. We shall further identify $\Mellin_a^b[f]$ with the analytic continuation of this function in the complex plane when there is no ambiguity in the domain. 
    
    When $a=0$, we write $\Mellin_a^b=\Mellin^b$. 
\end{Definition}
In the present work, we shall consider only functions which are meromorphic without any accessible logarithmic singularities, so we shall not encounter any ambiguities in the domain induced by the need for branch cuts. For more information regarding (truncated) Mellin transforms and their properties, see Appendix~\ref{app:Mellin}. 

Of particular use presently is the scaling property of this Mellin transform.\begin{Proposition}[Scaling Property of Mellin Transforms]
    \label{prop:MellinScaling}
    Let $\Mellin^\delta$ denote a truncated Mellin transform with $\delta>0$. Then for any $\lambda>0$ and for $f$ and $s\in\CC$ such that the transform converges,
    \[ 
        \Mellin^\delta[f(t/\lambda^2)](s) = \lambda^{2s}\Mellin^{\delta/\lambda^2}[f](s). 
    \]
\end{Proposition}
\noindent This proposition is a simple corollary of the change of variables formula, but notably the upper cutoff changes. The difference between cutoffs, notably, turns out to be a function which is entire with respect to the complex variable $s$. Note also that we are using a scaling factor of $\lambda^2$ owing to the scaling properties of heat content, viz. Proposition~\ref{prop:heatContentScaling}.

Putting the notions of a tube function and a truncated Mellin transform together, we may now formally state the definition of the relative tube zeta function.\begin{Definition}[Relative Tube Zeta Function]
    \label{def:tubeZeta}
    Let $(X,\Omega)$ be a relative fractal drum and let $\delta>0$. The \textbf{relative tube zeta function} $\tubezeta_{X,\Omega}$ of $X$ relative to $\Omega$ is given by
    \[ \tubezeta_{X,\Omega}(s;\delta) := \Mellin^\delta[t^{-\dimension} V_{X,\Omega}](s), \]
    for all $s\in\CC$ for which the analytic continuation of the integral transform is well defined. 
\end{Definition}
\noindent This tube zeta function will typically possess a meromorphic continuation in the complex plane, and its poles are of great importance to the geometry of the fractal. 

The singularities of this fractal zeta function are called complex dimensions \cite{LapvFr13_FGCD,LRZ17_FZF}. As we show in Section~\ref{sec:heatAnalysis}, they will play the role of exponents in explicit formulae for heat content. To formally define complex dimensions, we must specify the window $W$, a subset of the complex plane to which the function permits a meromorphic extension. Additionally, we note that the singularities of the tube zeta function are independent of the cutoff parameter $\delta$, that they are geometric invariants of a given fractal set, and further that the singularities of different variants of fractal zeta functions, most notably the tube and distance zeta functions, are the same \cite{LRZ17_FZF}.\begin{Definition}[Complex Dimensions of a Set]
    \label{def:cDims}
    Let $(X,\Omega)$ be a relative fractal drum and let $\tubezeta_{X,\Omega}$ be the relative tube zeta function of $X$. If $W\subset \CC$, then the \textbf{complex dimensions} of $X$ relative to $\Omega$ contained in the window $W$, denoted by $\Dd_X(W)$, are the poles of $\tubezeta_{X,\Omega}$ contained within $W$. 
\end{Definition}
Strictly speaking, other types of singularities are of interest and should be considered as complex dimensions. For the self-similar sets we consider here, however, only poles will occur, and thus we may restrict our definition, for simplicity.

\subsection{Languid Growth}
\label{ssc:languidity}

\subsubsection{Background}

Languidity refers to growth estimates needed to express the contour integral of an inverse Mellin transform in terms of a sum over the residues of the integrand at its poles in a vertical strip in the complex plane. The notion of languid growth was introduced in \cite{LF00} and then refined in \cite{LapvFr13_FGCD} and \cite{LRZ17_FZF} in the context of proving explicit tube formulae and other geometric, dynamical, and spectral formulae for generalized fractal harps or relative fractal drums, respectively. Strictly speaking, these definitions apply to the geometric object (the harp or the drum) but involve conditions on their associated zeta functions. As in \cite{LRZ17_FZF}, we extend these definitions to apply to a (zeta) function itself, and the corresponding geometric object, including now a self-similar system, may be defined to be languid when its associated zeta function is languid.

There are two types of languid growth, standard (or weak) languid growth and strong languid growth. We will provide the definitions of both, as the scaling zeta functions of a self-similar system will in fact be strongly languid. The formulae which we will obtain for tube and heat zeta functions by means of solving a general scaling functional equation, however, will be only languid, owing to the nature of the contribution from a remainder term in the (approximate) scaling functional equation. Languid growth is the content of \cite[Definition~5.1.3]{LRZ17_FZF} and \cite[Definition~5.2]{LapvFr13_FGCD} and strongly languid growth is the content of \cite[Definition~5.1.4]{LRZ17_FZF} and \cite[Definition~5.3]{LapvFr13_FGCD}. 

As the names would suggest, strong languid growth implies (weak) languid growth. However, there is an important caveat to this implication. A strongly languid function is only languid with respect to \textit{some} screen (to be defined), not any \textit{specified} screen. In order to compare the languid growth of two different functions, we introduce a notion of \textit{joint languidity} where the \textit{same} screen is used for both functions. This notion will precisely define the joint estimates for the truncated Mellin transform of a remainder term in a scaling functional equation and the associated scaling zeta function needed in Section~\ref{ssc:SFEsolutions} to prove our main results.

\subsubsection{Languidity Hypotheses}

In what follows, let $\zeta_f$ denote a complex function which is holomorphic in the open right half-plane $\HH_D$
and extended by analytic continuation to a domain in $\CC$ containing the closure of $\HH_{D}$. When $\zeta_f(s)=\Mellin_\alpha^\beta[f](s)$ is the truncated Mellin transform of a function $f$, we take $D$ to be its abscissa of absolute convergence $\sigma_{ac}$. (See Appendix~\ref{app:Mellin} for more detail in this case.) We assume that $\zeta_f$ admits analytic continuation to an open connected neighborhood of the set $W\setminus\Dd_f(W)$, where $W\subset\CC$ and where $\Dd_f(W)$ is a discrete subset of $W$ at which $\zeta_f$ may be singular. In the functions we analyze in this work, $\zeta_f$ will possess a necessarily unique meromorphic continuation in an open connected neighborhood of $W\subset\CC$, with possible singularities at the points of $\Dd_f(W)$.

Next, let $S:\RR\to\RR$ be a bounded, Lipschitz continuous function. We say that a \textbf{screen} $\Ss$ is a set of the form 
\[ \Ss = \set{S(\tau)+i\tau\suchthat \tau\in\RR}\subset\CC. \]
By a slight abuse of notation, we say that the function $S$ itself is a screen and identify its graph with the set $\Ss$, which is a rotated embedding of its graph in $\RR^2$. We say that a set $W=W_S\subseteq\CC$ is a \textbf{window associated to a screen} when 
\[ W=\set{s\in\CC\suchthat \Re(s)\geq S(\Im(s))} \] 
for some screen $S$. Note that $W_S$ is closed, thus when we consider analytic continuation to this window, we strictly speaking are considering analytic continuation to an open connected neighborhood containing $W_S$.

Given a function $\zeta_f$, we will impose the following constraints regarding an associated screen and window. Firstly, we assume that $S:\RR\to[-\infty,D]$, where $D\in\RR$ is such that $\zeta_f$ is holomorphic in $\HH_D$. When $\zeta_f$ is a Mellin transform or a tamed Dirichlet-type integral (see \cite[Definition A.1.2]{LRZ17_FZF} or Appendix~\ref{app:Mellin}), we assume that $D=\sigma_{ac}$ is its abscissa of absolute convergence. Secondly, we assume that $W_S$ is contained in the connected open set in $\CC$ for which the holomorphic continuation of $\zeta_f$ is defined, except perhaps for a discrete set $\Dd_f(W_S)$ of exceptional points (viz. where the meromorphic continuation of $\zeta_f$ has poles). Thirdly, we assume that the image of the screen does not intersect this discrete set of possible singularities of $\zeta_f$; in particular, $\zeta_f$ must be meromorphic on a connected open neighborhood of (the image of) the screen $S$ without poles on the screen. 

For brevity, let us say when these assumptions hold for a given function $\zeta_f$ (which is assumed to be holomorphic in some open right half-plane and admitting an analytic continuation to a connected open subset of $\CC$ containing the closure of $\HH_D$) and a given screen $S$ (with window $W_S$), that $\zeta_f$ is \textbf{extended with respect to the screen} $S$. In the event that $\zeta_f$ admits a necessarily unique analytic continuation to $\CC\setminus\Dd_f(\CC)$ for some discrete set $\Dd_f(\CC)$ (e.g. a meromorphic continuation to all of $\CC$), we say that it may be extended with respect to the (formal) screen $S(\tau)\equiv -\infty$. 

Lastly, let $\set{\tau_n}_{n\in\ZZ}$ be a doubly infinite sequence with the following properties:
\begin{align}
    \label{eqn:admissibleHeights}
    \lim_{n\to\infty} \tau_n = \infty, \text{ } \lim_{n\to-\infty}\tau_n =-\infty, \text{ and } 
        \tau_{-n}<0<\tau_n \text{ for all }n\geq1.
\end{align}
For brevity, we will say that $\set{\tau_n}_{n\in\ZZ}$ is a \textbf{sequence of admissible heights}.

Languidity consists of two hypotheses, the first of which concerns power-law or polynomial growth along a sequence of horizontal lines. The horizontal lines each pass from a point on a screen $S$ to some point in a half-plane in which the function is holomorphic.
\begin{Definition}[Languidity Hypothesis \textbf{L1}]
    \label{def:languidL1}
    \index{Languidity!Hypothesis \textbf{L1}}
    We say that $\zeta_f$ satisfies \textbf{languidity hypothesis L1} with exponent $\kappa\in\RR$ and with respect to the screen $S$ if the following hold. 
    \medskip 
    
    Firstly, there must exist some half-plane $\HH_D$ in which $\zeta_f$ is holomorphic and $\zeta_f$ must admit an extension to the screen $S$. Secondly, there must exist a positive constant $C>0$, a constant $\beta>D$, and a sequence of admissible heights $\set{\tau_n}_{n\in\ZZ}$ (in the sense of \eqref{eqn:admissibleHeights}) such that for any $\sigma\in[S(\tau_n),\beta]$,
    \[ |\zeta_f(\sigma+i\tau_n)| \leq C(|\tau_n|+1)^\kappa. \]
\end{Definition}
In other words, on every horizontal contour of the form $I_n = [S(\tau_n) +i\tau_n,\,\beta +i\tau_n ]$, $\zeta_f$ has at most power-law $\kappa$ growth with respect to the magnitude of its imaginary part $\tau_n$ as $|\tau_n|\to\infty$. Alternatively, on each horizontal contour $I_n$, $\zeta_f$ is uniformly bounded by the constant $C_n$, where $C_n=O((|\tau_n|+1)^\kappa)$ as $|n|\to\infty$.

The second hypothesis concerns power-law/polynomial growth along a vertical curve, the screen $S$. The standard (or weak) version of languidity concerns growth along a single curve, and strong languidity concerns growth along a sequence of such curves moving to the left. These curves are precisely the screens of the window(s) to which $\zeta_f$ admits an extension. 
\begin{Definition}[Languidity Hypothesis \textbf{L2}]
    \label{def:languidL2}
    \index{Languidity!Hypothesis \textbf{L2}}
    A function $\zeta_f$ is said to satisfy \textbf{languidity hypothesis L2} with exponent $\kappa\in\RR$ and with respect to the screen $S$ if there exists a positive constant $C>0$ such that for all $\tau \in\RR$ with $|\tau|\geq 1$, $|\zeta_f(S(\tau)+i\tau)| \leq C|\tau|^\kappa$.
\end{Definition}

A \textit{function} which satisfies both hypotheses \textbf{L1} and \textbf{L2} with respect to the same screen $S$ and exponent $\kappa$ is said to be languid, or equivalently to have languid growth. A geometric structure (such as a generalized fractal harp, a relative fractal drum, or a self-similar system) may be called languid if its associated zeta function (respectively, its geometric zeta function, its relative zeta function, or its scaling zeta function) is languid, i.e. it satisfies languidity hypotheses \textbf{L1} and \textbf{L2}. 

\begin{Definition}[(Standard/Weak) Languidity]
    \label{def:languid}
    \index{Languidity}
    We say that $\zeta_f$ is \textbf{languid} with exponent $\kappa\in\RR$ if there exists a screen $S$ such that $\zeta_f$ satisfies both languidity hypotheses \textbf{L1} (Definition~\ref{def:languidL1}) and \textbf{L2} (Definition~\ref{def:languidL2}) with exponent $\kappa$ and with respect to the screen $S$. 
    \medskip 

    Given a particular screen $S$, we say that a function $\zeta_f$ is \textbf{languid with respect to} $S$ (with exponent $\kappa$) if it satisfies these hypotheses (with exponent $\kappa$) with respect to the given screen.
\end{Definition}
We will not have use of this presently, but one may also say that a function $\zeta_f$ is languid if there exists some $\kappa\in\RR$ and some screen $S$ for which $\zeta_f$ is languid with exponent $\kappa$ and with respect to $S$.

\subsubsection{Strong Languidity}

Strong languidity may be seen as a sequence of languidity conditions. Namely, it concerns languidity along \textbf{a sequence of screens} $\set{S_m}_{m\in\NN}$ \textbf{converging to} $-\infty$. To be more precise, let $\sup S$ denote the supremum of the elements of the range of $S$, i.e. $\sup S:=\sup\set{S(\tau)\suchthat\tau\in\RR}$. Then a sequence of screens $\set{S_m}_{m\in\NN}$ is said to converge to $-\infty$ if $\lim_{m\to\infty}\sup{S_m}=-\infty$. To be strongly languid, a given function $\zeta_f$ must necessarily admit an analytic continuation to $\CC$ (excepting for a possible discrete set of singularities). In this work, a strongly languid function will be assumed to admit a meromorphic continuation to $\CC$ and this discrete set is the set of its poles. 

For this sequence $\set{S_m}_{m\in\NN}$ of screens, we will also impose the following. For each screen $S_m$, which is by definition Lipschitz continuous, its \textbf{Lipschitz constant} $K_m$ is the smallest constant such that for all $x,y\in\RR$, we have that $|S_m(x)-S_m(y)|\leq K_m|x-y|$. We say that the sequence $\set{S_m}_{m\in\NN}$ has \textbf{uniformly bounded Lipschitz constants} when there exists a constant $K$ such that for every Lipschitz constant $K_m$, $m\in\NN$, of $S_m$, we have that $K_m\leq K$ for every $m\in\NN$. (In other words, the set of all Lipschitz constants, $\set{K_m}_{m\in\NN}$ is bounded by some finite constant $K$.)

\begin{Definition}[Strong Languidity of a Function]
    \label{def:stronglyLanguid}
    Let $\zeta_f$ be a holomorphic function possessing an analytic continuation (viz. meromorphic continuation) to the whole complex plane except possibly for a discrete set of singular points (viz. the set of poles). Then $\zeta_f$ is said to be \textbf{strongly languid} with exponent $\kappa\in\RR$ if there exists a sequence $\set{S_m}_{m\in\NN}$ of screens converging to $-\infty$ with uniformly bounded Lipschitz constants such that:
    \begin{itemize}
        \item[\textbf{L1}] The function $\zeta_f$ satisfies languidity hypothesis \textbf{L1} with respect to each screen $S_m$ and the (fixed) constant $\kappa$. (Equivalently, it satisfies \textbf{L1} with respect to the formal screen $S(\tau)\equiv -\infty$.)
        \item[\textbf{L2}$'$] There exist positive constants $C$ and $B$ such that for all $\tau\in\RR$ and $m\geq 1$, $\zeta_f$ satisfies 
        \[ |\zeta_f(S_m(\tau)+i\tau)|\leq CB^{|S_m(\tau)|}(|\tau|+1)^\kappa. \]
    \end{itemize}
    
\end{Definition}
Note that the strong languidity condition allows for a prefactor with exponential growth related to $\norm{S_m}_\infty$, but is otherwise analogous to \textbf{L2}. Condition \textbf{L2}$'$ implies \textbf{L2} for each of the screens with finite supremum. Also, without loss of generality, the constant $C$ may be chosen to be the same in both hypotheses.

\subsubsection{Strong Languidity of Self-Similar Systems}

Self-similar systems and their associated scaling zeta functions will play a critical role in the analysis to follow. As it will turn out, a self-similar system is always strongly languid in the sense that its associated scaling zeta function $\zeta_\Phi$ is strongly languid with exponent $\kappa=0$. We first state this for scaling operators having positive, integral multiplicities and scaling ratios in $(0,1)$.
\begin{Proposition}[Languidity of Scaling Zeta Functions]
    \label{prop:languidSZF}
    \index{Languidity!of scaling zeta functions}
    Let $L:=\sum_{i=1}^m a_i M_{\lambda_i}$ be a scaling operator with distinct scaling ratios $\lambda_i\in(0,1)$ and positive integral multiplicities $a_i$. Then its associated scaling zeta function $\zeta_L$ is strongly languid with exponent $\kappa=0$ as in Definition~\ref{def:stronglyLanguid}. 
\end{Proposition}

\begin{proof}
    This result is a corollary of \cite[Theorem~3.26]{LapvFr13_FGCD} and the discussion in \cite[Section~6.4]{LapvFr13_FGCD}, as $\zeta_L$ is essentially the same as that of a self-similar fractal string (or harp). (Let the length and gap parameters be one.) Note in particular the explicit estimate in \cite[Equation~6.36]{LapvFr13_FGCD} used to establish the uniform bounds on screens with arbitrarily small real parts.
\end{proof}

For a self-similar system $\Phi$, its associated scaling operator $L_\Phi$ will always have scaling ratios in $(0,1)$ and positive, integral multiplicities. Thus, it is an immediate corollary of Proposition~\ref{prop:languidSZF} that the scaling zeta function $\zeta_\Phi=\zeta_{L_\Phi}$ is strongly languid. Extending the convention that a geometric structure itself may be dubbed languid when its associated zeta function is languid, we may say that self-similar systems are languid.  
\begin{Corollary}[Languiditiy of Self-Similar Systems]
    \label{cor:languidSystem}
    \index{Languidity!of self-similar systems}
    Let $\Phi$ be a self-similar system on $\RR^\dimension$. Then $\Phi$ is strongly languid (with exponent $\kappa=0$) in the sense that its associated scaling zeta function $\zeta_\Phi$ is strongly languid (with exponent $\kappa=0$) as in Definition~\ref{def:stronglyLanguid}.
\end{Corollary}

\subsubsection{Joint Languidity}

In what follows, we will require the existence of screens $S$ for which two functions $\zeta_f$ and $\zeta_g$ are both languid with respect to the \textit{same} screen $S$. We shall call this \textit{joint languidity}.

\begin{Definition}[Joint Languidity]
    \label{def:jointlyLanguid}
    Let $\zeta_f$ and $\zeta_g$ be functions which are both holomorphic in a half-plane $\HH_{\sigma_1}$ and both admitting a necessarily unique analytic continuation to a half-plane $\HH_{\sigma_0}$, where $\sigma_0<\sigma_1$, except perhaps for a discrete set $\Dd_{f,g}(\HH_{\sigma_0})$ at which either $\zeta_f$ or $\zeta_f$ may be singular.     
\medskip 

    We say that $\zeta_f$ and $\zeta_g$ are \textbf{jointly languid} with exponent $\kappa\in\RR$ if there exists a screen $S$ (contained in $\HH_{\sigma_0}\setminus\Dd_{f,g}(\HH_{\sigma_0})$) such that the following hold. 
    \begin{itemize}
        \item There is a sequence of admissible heights for which $\zeta_f$ and $\zeta_g$ both satisfy languidity hypothesis \textbf{L1} with exponent $\kappa$ and with respect to $S$ using this sequence.
        \item Both $\zeta_f$ and $\zeta_g$ satisfy languidity hypothesis \textbf{L2} with exponent $\kappa$ and with respect to $S$. 
    \end{itemize}
\end{Definition}
In this work, the functions we consider will have a necessarily unique meromorphic continuations to $\HH_{\sigma_0}$ and $\Dd_{f,g}(\HH_{\sigma_0})$ is the union of the sets of poles of $\zeta_f$ and $\zeta_g$ in $\HH_{\sigma_0}$. In the event that $\zeta_g$ is holomorphic in $\HH_{\sigma_0}$, then $\Dd_{f,g}(\HH_{\sigma_0})=\Dd_f(\HH_{\sigma_0})$ consists only of poles of $\zeta_f$.

\section{Solutions of Scaling Functional Equations}
\label{sec:SFE}

Scaling operators, scaling zeta functions, and scaling functional equations were studied in \cite{Hof25} for the purpose of analyzing tube functions and complex dimensions of (osculant) relative fractal drums of self-similar fractals. The notion of scaling zeta functions in an earlier form, during the development of what would become fractal zeta functions and complex dimensions, also appears in the work of the second author with E. Pearse and S. Winter \cite{LP06,LP10,LPW11}. See also \cite{HL06,LapvFr13_FGCD,LRZ17_FZF}. This framework is essentially a multiplicative analogue of the renewal theory developed by Feller in the context of queuing problems in probability theory \cite{Fel50}, which may be regarded as the earliest incarnation of solving additive functional equations, where a function is expressed in terms of shifts of its input. 

These methods have been applied with great success in the realm of fractal geometry, including but not limited to the work of Strichartz on self-similar measures \cite{Str1,Str2,Str3}, the work of the second author on the vibrations of fractal drums \cite{Lap93_Dundee}, and the work of Kigami and the second author on the Weyl problem for Laplacians on self-similar sets \cite{KL93}. Additionally, in the case of von Koch fractals, the renewal theorem has ben used to study the heat content of the von Koch snowflake \cite{FLV95a} as well as of generalized von Koch fractals \cite{vdBGil98,vdBHol99,vdB00_generalGKF,vdB00_squareGKF}. Furthermore, the notion of a (scaling) functional equation was employed by Deniz, Ko\c cak, \"Ozdemir, and \"Ureyen in \cite{DKOU15} to provide a new proof of tube formulae for self-similar sprays (originally established in the work of Lapidus and Pearse \cite{LP10,LPW11}), a class of disjoint self-similar fractals. 

\subsection{Scaling Functional Equations}
\label{ssc:SFEs}

\subsubsection{Scaling Operators}

First, we define scaling operators which act on $C^0(\RR^+)$, where $\RR^+:=(0,\infty)$. A \textbf{pure scaling operator} $M_\lambda$, where $\lambda\in\RR^+$, shall act by precomposition of the function with scaling, namely \begin{equation}
    \label{eqn:defPureScalingOp}
    M_\lambda[f](x) := f(x/\lambda).
\end{equation}
This convention of division by the scaling factor shall be convenient for our applications. A general \textbf{scaling operator} $L$ is any finite linear combination of such pure scaling operators with real coefficients, i.e. $L=\sum_{k=1}^K a_k M_{\lambda_k}$, where $\lambda_k\in\RR^+$, and $a_k\in\RR$ for each $k=1,...,K$. Later, we will add the constraint that the multiplicities $a_k$ be positive and integral.

\subsubsection{Scaling Zeta Functions and Similarity Dimensions}

Given such a scaling operator $L$, we define a \textit{scaling zeta function} associated to $L$ as in \cite{Hof25}. Given a scaling operator $L=\sum_{k=1}^K a_k M_{\lambda_k}$, the \textbf{scaling zeta function} $\zeta_L$ associated to $L$ is the analytic continuation in $\CC$ of the function defined by 
\[ \zeta_L(s) := \cfrac1{1-\sum_{k=1}^K a_k \lambda_k^s} \]
for all $s\in\CC\setminus\Dd_L$, where $\Dd_L$ the (discrete) set of singularities of $\zeta_L$. This function plays a key role in describing solutions to scaling functional equations, owing to its relation to the Mellin transform of such functions and the role of its singularities. For example, the complex dimensions of an osculant, self-similar fractal drum (with appropriate estimates) is a subset of the poles of such an associated scaling zeta function (viz. \cite[Theorem 5.5]{Hof25}). 

Given a self-similar system $\Phi$, we can define a scaling operator $L_\Phi$ associated to $\Phi$ (and thus also a scaling zeta function associated to $\Phi$). Letting $\lambda_\ph$ denote the scaling ratio of a similitude $\ph\in\Phi$, we define the \textbf{scaling operator} $L_\Phi$ \textbf{of the self-similar system} $\Phi$ by 
\begin{equation}
    \label{eqn:defSystemScalingOp}
    L_\Phi := \sum_{\ph\in\Phi} M_{\lambda_\ph}.
\end{equation}
The scaling zeta function associated to $\Phi$ is simply the scaling zeta function associated to this operator $L_\Phi$, i.e. $\zeta_{L_\Phi}$, which we will denote by $\zeta_\Phi$. We note that the multiplicities of $L_\Phi$ are positive and integral, and that the scaling ratios $\lambda_\ph$ lie in $(0,1)$ since the mappings in $\Phi$ are nontrivial and contractive. 

For such scaling zeta functions, we can give a precise bounds for the locations of their poles. For this purpose, we define \textit{upper and lower similarity dimensions} of a self-similar system $\Phi$ which will define a vertical strip in the complex plane containing any such singularity. 

Firstly, an upper similarity dimension is an extension of the definition of the similarity dimension of a set, defined by means of a Moran scaling equation.  
\begin{Definition}[(Upper) Similarity Dimension]
    \label{def:upperSimDim}
    Let $\Phi$ be a self-similar system and let $\set{\lambda_\ph}_{\ph\in\Phi}$ denote the set of scaling ratios of the similitudes $\ph\in\Phi$. Then there is a unique real solution $D$ to Moran's equation,
    \begin{equation}
        \label{eqn:Moran2}
        1 = \sum_{\ph\in\Phi} \lambda_\ph^D. 
    \end{equation}
    This number $D=\simdim(\Phi)$ is called the \textbf{(upper) similarity dimension} of $\Phi$. 
\end{Definition}
Under the imposition of the open set condition (Definition~\ref{def:OSC}), it follows from Moran's theorem \cite{Mor46} that the similarity dimension for a self-similar system of mappings in $\RR^\dimension$ coincides with the Minkowski dimension (and with the Hausdorff dimension) of the invariant set of $\Phi$. This invariant set is called the \textit{self-similar set} associated with $\Phi$. Consequently, it follows that $D=\simdim(\Phi)$ satisfies the bounds $0\leq D\leq\dimension$. Further, supposing that $|\Phi|\geq2$, it follows that $D>0$. 

We will also denote $\simdim(\Phi)=\overline{\dim}_S(\Phi)$ when we wish to explicitly differentiate the upper similarity dimension from its lower counterpart (to be defined presently). Under the imposition of the open set condition (Definition~\ref{def:OSC}), the (upper) similarity dimension of a self-similar system is exactly the similarity dimension of its invariant set by Moran's theorem \cite{Mor46}. It is called an \textit{upper} similarity dimension since it equals the largest real part of the poles of $\zeta_\Phi$, with all its other poles having equal or smaller real parts, which follows by \cite[Theorem 3.6]{LapvFr13_FGCD}. 

Next, we define the \textit{lower similarity dimension} of a self-similar system. The (upper) similarity dimension is a pole of $\zeta_\Phi$ with largest real part (per \cite[Theorem 3.6]{LapvFr13_FGCD}), so the lower similarity dimension will be the infimum of these poles. 
\begin{Definition}[Lower Similarity Dimension]
    \label{def:lowerSimDim}
    Let $\Phi$ be a self-similar system and let $\zeta_\Phi$ be its associated scaling zeta function (i.e. $\zeta_\Phi(s) = (1-\sum_{\ph\in\Phi}\lambda_\ph^s)^{-1}$, where $\lambda_\ph$ is the scaling ratio of $\ph$). Then the \textbf{lower similarity dimension} of $\Phi$, denoted by $\lowersimdim(\Phi)$, is defined by 
    \begin{equation}
        \label{eqn:defLowerSimDim}
        \lowersimdim(\Phi):=\inf\set{\Re(\omega)\suchthat \omega\in\Dd_\Phi(\CC)},
    \end{equation}
    where $\Dd_\Phi(\CC)$ denotes the set of poles of $\zeta_\Phi$ in $\CC$ (given by solutions of the equation $\zeta_\Phi(\omega)^{-1}=0$).  
\end{Definition}
If $\Dd_\Phi(\CC)$ were to contain points with arbitrarily small real parts, then we would have that $\lowersimdim(\Phi)=-\infty$. However, we have the following lower bound for $\lowersimdim(\Phi)$, obtained in \cite{LapvFr13_FGCD}. Given a self-similar system $\Phi$, let $\set{r_k}_{k=1}^M$ denote the \textit{distinct} scaling ratios of the mappings in $\Phi$ with corresponding multiplicities $m_k$. Suppose that the scaling ratios are ordered by size, i.e. $r_1> r_2> ...> r_M$. Then there is a unique real solution $D_\ell$ to the equation 
\begin{equation}
    \label{eqn:defLowerDimBound}
    \frac1{m_M}(r_M^{-1})^{D_\ell}+\sum_{k=1}^{M-1}\frac{m_k}{m_M}\left(\frac{r_k}{r_M}\right)^{D_\ell} = 1,
\end{equation}
and by \cite[Theorem 3.6]{LapvFr13_FGCD}, we have that $D_\ell\leq \lowersimdim(\Phi)$. We summarize these properties of the similarity dimensions as Proposition~\ref{prop:simDimBounds}, which is an immediate corollary of \cite[Theorem 3.6]{LapvFr13_FGCD}, noting that $\zeta_\Phi$ is the reciprocal of a Dirichlet polynomial.
\begin{Proposition}[Similarity Dimension Bounds]
    \label{prop:simDimBounds}
    Let $\Phi$ be a self-similar system and let $\zeta_\Phi$ be is associated scaling zeta function. Let $D_\ell$ be the unique real solution to $\eqref{eqn:defLowerDimBound}$ corresponding to $\Phi$. Then for any $\omega\in\Dd_\Phi(\CC)$, the set of poles of $\zeta_\Phi$ in $\CC$, we have that 
    \begin{equation}
        \label{eqn:simdimBounds}
        -\infty<D_\ell \leq \lowersimdim(\Phi) \leq \Re(\omega) \leq \simdim(\Phi). 
    \end{equation}
    If $|\Phi|\geq2$, then $\simdim(\Phi)\geq D_\ell>0$. Furthermore, under the imposition of the open set condition (Definition~\ref{def:OSC}), we have that $\simdim(\Phi)\leq\dimension$. 
\end{Proposition}
Note that the upper bound in \eqref{eqn:simdimBounds} is achieved, since $D=\simdim(\Phi)$ is a pole of $\zeta_\Phi$, but that the lower bound $D_\ell\leq\lowersimdim(\Phi)$ need not be achieved in general. A sufficient condition for this lower similarity dimension to be the infimum of the real parts of the poles of $\zeta_\Phi$ is the full rank nonlattice case (called the generic nonlattice case in \cite{LapvFr13_FGCD}), i.e. when the multiplicative group $G=\prod_{k=1}^M r_k^\ZZ$ has rank $M$. Note that this occurs for Lebesgue almost every tuple of distinct scaling ratios $(r_1,...,r_M)\in(0,1)^M$.

\subsubsection{Scaling Functional Equations}

Given functions $f,R\in C^0(\RR^+)$ and a scaling operator $L$, a scaling functional equation for $f$ with remainder $R$ is an identity of the form $f=L[f]+R$. If $R\equiv 0$, the scaling functional equation is said to be exact, and if not then it is said to be an approximate scaling functional equation. Given such a scaling functional equation, it can be solved directly by means of truncated Mellin transforms \cite{Hof25} or it can be converted into an additive functional equation through changes of variables and then solved by the renewal theorem of Feller \cite{Fel50}. 

A scaling functional equation can be established for a quantity $f_A(t)=f(t;A)$, where $t\in\RR^+$ and $A$ is a geometric object such as a relative fractal drum or a subset of Euclidean space, through two elements: a \textit{decomposition induced by a self-similar system} $\Phi$ and a \textit{scaling law} satisfied by $f$. To the former, a decomposition of $f$ induced by $\Phi$ is an identity of the form 
\[ f(t;A) = \sum_{\ph\in\Phi} f(t;\ph[A]) + R(t), \]
where $R$ is some error quantity called a \textit{decomposition remainder}. A scaling law for a family of functions $\set{f_A}_{A\in\Aa}$ is an identity of the form $f(t;\ph[A])=f(t/\lambda_\ph^\alpha;A)$, where $\alpha$ is fixed (for tube functions $\alpha=1$ and for heat content, per Corollary~\ref{cor:heatScalingLaw}, $\alpha=2$) and $\lambda_\ph$ is the scaling ratio of $\ph$. Note that we assume for each $A\in\Aa$, so too is $\ph[A]\in\Aa$. Letting $L_\Phi^\alpha=\sum_{\ph\in\Phi}M_{\lambda_\ph^\alpha}$, it follows that $f_A$ satisfies the scaling functional equation $f_A=L_\Phi^\alpha[f_A]+R$. 

More precisely, let $\Phi$ be a self-similar system on $\RR^\dimension$ (or, more generally, on a complete metric space) and let $\Aa$ be a collection of objects with the following closure property: if $A\in\Aa$, then $\ph[A]$ is well-defined and $\ph[A]\in \Aa$ for each $\ph\in\Phi$. For brevity, we will say that $\Aa$ is $\Phi$-closed. Next, let $\set{f_A}_{A\in\Aa}$ be a family of functions $f_A:I\to G$ for some set $I$ and an additive semigroup $(G,+)$. We will identify $f$ with the family and write $f(t;A)=f_A(t)$ for $t\in I$ and $A\in\Aa$.

Let $X\subset\RR^\dimension$ be a self-similar set, let $\Phi$ be a self-similar system having $X$ as its invariant set, and let $f_A(t)=f(t;A)$ be a function defined on $\RR$ and certain subsets $A\subseteq \RR^\dimension$. Here, we will only require that if $f_A$ is defined on $A$, then $f_{\ph[A]}$ is defined for any image $\ph[A]$ of $A$ under the maps $\ph\in\Phi$. 

\begin{Definition}[Induced Decomposion]
    \label{def:inducedDecomp}
    \index{Decomposition!Induced decomposion}
    Let $\Phi$ be a self-similar system on $\RR^\dimension$, let $\Aa$ be a $\Phi$-closed collection of objects, and let $\Ff=\set{f_A}_{A\in\Aa}$ be a family of functions $f_A:I\to G$ for some set $I$ and an additive semigroup $(G,+)$. Let also $R:I\to G$ be a function. 
    \medskip 

    A self-similar system $\Phi$ is said to \textbf{induce a decomposition} of $f$ on $I$ with remainder $R$ if for all $t\in I$ and $f_A\in\Ff$,
    \begin{equation}
        \label{eqn:inducedDecomp}
        f(t;A) = \sum_{\ph\in\Phi} f(t;\ph[A]) + R(t).
    \end{equation}
\end{Definition}
In this work, $I$ is always a subset of $\RR$ and the codomain is the field $\RR$ with its standard addition operation. Further, $A$ is some geometric structure, either a bounded subset of $\RR^\dimension$ or, more generally, a relative fractal drum $(X,\Omega)$ in $\RR^\dimension$. If $A\subset\RR^\dimension$, $\ph[A]$ is its pointwise image. For a relative fractal drum, we define $\ph[(X,\Omega)]:=(\ph[X],\ph[\Omega])$, noting that this also defines a relative fractal drum. 

An induced decomposition will be paired with a scaling law satisfied by the function $f(t;A)$. A linear scaling (invariance) law would be of the form $f(\lambda t;\lambda A)=f(t;A)$, for each $\lambda\in\RR^+$. However, depending on the context, the real and set parameters may scale differently. In the case of heat contents, the function will obey a quadratic scaling law: $f(\lambda^2\,t;\lambda A)=f(t;A)$. We provide a general notion for defining the scaling relation which allows one to specify the parameter $\alpha$ dictating this relationship. 
\begin{Definition}[$\alpha$-Scaling Law]
    \label{def:scalingLaw}
    \index{Scaling law}
    Let $\Ff=\set{f(\cdot;A)}_{A\in\Aa}$ be a family of functions $f_A:\RR^+\to\RR$ where $\Aa$ is a $\set{\ph}$-closed collection of objects for any similitude $\ph$ of $\RR^\dimension$. 
    \medskip 
    
    Given $\alpha\in\RR$ (and in this work, $\alpha>0$), we say that $\Ff$ has an $\alpha$\textbf{-scaling law} if for any similitude $\ph$ with scaling ratio $\lambda>0$ and for any $f_A\in\Ff$, $f(t;\ph[A])=f(t/\lambda^\alpha;A)$. 
\end{Definition}

We conclude this section with the main idea: a scaling law and an induced decomposition for a family of functions together lead to the scaling functional equation of a single function.
\begin{Proposition}[Scaling Functional Equation (SFE)]
    \label{prop:inducedSFE}
    \index{Scaling functional equation (SFE)}
    Let $\Ff=\set{f_A}_{A\in\Aa}$ be a family of functions on $I$ satisfying an $\alpha$-scaling law. Suppose that $\Phi$ is a self-similar system which induces a decomposition of $\Ff$ on $I$ with remainder term $R$. Define the scaling operator $L_\Phi^\alpha$ by $L_\Phi^\alpha:=\sum_{\ph\in\Phi}M_{\lambda_\ph^\alpha}$. 
    \medskip 

    Then each function $f_A\in\Ff$ in the family satisfies the \textbf{scaling function equation} $f_A=L_\Phi^\alpha[f_A]+R$ on $I$ with error term $R$, which is to say that for all $t\in I$,
    \[ f_A(t)=L_\Phi^\alpha[f_A](t)+R(t). \]
    We say that $\Phi$ induces a scaling functional equation for $f_A$ with operator $L_\Phi^\alpha$.
\end{Proposition}
Proposition~\ref{prop:inducedSFE} is, in essence, an untangling of definitions meant to lead to a single key concept: \textit{a self-similar system induces a scaling functional equation} for functions with scaling laws. It is, however, a consequence of the definitions that any function in this family satisfies an induced scaling functional equation. The proof is merely to apply the $\alpha$-scaling law to each term of an induced decomposition and then use the definition of the scaling operator $L_\Phi^\alpha$ to rewrite the decomposition in terms of a single function.

\subsubsection{Scaling Functional Equations for Heat Content}

Proposition~\ref{prop:heatContentScaling} implies that the total heat content $\heatContent(t)=E(t;\Region)$, viewed as a family of functions, satisfies a $2$-scaling law in the sense of Definition~\ref{def:scalingLaw}, provided a fixed set of boundary conditions for which Proposition~\ref{prop:heatContentScaling} holds. We state this in our specific case, i.e. for Problem~\ref{prob:specificHeatProblem}.

\begin{Corollary}[$2$-Scaling Law of Heat Content]
    \label{cor:heatScalingLaw}
    \index{Scaling law!of heat content}
    Given any bounded open set $A\subset\RRN$ and any similitude $\ph$ on $\RRN$, let $E_A(t)$ be the total heat content for the PWB solution $u_\Region$ of Problem~\ref{prob:specificHeatProblem} for each of the sets $\Region=A$ and $\Region=\ph^{\circ k}[A]$, $k\in\NN$. Then the family of normalized functions $t^{-\dimension/2}E_A(t)$ (with respect to the $\Phi$-closed set $\Aa=\set{\ph^{\circ k}[A]\suchthat k\in\NN_0}$) satisfies a $2$-scaling law in the sense of Definition~\ref{def:scalingLaw}, viz. for every $A\in\Aa$ and $t>0$,
    \[ (t/\lambda^2)^{-\dimension/2}E_A(t/\lambda^2) = t^{-\dimension/2}E_{\ph[A]}(t). \]
\end{Corollary}
\begin{proof}
    The constant, nonnegative boundary conditions of Problem~\ref{prob:specificHeatProblem} ensure that $u_\Region$ is unique and they are clearly scale invariant and parabolically scale invariant in their respective variables. Because $\ph$ is a similitude, every set $\ph^{\circ k}[A]$ is bounded---using the bound $\lambda^kC$, where $C$ is a bound such that $A$ is contained in a ball of radius $C$---and open, which follows because $\ph$ is injective and its inverse is also a similitude, hence continuous. 

    Applying Proposition~\ref{prop:heatContentScaling} to an arbitrary set $A\in\Aa$, we have that 
    \begin{align*}
        t^{-\dimension/2}E_{\ph[A]}(t) = t^{-\dimension/2}\lambda^\dimension E_{A}(t/\lambda^2) = (t/\lambda^2)^{-\dimension/2}E_A(t/\lambda^2).
    \end{align*}
\end{proof}

Suppose now that our region $\Region$ has a self-similar boundary $\Regionbd$ corresponding to a self-similar system $\Phi$ and that $\Omega$ is an osculating set (see Definition~\ref{def:oscRFD}) for $\Phi$. In order to use the scaling property to obtain a scaling functional equation, we will need an induced decomposition in the sense of Definition~\ref{def:inducedDecomp}. For later use, we define the remainder which occurs as the difference of the heat content and the sum of scaled copies. Given a suitable estimate of this quantity, we will be able to obtain explicit formulae for the heat content.

\begin{Definition}[Decomposition Remainder]
    \label{def:heatDecompRem}
    \index{Decomposition!Decomposition remainder}
    Let $u_\Region$ be the PWB solution to Problem~\ref{prob:specificHeatProblem} on the bounded open set $\Region\subset\RRN$. Suppose that $\Regionbd$ is the attractor of a self-similar system $\Phi$ and that the relative fractal drum $(\Regionbd,\Region)$ is osculant. 
    \medskip 

    We define the \textbf{decomposition remainder} of $\heatContent$ to be the quantity 
    \begin{equation}
        \label{eqn:defDecompRem}
        \decompRem(t) := \heatContent(t) - \sum_{\ph\in\Phi} E_{\ph[\Region]}(t).
    \end{equation}
    The normalized remainder is the quantity $R(t):=t^{-\dimension/2}\decompRem(t)$ for $t>0$.    
\end{Definition}

Note that in the case of Problem~\ref{prob:specificHeatProblem}, we can deduce that $\decompRem$ is bounded for all $t\geq 0$ as a corollary of Proposition~\ref{prop:heatContentBounded}. This will provide an upper bound for the abscissa of absolute convergence of its truncated Mellin transform, $\Mellin^\delta[t^{-\dimension}\decompRem(t)]$, as by Lemma~\ref{lem:MellinHolo}, we may deduce that this transform, as a function of the complex variable $s$, is holomorphic in the right half plane $\HH_{\dimension/2}$. This helps establish some technical preliminaries for later proofs, allowing us to concern ourselves solely with asymptotic estimates of the form $\decompRem(t)=O(t^{\rempow/2})$ as $t\to0^+$, for some $\sigmaRem\in\RR$. When $\sigmaRem$ is the smallest such parameter, it corresponds to $\sigma_0=2\sigma_{ac}$, where $\sigma_{ac}$ is the abscissa of absolute convergence of the normalized remainder. (Division by two is for normalization owing to the quadratic nature of the heat content scaling law.)

The most precise results occur when there is no error term $\decompRem$, viz. when $\decompRem\equiv 0$. In this case, the results of this work apply where estimates for the remainder terms may be chosen of the form $O(t^{n})$ as $t\to0^+$, for any $n>0$. This occurs when the set $X$ partitions into disjoint, self-similar copies (up to sets of measure zero). The novelty of this method, though, lies in its ability to handle cases where this is not true, but with explicit estimates for the degree to which the decomposition is not exact, such as in the case of generalized von Koch fractals. The key part of using this scaling function approach is obtaining good estimates for the remainder term (viz. for small values of $\sigmaRem$), as the expansion will be explicit up to an order determined by these estimates. For the reader wishing to establish such functional equations for a particular class of self-similar sets, when studying heat content, see \cite{FLV95a,vdB00_squareGKF}.

\subsection{Admissible Remainders}
\label{ssc:admissibleRem}

\subsubsection{Admissibility}

As part of our analysis of scaling functional equations, we will be required to show that certain functions are languid (see Section~\ref{ssc:languidity}). In so doing, we will need to constrain the types of remainders which can appear in order for these results to be established. In this section, we introduce terminology for such \textit{admissible remainders} relative to a given self-similar system as well as some sufficient conditions for admissibility. 

The admissibility of a remainder is essentially tied to a need for \textit{joint languidity}. Two functions, the Mellin transform of the remainder $\zeta_R$ and the scaling zeta function $\zeta_\Phi$, must share the same screen and the same horizontal contours on which they are both uniformly bounded. They must both be languid (with exponent $\kappa$) on a shared screen. It is for this reason that we introduced the notion of \textit{joint languidity} (Definition~\ref{def:jointlyLanguid}).

\begin{Definition}[Admissible Remainders and Screens]
    \label{def:admissibleRem}
    \index{Scaling functional equation (SFE)!Admissible remainders}
    Let $R$ be a function such that its truncated Mellin transform $\zeta_R(s;\delta)=\Mellin^\delta[R](s)$ is holomorphic in the half-plane $\HH_{\sigmaRem}$.\footnote{This occurs, for instance, if $R$ is continuous on $(0,\delta]$ and $R(t)=O(t^{-\sigmaRem})$ as $t\to0^+$ (per Lemma~\ref{lem:MellinHolo}).}
    \medskip

    Let $L$ be a scaling operator with associated scaling zeta function $\zeta_L$. We say that $R$ is an \textbf{admissible remainder} for $L$ if there exists a screen $S$ such that $\zeta_L$ and $\zeta_R$ are jointly languid with exponent $\kappa=0$ with respect to $S$ in the sense of Definition~\ref{def:jointlyLanguid}. Any such screen $S$ is called an \textbf{admissible screen} for $R$ and $L$ (or their respective zeta functions). 
    \medskip 

    Given a self-similar system $\Phi$, we say that $R$ is an admissible remainder for $\Phi$ if it is admissible for the associated operator $L_\Phi$ (and similarly for admissible screens). 
\end{Definition}

\subsubsection{Lower Dimension Criterion}

The easiest way to obtain a nontrivial admissible screen occurs when $\zeta_R$ is holomorphic in a half-plane to the left of the vertical strip containing all of the poles of $\zeta_L$. This ensures that a screen may be chosen to the left of any pole of $\zeta_L$ but also in a region where $\zeta_R$ is holomorphic. Supposing that $L=L_\Phi$ is the scaling operator associated to a self-similar system, we have that this strip is bounded explicitly by the similarity dimensions of $\Phi$: if $\omega\in\Dd_\Phi$, then $\lowersimdim(\Phi)\leq\Re(\omega)\leq\simdim(\Phi)$. So, the first criterion is to ensure an estimate for $R$ that guarantees that the abscissa of absolute convergence of $\zeta_R$ is strictly smaller than the explicit bound $D_\ell$ for the lower similarity dimension, which is used to estimate $\zeta_\Phi$.

\begin{Theorem}[Lower Dimension Criterion for Admissibility]
    \label{thm:lowerDimAdmissibility}
    Let $\Phi$ be a self-similar system and let $D_\ell\leq\underline{\dim}_S(\Phi)$ be the lower similarity dimension bound as in Proposition~\ref{prop:simDimBounds}. For any $R\in C^0(\RR^+)$, if there exists $\sigmaRem<D_\ell$ such that as $t\to0^+$, $R(t)=O(t^{-\sigmaRem})$, then $R$ is an admissible remainder for $\Phi$ and any screen of the form $S_\e(\tau)\equiv \sigmaRem+\e$, with $0<\e<D_\ell-\sigmaRem$, is admissible. 
\end{Theorem}
\begin{proof}
    Let $D_\ell\leq\underline{\dim}_S(\Phi)$ be the lower similarity dimension bound for $\Phi$ and let $S_\e(\tau)\equiv \sigmaRem+\e$ be a constant screen in $\HH_{\sigmaRem}$, where $\e>0$ is such that $\sigmaRem+\e<D_\ell$. We will show that $\zeta_R$ and $\zeta_\Phi$ are jointly languid with exponent $\kappa=0$ on $S_\e$.  

    To start, we have that $\zeta_\Phi$ is strongly languid with exponent $\kappa=0$ by Corollary~\ref{cor:languidSystem}. This guarantees that for any $\sigma_-$, there exists a sequence of admissible heights $\set{T_n}_{n\in\ZZ}$ (i.e. with $T_{-n}\to-\infty$ and $T_n\to\infty$ as $n\to\infty$ and with $T_n>0>T_{-n}$ for each $n\geq 1$) and some screen $S_-$ with $\sup S_- < \sigma_-$ such that $\zeta_\Phi$ is uniformly bounded on horizontal contours of the form $[S_-(T_n)+i T_n,\sigma_++i T_n]$, where $\sigma_+>\max(\sigma_-,\dim_S(\Phi))$. In particular, $\zeta_\Phi$ will be uniformly bounded on the subsets $I_n:=[\sigmaRem+\e+iT_n,\sigma_++iT_n]$. This establishes languidity hypothesis \textbf{L1} for $\zeta_\Phi$. 

    Next, we note that the estimate on $R$ implies that the function $\zeta_R$ is holomorphic in the open right half-plane $\HH_{\sigmaRem}$ by Lemma~\ref{lem:MellinHolo}. By Corollary~\ref{cor:MellinBounds}, we also have that $\zeta_R$ is bounded on the screen $S_\e(\tau)\equiv \sigmaRem+\e>\sigmaRem$ as well as on any vertical strip of the form $\HH_{\sigmaRem+\e}^{\sigma_+}$. This establishes both languidity hypotheses \textbf{L1} and \textbf{L2} with exponent $\kappa=0$ and on the screen $S_\e$ with respect to the intervals $I_n\subset \HH_{\sigmaRem+\e}^{\sigma_+}$, with shared sequence of admissible heights.
    
    It remains to show that $\zeta_\Phi$ is bounded on the screen $S_\e$. To this end, we enforce that $\sigmaRem+\e<D_\ell$, in which case for any $s$ on the screen, $\Re(s)=\sigmaRem+\e<D_\ell$. Let $\set{r_k}_{k=1}^m$ denote the set of unique scaling ratios of $\Phi$, arranged in decreasing order $r_1\geq ...\geq r_M$, and let $m_k$ denote the multiplicity of $r_k$. Define the function 
    \begin{equation}
        p(t) = \frac1{m_M}(r_M^{-1})^t+\sum_{k=1}^{M-1}\frac{m_k}{m_M}\left(\frac{r_k}{r_M}\right)^t,
    \end{equation}
    which is readily seen to be strictly increasing with range $(0,\infty)$. Note that $p(D_\ell)=1$ by definition. We will obtain a bound for $\zeta_\Phi(s)$ when $\Re(s)=\sigmaRem+\e<D_\ell$ using this function. 
    
    To that end, let $f(s)=\zeta_\Phi(s)^{-1}=1-\sum_{k=1}^M m_k r_k^s$ be the denominator and let $\sigma=\Re(s)$. Then 
    \begin{align*}
        |r_M^{-s}/m_M f(s)+1| &= \left| r_M^{-s}/m_M - \sum_{k=1}^{M-1} m_k/m_M(r_k/r_M)^s \right| \\
            &\leq p(\sigma) < p(D_\ell) = 1.
    \end{align*}
    By the reverse triangle inequality, we have that 
    \begin{align*}
        1 > p(\sigma) \geq \left| |r_M^{-s}/m_M f(s)| - 1 \right| = \left| r_M^{-\sigma}/m_M|f(s)| - 1 \right|,
    \end{align*}
    which may be rewritten as $| |f(s)|-m_M r^{\sigma} | \leq m_M r^{\sigma}p(\sigma)<m_M r_M^\sigma$. Once again using the reverse triangle inequality, we find that the lower bound for $|f|$ is furnished by
    \begin{align*}
        |f(s)| \geq m_M r_M^\sigma - | |f(s)|-m_M r^{\sigma} | \geq m_M r_M^\sigma - m_M r^{\sigma}p(\sigma) >0.
    \end{align*}
    It follows that when $\Re(s)=\sigmaRem+\e$ is fixed, $|\zeta_\Phi(s)|\leq C_\Phi$, with $C_\Phi = (m_M r_M^\sigma - m_M r^{\sigmaRem+\e}p(\sigmaRem+\e))^{-1}$. This establishes hypothesis \textbf{L2} for $\zeta_\Phi$, and thus $\zeta_\Phi$ and $\zeta_R$ are jointly languid on the screen $S_\e$ (with exponent $\kappa=0$).
\end{proof}

In order to accommodate $\alpha$-scaling laws with $\alpha>0$ (in the sense of Definition~\ref{def:scalingLaw}), we will consider scaling operators of the form $L_\Phi^\alpha:=\sum_{\ph\in\Phi}M_{\lambda_\ph^\alpha}$, where $\Phi$ is a self-similar system. In this case, the zeta function associated to $L_\Phi^\alpha$ is exactly the function $\zeta_\Phi(\alpha s)$, where $\zeta_\Phi(s)$ is the scaling zeta function associated to $\Phi$ itself. 

This change of variables amounts to a simple rescaling of the bound with respect to the lower similarity dimension of $\Phi$ a remainder must satisfy in order to be admissible by Theorem~\ref{thm:lowerDimAdmissibility}. Namely, when $R(t)=O(t^{-\sigma_R})$ as $t\to0^+$ for some $\sigma_R<D_\ell/\alpha\leq\lowersimdim(\Phi)/\alpha$, we have that $\zeta_\Phi(\alpha s)$ has poles in the strip $\lowersimdim(\Phi)/\alpha \leq \Re(s) \leq \simdim(\Phi)/\alpha$ and $\sigma_R$ lies strictly to the left of this bound. Equivalently, if we state the estimate for $R$ in the form $R(t)=O(t^{-\sigmaRem/\alpha})$ (as $t\to0^+$), where $\sigmaRem/\alpha=\sigma_R$, then $\sigmaRem$ must satisfy the bound $\sigmaRem<D_\ell$ directly. The estimates for $\zeta_\Phi$ apply to its rescaled analogue, and thus the proof of Theorem~\ref{thm:lowerDimAdmissibility} yields the following corollary.

\begin{Corollary}[Rescaled Lower Dimension Criterion]
    \label{cor:lowerDimAdmissibilityRescaled}
    Let $\Phi$ be a self-similar system and let $D_\ell\leq\underline{\dim}_S(\Phi)$ be the lower similarity dimension bound as in Proposition~\ref{prop:simDimBounds}. Given $\alpha>0$, define the scaling operator $L_\Phi^\alpha:=\sum_{\ph\in\Phi}M_{\lambda_\ph^\alpha}$. 
    \medskip     
    
    For any $R\in C^0(\RR^+)$, if there exists $\sigma_R<D_\ell/\alpha$ such that as $t\to0^+$, $R(t)=O(t^{-\sigma_R})$, then $R$ is an admissible remainder for $L_\Phi^\alpha$ and any screen of the form $S_\e(\tau)\equiv \sigma_R+\e$, with $0<\e<D_\ell/\alpha-\sigma_R$, is admissible. 
\end{Corollary}

\begin{proof}
    By Lemma~\ref{lem:MellinHolo}, we have that $\zeta_R(s)$ is holomorphic in $\HH_{\sigma_R}$ and thus $\zeta_R(z/\alpha)$ is holomorphic when $\Re(z)>\alpha\,\sigma_R$. Further, $\zeta_R(s)$ is bounded on any vertical strip of the form $\HH_a^b$, $\sigma_R<a\leq b<\infty$, so $\zeta_R(z/\alpha)$ is bounded on the corresponding strips $\HH_{a'}^{b'}$, $\alpha\,\sigma_R<a'\leq b'<\infty$. 
    
    Since $\alpha\,\sigma_R<D_\ell$, by repeating the proof of Theorem~\ref{thm:lowerDimAdmissibility} (with $\sigmaRem=\alpha\,\sigma_R$) we obtain that $\zeta_\Phi(z)$ and $\zeta_R(z/\alpha)$ are jointly languid on screens of the form $S_\e(\tau)\equiv \alpha\,\sigma_R+\e$ when $0<\e<D_\ell -\alpha\,\sigma_R$. Taking $z=\alpha s$, we see that the functions $\zeta_\Phi(\alpha s)$ and $\zeta_R(s)$ are jointly languid on screens of the form $S'_{\e'}(\tau)\equiv \sigma_R+\e'$ where $\e'=\e/\alpha\in(0,D_\ell/\alpha-\sigma_R)$. 
\end{proof}

\subsubsection{Lattice Criterion}

The next criterion is related to when we have explicit knowledge of the locations of the singularities of $\zeta_\Phi$. In particular, in the lattice case (see Definition~\ref{def:latticeDichotomy}), we can explicitly show that all of the poles lie on one of finitely many vertical lines (and are distributed periodically along these lines with a shared period for each line); see for instance \cite[Theorem 3.6]{LapvFr13_FGCD}. Thus, we can easily choose screens within this region which will never encounter singularities of $\zeta_\Phi$, with distance to any pole bounded by the distance of the real part of the screen to the real part of the closest of the finitely many exceptional points.

\begin{Theorem}[Lattice Criterion for Admissiblility]
    \label{thm:latticeCaseAdmissibility}
    Let $\Phi$ be a self-similar system and suppose that its distinct scaling ratios $\set{\lambda_\ph}_{\ph\in\Phi}$ are arithmetically related (see Definition~\ref{def:latticeDichotomy}). Let $R$ be a continuous function on $\RR^+$ with the estimate that $R(t)=O(t^{-\sigma_0})$ as $t\to0^+$, for some $\sigma_0\in\RR$. 
    \medskip
    
    Then for all but finitely many $\sigma>\sigma_0$, $S_\sigma(\tau)\equiv \sigma$ is an admissible screen. Consequently, there are admissible screens of the form $S_{\sigma_0+\e}$ for any $\e>0$ sufficiently small.
\end{Theorem}
\begin{proof}
    By definition, in the lattice case there exists some $\lambda_0\in\RR^+$ such that for each $\ph\in\Phi$, there is a positive integer $k_\ph$ such that $\lambda_\ph=\lambda_0^{k_\ph}$. Under this assumption, we may explicitly write the denominator of $\zeta_\Phi$ as the Dirichlet polynomial 
    \[ P(s) = 1-\sum_{\ph\in\Phi} \lambda_\ph^s = 1- \sum_{\ph\in\Phi} \lambda_0^{k_\ph s}. \]
    Under the change of variables $s=\log_{\lambda_0}z$ (so that $\lambda_0^s=z$), we have that 
    \[ P(\log_{\lambda_0}z) = 1- \sum_{\ph\in\Phi} z^{k_\ph}. \]
    This is precisely a polynomial in the variable $z$ (since $|\Phi|=n<\infty$), so by the fundamental theorem of algebra it has finitely many roots (exactly $K=\max(\set{k_\ph}_{\ph\in\Phi})$) in $\CC$. 
    
    Denote these roots by $Z=\set{z_j}_{j=1}^K$. Any solution of $P(\omega)=0$ must then be of the form $\lambda_0^\omega=z_j$ for some $z_j\in Z$. We note that $0\notin Z$ since the polynomial equals one when $z=0$, so there exists a branch of the logarithm for which $\log z_j$ is well defined for each $z_j\in Z$ and without loss of generality we may find a single branch defined for each $z_j\in Z$ and for $\lambda_0$ since $Z$ is finite. The logarithm is multivalued, however, with $\lambda_0^\omega=e^{\omega\log\lambda_0}=e^{\omega\log\lambda_0+2\pi i m}$ for any $m\in\ZZ$. So, we will obtain as solutions to $P(\omega)=0$ exactly the points $\omega_{j,m}$ of the form $\omega_{j,m}= \log(z_j)/\log(\lambda_0) + 2\pi i m/\log(\lambda_0)$, where $m\in\ZZ$ is arbitrary. 

    Observe that there are finitely many real parts, $\sigma_j=\Re(\log(z_j)/\log(\lambda_0))$, $j=1,...,K$, at which these poles occur and that the imaginary parts are all distributed with the same period, $2\pi/\log\lambda_0$. We will show, starting with hypothesis \textbf{L1}, that for any screen of the form $S_\sigma(\tau)\equiv \sigma$, $\sigma\neq\sigma_j$ for each $j=1,...,K$, $\zeta_\Phi$ is languid with exponent $\kappa=0$ with respect to $S_\sigma$. 

    The next two steps of the proof are the same as in the proof of Theorem~\ref{thm:lowerDimAdmissibility}. In short, by the strong languidity of $\zeta_\Phi$, for any $\sigma_-$, there exists a sequence of admissible heights $\set{T_n}_{n\in\ZZ}$ on which $\zeta_\Phi$ is bounded on horizontal intervals of the form $[\sigma_-+iT_n,\sigma_++iT_n]$, where $\sigma_-$ is arbitrarily small and $\sigma_+>\max(\simdim(\Phi),\sigma_-)$. Choosing $\sigma_-=\sigma$ establishes hypothesis \textbf{L1} with respect to the screen $S_\sigma$ for $\zeta_\Phi$. Secondly, we have that $\zeta_R$ satisfies hypotheses \textbf{L1} and \textbf{L2} for the sequence of heights as above on any screen of the form $S_\sigma$, $\sigma>\sigma_0$, because $\zeta_R$ is a Mellin transform and is holomorphic in $\HH_{\sigma_0}$. 

    It only remains to show that $\zeta_\Phi$ satisfies hypothesis \textbf{L2} on $S_\sigma$ when $\sigma$ is not one of the exceptional values $\sigma_j$, $j=1,...,K$. We will show that $f(\tau)$ is bounded from below by a strictly positive constant on all of $\RR$, and thus $\zeta_\Phi$ will be bounded on $S_\sigma$. As we have shown, $P(\sigma+i\tau)\neq0$ whenever $\sigma\notin\set{\sigma_j}_{j=1}^K$. Let $f(\tau)=|P(\sigma+i\tau)|$. Given any compact interval $[a,b]$, we have that $f(t)$ must be nonzero and bounded from below by a strictly positive constant on $[a,b]$. This follows by continuity and the intermediate value theorem, noting that if $f(\tau)=0$, then $P(\sigma+i\tau)=0$, which is a contradiction. 
    
    We will use the (multiplicative) periodicity of $f$ to show that $f(t)=|P(\sigma+i\tau)|$ cannot become arbitrarily small as $|\tau|\to\infty$. Recall that  
    \[
        P(\sigma+i\tau) = 1 - \sum_{j=1}^K \lambda_0^{k_j(\sigma_0+i\tau)} = 1- \sum_{j=1}^K \lambda_0^{k_j\sigma}\cdot e^{i (k_j\log\lambda_0)\tau}.
    \]
    Under the transformation $\tau\mapsto (2\pi m/\log\lambda_0) \,\tau$, for any $m\in\ZZ$, we have that each exponential is invariant since $k_j$ is an integer. Thus, $P(\sigma+i\tau)=P(\sigma+i(2\pi m/\log\lambda_0)\tau)$. Choosing $m=-1$ shows us that the function $f(\tau):=|P(\sigma+i\tau)|$ is multiplicatively periodic with period $p=-2\pi/\log\lambda_0=2\pi/\log\lambda_0^{-1}$ or a rational multiple thereof, depending on the greatest common divisor of the integers $k_j$, $j=1,...,K$. However, if they share a greatest common divisor $\text{GCD}$, it is possible to redefine $\lambda_0$ by $\lambda_0'=\lambda_0^{\text{GCD}}$ and obtain a new set of integers without this property. Note that $\log(\lambda_0^{-1})>0$.

    By the previous argument, $f(\tau)=|P(\sigma+i\tau)|$ is bounded from below on any compact interval. So, $f$ is nonzero on the interval $[-1,1]$. Now pick one full multiplicative period in $(0,\infty)$, say $[p^{m_0},p^{m_0+1}]$ (if $p>1$) or $[p^{m_0+1},p^{m_0}]$ (if $p<1$). The function $f$ must be bounded from below strictly away from zero as this is a compact interval. By periodicity, this same bound applies to any interval of the form $[p^{m},p^{m+1}]$ or $[p^{m+1},p^{m}]$, respectively, for any $m\in\ZZ$. Since the function $p\mapsto p^t$ is surjective onto $(0,\infty)$, this implies that $f$ is bounded uniformly from below by a strictly positive constant, the same as the first bound, when $\tau>0$. For $\tau<0$, we can use the starting interval $[-p^{m_0+1},-p^{m_0}]$ or $[-p^{m_0},-p^{m_0+1}]$, depending on whether $p>1$ or $p<1$, and the same argument to deduce that it is also bounded from below by a strictly positive constant on $(-\infty,0)$. Taking the minimum of these three bounds shows that $f(\tau)=|P(\sigma+i\tau)|>C>0$. 
    
    It follows that $\zeta_\Phi$ is bounded from above on $S_{\sigma_0}$, establishing hypothesis $\textbf{L2}$. Thus, we have established the joint languidity of $\zeta_\Phi$ and $\zeta_R$ on any screen of the form $S_\sigma$, where $\sigma>\sigma_0$ and where $\sigma\notin\set{\sigma_j}_{j=1}^K$. When choosing a screen $S_{\sigma_0+\e}$, choosing $\e$ with $0<\e<\min_{j=1,...,K}(|\sigma_j-\sigma_0|)$ is sufficient.
\end{proof}

\subsection{Solutions of Scaling Functional Equations}
\label{ssc:SFEsolutions}

In this section, we will obtain explicit formulae for functions which satisfy a scaling functional equation. Given a scaling operator $L$, a scaling functional equation is a relation of the form $f=L[f]+R$, where $R$ is a remainder term. We obtain formulae which are valid up to an asymptotic order determined by estimates of an \textit{admissible remainder} term (in the sense of Definition~\ref{def:admissibleRem}). We begin with the statement of our results followed by the discussion and estimates needed for their proof.

We focus on scaling functional equations which are induced by a self-similar system on functions which satisfy certain scaling laws (see Proposition~\ref{prop:inducedSFE}), as these are the main type of scaling functional equations that we will need for our application to heat content in Section~\ref{sec:heatAnalysis}. Explicitly, let $\Phi$ be a self-similar system, let $\alpha>0$, and define the scaling operator $L_\Phi^\alpha:=\sum_{\ph\in\Phi}M_{\lambda_\ph^\alpha}$. The scaling functional equations we study herein will be stated in terms of such scaling operators. Note, though, that any general scaling operator with scaling ratios in $(0,1)$ each having positive, integral multiplicity can be written in this form. 

There are two types of explicit formulae, namely pointwise and distributional (just as in \cite{LapvFr13_FGCD,LRZ17_FZF}). The former have the advantage of being simpler in their statements, but the disadvantage that they 
may only be valid for antiderivatives of the function satisfying the scaling functional equation and not the function itself. The distributional explicit formulae we obtain do not have this restriction, but they will require additional terminology to state and interpret in a weak formulation. 

\subsubsection{Pointwise Explicit Formulae}

The pointwise explicit formulae we establish in this work for solutions to a scaling functional equation, say $f\in C^0(\RR)$, will be valid for its antiderivatives. To that end, we introduce the following notation and convention. Firstly, define $f^{[0]}:=f$. For any $k>0$, we define $f^{[k]}$ recursively by integrating the previous antiderivative and imposing the convention that $f^{[k]}(0)=0$. Namely, for $k>0$,
\[
    f^{[k]}(t) := \int_0^t f^{[k-1]}(\tau)\,d\tau.
\] 
We will also denote $\zeta_f(s;\delta)=\Mellin^\delta[f](s)$. In our applications, typically we have some function $F(t)$ of interest which must be normalized in order to satisfy a scaling law in the sense of Definition~\ref{def:scalingLaw}, so we will have that $f(t) = t^{-\beta} F(t)$ for a given parameter $\beta$. For tube functions (cf. \cite{Hof25}), $\beta$ is the dimension $\dimension$ and for heat content in Section~\ref{sec:heatAnalysis}, it will be $\dimension/2$. When we state the explicit formula, we will do so in terms of the parameter $\beta$, noting that typically $\beta=\dimension$. Let $F^{[k]}$ be defined in the same manner as $f^{[k]}$, $k\geq 0$. 

As an additional preliminary to stating the result, we define the Pochhammer symbol $(z)_w:=\Gamma(z+w)/\Gamma(w)$ for $z,w\in\CC$. Note that when $w=k$ is a positive integer, this simplifies to $(z)_k=z(z+1)\cdots(z+k-1)$ or, when $w=0$, to $(z)_0=1$.

Lastly, we must give meaning to the summations which will appear in what follows. Let $\Dd\subset\CC$ be a discrete (and hence at most countably infinite) subset with the property that for any $m\in\NN$, 
\[ 
    \Dd_m:=\set{\omega\in\Dd\suchthat |\Im(\omega)|\leq m}<\infty.
\] 
Then we define 
\[ 
    \sum_{\omega\in\Dd} a_\omega := \lim_{m\to\infty} \sum_{\omega\in\Dd_m}a_\omega.  
\]
Note that this summation is a \textit{symmetric limit} of finite partial sums. That is to say, it is a sum over the elements of $\Dd$ with increasing large absolute values of imaginary parts, but with the upper and lower bounds taken at the same rate. As such, the convergence of these sums is comparatively more delicate; it may be that the limit, when taken independently, does not exist. 

\begin{Theorem}[Pointwise Explicit Formula]
    \label{thm:pointwiseFormula}
    Let $\Phi$ be a self-similar system, $\alpha>0$, $\beta\in\RR$, and $f(t)=t^{-\beta/\alpha}F(t)\in C^0(\RR^+)$. Suppose that $\Phi$ induces the scaling functional equation $f=L_\Phi^\alpha[f]+R$ on $[0,\delta]$ with admissible remainder term $R$ (in the sense of Definition~\ref{def:admissibleRem}) with corresponding screen $S$. Let $\sigma_R$ denote the abscissa of absolute convergence of $\zeta_R$, $D=\dim_S(\Phi)$ the (upper) similarity dimension of $\Phi$, and suppose $\beta/\alpha+1>\max(D/\alpha,\sigma_R)$. Lastly, suppose that $S$ is contained in the half-plane $\HH_{\sigma_R}$ and write $W_S$ for the window to the right of $S$.
\medskip 

    Then for every $k\geq 2$ in $\ZZ$ and every $t\in(0,\delta)$, we have that
    \[ 
        F^{[k]}(t) = \sum_{\omega\in\Dd_\Phi(\alpha W_S)} 
            \Res\Bigg(\cfrac{t^{(\beta-s)/\alpha+k}}{((\beta-s)/\alpha+1)_k}\zeta_f(s/\alpha;\delta);\omega\Bigg)
            + \Rr^k(t),
    \]
    where $\zeta_f$ is given by \eqref{eqn:zetaFormulaAlpha}. The error term satisfies $\Rr^k(t)=O(t^{\beta/\alpha-\sup(S)+k})$ as $t\to0^+$. 
\end{Theorem}

Note, in particular, that the order of the remainder directly controls the error of the approximation. As we are considering small values of $t$, the larger the exponent, the better the remainder estimate. The restriction of $k\geq2$ is related to the exponent $\kappa=0$ of the languid growth of $\zeta_\Phi$, and consequently of $\zeta_f$.

\subsubsection{Distributional Explicit Formulae}

Put simply, a \textit{distribution} is a functional on a prescribed space of functions called \textit{test functions}. A standard distribution is a functional on the space $C_c^\infty$ of smooth, compactly supported functions on a given domain. The type of distributions we consider here, called \textit{tempered distributions}, will be a subspace of these distributions. 

We define as the space of test functions the class of Schwartz functions, or functions of rapid decrease. For a finite domain $(0,\delta)$, just as in \cite[Chapter~5]{LapvFr13_FGCD} and \cite[Chapter~5]{LRZ17_FZF}, the space is defined by 
\begin{equation}
    \Ss(0,\delta):=\left\{\testfn\in C^\infty(0,\delta)\suchthat 
    \begin{aligned}
        &\forall m\in\ZZ,\,\forall q\in\NN,\text{ as }t\to0^+,\\ 
        &t^m\testfn^{(q)}(t)\to0 \text{ and }(t-\delta)^m\testfn^{(q)}(t)\to0\,
    \end{aligned}
    \right\}.
\end{equation}
Note that $C_c^\infty(0,\delta)\subseteq\Ss(0,\delta)$, with a continuous embedding, whence by duality the space of functionals on $\Ss(0,\delta)$ is a subset of the usual space of distributions, the dual of $C_c^\infty(0,\delta)$. Typically, for a distribution $F$ and a test function $\testfn$, we write $F(\testfn)=\bracket{F,\testfn}$, denoting by $\bracket{\cdot,\cdot}$ the natural pairing of $\Ss(0,\delta)$ with its dual space. 

In this distributional setting, the restriction of $k\geq 2$ may be relaxed in Theorem~\ref{thm:pointwiseFormula}. However, equality must be interpreted \textit{in the sense of distributions}: we say that $F=G$ in the sense of distributions if for any test function $\testfn\in\Ss(0,\delta)$, $\bracket{F,\testfn}=\bracket{G,\testfn}$. In addition to the standard definitions for the definitions of differentiation and integration of distributions, we note the following relevant identity regarding residues at a point:
\begin{equation}
    \bracket{\Res(t^\beta G(s);\omega),\testfn} = \Res(\Mellin[\testfn](\beta+1)G(s);\omega).
\end{equation}
See \cite[Equation~5.2.5]{LRZ17_FZF}, as well as the preceding discussion regarding the extension of $\testfn$ to all of $\RR^+$. 

Secondly, we provide the definition for a distribution $\Rr$ to satisfy the error estimate $O(t^\alpha)$ as $t\to0^+$ (defined as in \cite[Theorem 5.2.11]{LRZ17_FZF}). Given a test function $\testfn\in\Ss(0,\delta)$ and $a>0$, define the new test function $\testfn_a(t):= a^{-1}\testfn(t/a)$. We say that $\Rr(t)=O(t^\alpha)$ as $t\to0^+$ if for all $a>0$ and for all $\testfn\in\Ss(0,\delta)$,
\begin{equation}
    \label{eqn:distRemEst}
    \bracket{\Rr,\testfn_a(t)} = O(a^{\alpha}),
\end{equation}
as $t\to0^+$ in the usual sense. 

\begin{Theorem}[Distributional Explicit Formula]
    \label{thm:distFormula}
    Let $\Phi$ be a self-similar system, $\alpha>0$, $\beta\in\RR$, and $f(t)=t^{-\beta/\alpha}F(t)\in C^0(\RR^+)$. Suppose that $\Phi$ induces the scaling functional equation $f=L_\Phi^\alpha[f]+R$ on $[0,\delta]$ with admissible remainder term $R$ (in the sense of Definition~\ref{def:admissibleRem}) with corresponding screen $S$. Let $\sigma_R$ denote the abscissa of absolute convergence of $\zeta_R$, $D=\dim_S(\Phi)$ the (upper) similarity dimension of $\Phi$, and suppose $\beta/\alpha+1>\max(D/\alpha,\sigma_R)$. Lastly, suppose that $S$ is contained in the half-plane $\HH_{\sigma_R}$ and write $W_S$ for the window to the right of $S$.
    \medskip

    Then for any $k\in\ZZ$, we have that, in the sense of distributions, $F^{[k]}$ satisfies
    \begin{equation}
        \label{eqn:distributionalIdentity}
        \begin{split}
            F^{[k]}(t) &= \sum_{\omega\in\Dd_\Phi(\alpha W_S)} 
                \Res\Bigg(\cfrac{t^{(\beta-s)/\alpha+k}}{((\beta-s)/\alpha+1)_k}\zeta_f(s/\alpha;\delta);\omega\Bigg)
                + \Rr^{[k]}(t),
        \end{split} 
    \end{equation}
    as $t\to0^+$ for a remainder distribution term $\Rr^{[k]}$. See \eqref{eqn:bracketIdentity} for the explicit identity of action on test functions. 
    \medskip

    Here, $\zeta_f$ is as in Corollary~\ref{cor:structureOfPoles}. The distributional remainder term satisfies the estimate $\Rr^{[k]}(t)=O(t^{\beta/\alpha-\sup(S)+k})$ as $t\to0^+$, in the sense of \eqref{eqn:distRemEst}.
\end{Theorem}
Most precisely, \eqref{eqn:distributionalIdentity} means that for any $\testfn\in\Ss(0,\delta)$, 
\begin{equation}
    \label{eqn:bracketIdentity}
    \begin{split}
        \bracket{F^{[k]},\testfn} &= \sum_{\omega\in\Dd_\Phi(\alpha W_S)} 
        \Res\Bigg(\cfrac{\Mellin[\testfn]({(\beta-s)/\alpha+k+1})}{((\beta-s)/\alpha+1)_k}\zeta_f(s/\alpha);\omega\Bigg)
        + \bracket{\Rr^{[k]},\testfn}.
    \end{split}
\end{equation}
While this formulation is less direct compared to the pointwise expansion, it does have the advantage of requiring less regularity to leverage the expansion. Namely, it is valid when $k=0$, yielding a formula (in the sense of distributions) for the function $f$ itself. 
\begin{Corollary}[Distributional Explicit Formula, $k=0$]
    \label{cor:distFormulaZero}
    Let $\Phi$ be a self-similar system, $\alpha>0$, $\beta\in\RR$, and $f(t)=t^{-\beta/\alpha}F(t)\in C^0(\RR^+)$. Suppose that $\Phi$ induces the scaling functional equation $f=L_\Phi^\alpha[f]+R$ on $[0,\delta]$ with admissible remainder term $R$ (in the sense of Definition~\ref{def:admissibleRem}) with corresponding screen $S$. Let $\sigma_R$ denote the abscissa of absolute convergence of $\zeta_R$, $D=\dim_S(\Phi)$ the (upper) similarity dimension of $\Phi$, and suppose $\beta/\alpha+1>\max(D/\alpha,\sigma_R)$. Lastly, suppose that $S$ is contained in the half-plane $\HH_{\sigma_R}$ and write $W_S$ for the window to the right of $S$.
    \medskip

    Then in the sense of distributions we have that for $t\in[0,\delta)$,
    \begin{equation}
        \label{eqn:distributionalIdentityZero}
            F(t) = \sum_{\omega\in\Dd_\Phi(\alpha W_S)} 
                \Res\big(t^{{\beta-s}/\alpha}\zeta_f(s/\alpha;\delta);\omega\big) + \Rr(t).
    \end{equation}
    The distributional remainder term satisfies the estimate $\Rr(t)=O(t^{\beta/\alpha-\sup(S)})$ as $t\to0^+$, in the sense of \eqref{eqn:distRemEst}.
    \medskip

    If we further assume that the poles of $\zeta_\Phi$ in $W_S$ are simple, then \eqref{eqn:distributionalIdentityZero} simplifies to 
    \[
        f(t) = \sum_{\omega\in\Dd_\Phi(\alpha W_S)} 
            \Res(\zeta_f(s/\alpha;\delta);\omega)\,t^{(\beta-\omega)/\alpha} + \Rr(t).
    \]
\end{Corollary}
In this formulation, and in particular in the case when the self-similar system induces simple poles of its zeta function, the formula is a simple expansion with constant coefficients and powers determined by the poles $\omega$. The computation of the residues, however, is not a trivial matter and beyond the current scope of this work. Corollary~\ref{cor:distFormulaZero} is a direct application of Theorem~\ref{thm:distFormula}.

\subsubsection{Formulae for the Zeta Functions}

We will discuss the solution of scaling functional equations by means of truncated Mellin transforms (see Definition~\ref{def:truncatedMellinTransform} and Appendix~\ref{app:Mellin}). If what follows, let $\delta>0$ be fixed and let $\Mellin^\delta$ be the truncated Mellin transform on $(0,\delta)$. Given a function $f$ which is locally (Lebesgue) integrable on $(0,\delta)$, define $\zeta_f(s;\delta):=\Mellin^\delta[f](s)$ for all $s\in\CC$ for which the integral is convergent. We say that $\zeta_f(s;\delta)$ is a \textbf{tamed Dirichlet-type integral (DTI)} (viz. \cite[Definition A.1.2]{LRZ17_FZF}) if $d\mu(t)=f(t)dt$ is a local complex measure (i.e. its restriction to any compact subset is a complex measure) on $[0,\delta]$ and if there exists $\sigma\in\RR$ such that the defining integral is convergent (and hence holomorphic) on $\HH_\sigma$. See Appendix~\ref{app:Mellin} for more information. For example, if $f$ is integrable, bounded away from $0$, and $f(t)=O(t^{-\sigma})$ as $t\to0^+$ then $\zeta_f$ is holomorphic on $\HH_\sigma$ (viz. Lemma~\ref{lem:MellinHolo}). 

Given a scaling operator $L:=\sum_{k=1}^m a_k M_{\lambda_k}$, define 
\begin{equation}
    \label{eqn:defPartialZeta}
    \xilf(s;\delta) := \sum_{k=1}^m a_k\lambda_k^s\Mellin_\delta^{\delta/\lambda_k}[f](s). 
\end{equation}
If $\Phi$ is a self-similar system, then define $\xilfphi:=\xi_{L_\Phi,f}$, where $L_\Phi$ is the scaling operator associated to $\Phi$.  Lastly, we recall that $\zeta_L$ is the scaling zeta function of $L$, with singular set $\Dd_L$ (see Section~\ref{ssc:SFEs}).

The following is essentially \cite[Theorem 4.8]{Hof25}, although we take the opportunity to state the theorem with slightly more precise and slightly more general hypotheses. 
\begin{Theorem}[Zeta Functions of Solutions to General SFEs]
    \label{thm:zetaFormula}
    Let $L:=\sum_{k=1}^m a_k M_{\lambda_k}$ be a scaling operator and $\Lambda=\max\set{\lambda_k^{-1}}_{k=1}^m$. 
    Fixing $\delta>0$, let $f$ and $R$ be integrable functions on $(0,\Lambda\delta)$ and $(0,\delta)$, respectively, and suppose that $\zeta_f(s;\Lambda\delta):=\Mm^{\Lambda\delta}[f](s)$ and $\zeta_R(s;\delta):=\Mm^\delta[R](s)$ are tamed DTIs (viz. \cite[Definition A.1.2]{LRZ17_FZF}), with respective abscissae of absolute convergence $\sigma_f$ and $\sigma_R$, respectively. 
Let $L:=\sum_{k=1}^m a_k M_{\lambda_k}$ be a scaling operator for which $f$ satisfies the scaling functional equation $f=L[f]+R$ for all $t\in (0,\delta)$. 
    \medskip
 
    Then for all $s\in\HH_{\sigma_R}\setminus \Dd_L(\CC)$, 
    \begin{equation}
        \label{eqn:ZetaFormula}
        \zeta_f(s;\delta)= \zeta_L(s)(\xilf(s;\delta)+\zeta_R(s;\delta)),
    \end{equation}
    where $\xilf$, defined as in \eqref{eqn:defPartialZeta}, is an entire function. Further, $\zeta_f$ is holomorphic in the half-plane $\HH_{\max(D,\sigma_R)}$, where where $D$ is the unique real pole of $\zeta_L$, and admits a meromorphic continuation to $\HH_{\sigma_R}$ with any poles contained in $\Dd_L(\HH_{\sigma_R})$. Further, these poles are independent of the choice of $\delta$, i.e. $\zeta_f(s;\delta)$ and $\zeta_f(s;\delta')$ have the same poles for any $\delta'\in(0,\delta)$.
\end{Theorem}
\begin{proof}
    As tamed DTIs, there exist $\sigma_f,\sigma_R\in[-\infty,\infty)$ for which $\zeta_f(s;\Lambda\delta)$ and $\zeta_R(s;\delta)$ are holomorphic in $\HH_{\sigma_f}$ and $\HH_{\sigma_R}$, respectively, with the convention that $\HH_{-\infty}=\CC$. Consequently, they are both holomorphic in $\HH_{\sigma_M}$ where $\sigma_M:=\max(\sigma_f,\sigma_R)$. 
Additionally, we have that by Lemma~\ref{lem:MellinScaling}, for each $M_{\lambda_k}$, 
    \[ \Mellin^\delta[M_{\lambda_k}[f]](s) = \lambda_k^s \Mellin^{\delta/\lambda_k}[f](s). \]
    Note that each function $\Mellin^{\delta/\lambda_k}[f]$, $k=1,...,m$, is holomorphic in $\HH_{\sigma_f}$ since $\zeta_f(s;\Lambda\delta)$ is assumed to be a tamed DTI for the largest interval of the form $(0,\delta/\lambda_k)$, $k=1,...,m$. (It follows from Lemma~\ref{lem:MellinHolo} that the difference between $\zeta_f(s;\delta_1)$ and $\zeta_f(s;\delta_2)$ is an entire function for any $\delta_1,\delta_2\in(0,\Lambda\delta)$, so integration over a subset yields a function with the same holomorphicity properties.) 

    With these holomorphicity properties, the proof of \cite[Theorem 4.8]{Hof25} is applicable without further modification, noting that $E(s)=\xilf(s;\delta)$ and $\sigma_0=\sigma_R$. The independence of the poles of $\zeta_f$ on the value of $\delta$ may be seen as a corollary of Lemma~\ref{lem:MellinHolo}.

\end{proof}

We now specialize to operators of the form $L_\Phi^\alpha:=\sum_{\ph\in\Phi}M_{\lambda_\ph^\alpha}$, where $\Phi$ is a self-similar system and $\alpha>0$. We require that $\alpha>0$, so that $\lambda_\ph^\alpha\in(0,1)$ for each $\ph\in\Phi$. When $\alpha=1$, this is exactly the scaling operator $L_\Phi$ associated to $\Phi$. For other values of $\alpha$, the effect on the associated zeta function is simply a rescaling of the input relative to $\zeta_\Phi$, namely $\zeta_{L_\Phi^\alpha}(s)=\zeta_\Phi(\alpha s)$. (This is an immediate consequence of the definition of these functions and elementary properties of exponents.) Note that it follows that if $D=\dim_S(\Phi)$ is the similarity dimension of $\zeta_\Phi$, an upper bound for the real parts of its poles by Proposition~\ref{prop:simDimBounds}, then $\zeta_{L_{\Phi}^\alpha}$ is holomorphic in the half plane $\HH_{D/\alpha}$. Further, the sets of poles are directly related: any pole $\omega\in\Dd_\Phi$ exactly corresponds to the pole $\omega/\alpha\in\Dd_{L_\Phi^\alpha}$. 
These observations together yield the following corollary of \cite[Theorem 4.8]{Hof25}. \begin{Corollary}[Zeta Functions of Solutions to Self-Similar SFEs]
    \label{cor:structureOfPoles}
    Let $f,R\in C^0(\RR^+)$ and suppose that there exist $\sigma_R,\sigma_f\in\RR$ such that $R(t)=O(t^{-\sigma_R})$ and $f(t)=O(t^{-\sigma_f})$ as $t\to0^+$. Let $\Phi$ be a self-similar system, let $\alpha>0$, and let ${L_\Phi^\alpha}:=\sum_{\ph\in\Phi}M_{\lambda_\ph^\alpha}$. If $f$ satisfies the scaling functional equation $f={L_\Phi^\alpha}[f]+R$ for all $t\in (0,\delta)$, then
    \begin{equation}
        \label{eqn:zetaFormulaAlpha}
        \zeta_f(s;\delta) = \zeta_\Phi(\alpha s)(\xilfphia(s;\delta)+\zeta_R(s;\delta))
    \end{equation}
    and is holomorphic in $\HH_{\max(D/\alpha,\sigma_R)}$, where $D=\dim_S(\Phi)$. Further, $\zeta_f(s;\delta)$ is meromorphic in $\HH_{\sigma_R}$, it has poles in a subset of $\alpha^{-1}\Dd_\Phi(\HH_{\sigma_R})$, and these poles are independent of $\delta$. 
\end{Corollary}
We note that the assumptions on $f$ and $R$ are sufficient to ensure that $\zeta_f$ and $\zeta_R$ are tamed DTIs (viz. as a corollary of Lemma~\ref{lem:MellinHolo}), as they are holomorphic in the half-planes $\HH_{\sigma_f}$ and $\HH_{\sigma_R}$, respectively. These values of $\sigma_f$ and $\sigma_R$ give upper bounds for the abscissa of absolute convergence of $\zeta_f$ and $\zeta_R$, respectively. So long as $\sigma_f$ and $\sigma_R$ are chosen to be the minimal such exponents for which these estimates hold, they will equal their respective abscissae of absolute convergence. (See Appendix~\ref{app:Mellin} for more detail.)

When $L=L_\Phi^\alpha$, we note that \eqref{eqn:defPartialZeta} is explicitly the entire function 
\begin{equation}
    \label{eqn:defPartialZetaAlpha}
    \xilfphia(s;\delta) = \sum_{\ph\in\Phi} \lambda_\ph^{\alpha s}\Mellin_\delta^{\delta/\lambda_\ph^\alpha}[f](s). 
\end{equation}
In this setting, it will be convenient to state the estimate for $R$ as $t\to0^+$ in the form $R(t)=O(t^{-\sigmaRem/\alpha})$, where $\sigma_R=\sigmaRem/\alpha$. Under the change of variables $s'=s/\alpha$, a pole $\omega'=\omega/\alpha\in\alpha^{-1}\Dd_\Phi(\HH_{\sigma_R})$ of $\zeta_f(s';\delta)$ corresponds to the pole $\omega\in\Dd_{\Phi}(\HH_{\sigmaRem})$ of $\zeta_f(s/\alpha;\delta)$. Thus, in Section~\ref{ssc:SFEsolutions} when we consider poles of $\zeta_f$ in $\HH_{\sigma_R}$, under these conventions we will be considering the poles of $\zeta_\Phi$ in $\HH_{\sigma_0}$. In Section~\ref{ssc:cDims}, we will see that this change of variables ultimately clarifies the direct connection of spectral complex dimensions (regarding heat content and heat zeta functions) to the geometric complex dimensions (defined by fractal zeta functions such as the tube zeta function). 

The next step for solving the scaling functional equation $f=L_\Phi^\alpha[f]+R$ is to apply the Mellin inversion theorem to \eqref{eqn:zetaFormulaAlpha}, which is the content of \cite[Theorem 4.9]{Hof25}. Note the convention regarding the contour integral, in that 
\[ \int_{c-i\infty}^{c+i\infty} f(z)\,dz := \lim_{T\to+\infty}\int_{c-iT}^{c+iT}f(z)\,dz, \]
which may also be considered a principal value integral (e.g. as in \cite{Gra10}). We state the formula using the notation of Corollary~\ref{cor:structureOfPoles} (i.e. in terms of $\zeta_\Phi$, $\zeta_R$, and $\xilfphia$) here as a proposition for later reference. 
\begin{Proposition}[Mellin Inversion Formula]
    \label{prop:MellinInversion}
    Let $f,R\in C^0(\RR^+)$. Let $\Phi$ be a self-similar system, let $\alpha>0$, and let ${L_\Phi^\alpha}:=\sum_{\ph\in\Phi}M_{\lambda_\ph^\alpha}$. Suppose that $f$ satisfies the scaling functional equation $f={L_\Phi^\alpha}[f]+R$ for all $t\in [0,\delta]$ and that $\zeta_R(s;\delta)=\Mellin^\delta[R](s)$ is holomorphic in $\HH_{\sigma_R}$. Lastly, let $D=\dim_S(\Phi)$ and $c>\max(D/\alpha,\sigma_R)$. Then for any $t\in(0,\delta)$, $f$ is given by:
    \begin{align*}
        f(t)    &= \Mm^{-1}[\zeta_\Phi(\alpha s)(\xilfphia(s;\delta)+\zeta_R(s;\delta))](t)\\
                &= \frac{1}{2\pi i} \int_{c-i\infty}^{c+i\infty} 
                    t^{-s} \zeta_\Phi(\alpha s)(\xilfphia(s;\delta)+\zeta_R(s;\delta))\,ds.
    \end{align*}
\end{Proposition} 
It remains to evaluate this contour integral by means of the residue theorem. Strictly speaking, it requires a sort of unbounded residue theorem, using a sequence of regions for which the integrand has finitely many poles (on which the residue theorem applies) that converge to the region containing all of the countably many singularities. The conditions of languidity are necessary in this process, used to estimate contour integrals over pieces of the boundaries of regions involved in this process. This method developed to establish explicit formulae for fractal tube functions in \cite[Chapter 5 and Chapter 8]{LapvFr13_FGCD} and then, with refinements to the proof, used to prove explicit formulae for relative tube functions in \cite[Chapter 5]{LRZ17_FZF}.

\subsubsection{Languidity from Scaling Functional Equations}

We will need growth estimates for $\zeta_f(s;\delta)=\Mellin^\delta[f](s)$ when $f$ is the solution of a scaling functional equation. In particular, using the formula in Corollary~\ref{cor:structureOfPoles}, we show that $\zeta_f$ is languid provided some control over the remainder term itself. 
\begin{Theorem}[Languidity of Zeta Functions from SFEs]
    \label{thm:languidityFromSFEs}
    Let $\Phi$ be a self-similar system, let $\alpha>0$, and $f\in C^0(\RR^+)$. Suppose that $\Phi$ induces the scaling functional equation $f=L_\Phi^\alpha[f]+R$ on $[0,\delta]$ with admissible remainder term $R$ for $L_\Phi^\alpha$ (in the sense of Definition~\ref{def:admissibleRem}) and let $S$ be an admissible screen.
    \medskip 
    
    Then the zeta function $\zeta_f(s;\delta)=\Mellin^\delta[f](s)$ is also languid with exponent $\kappa=0$ with respect to $S$. 
\end{Theorem}
\begin{proof}
    Because $R$ is an admissible remainder for $L_\Phi^\alpha$ with admissible screen $S$, we have that $\zeta_\Phi(\alpha s)$ and $\zeta_R(s;\delta)$ are jointly languid (with exponent $\kappa=0$) with respect to this screen $S$. We have that $\xilfphia$ is an entire function by Corollary~\ref{cor:structureOfPoles} and explicitly we show that $\xilfphia$ is bounded on any vertical strip of the form $\HH_a^b=\set{s\in\CC\suchthat a<\Re(s)<b}$.
    
    In order to do so, it suffices to find positive constants $C_\xi,A_\xi$ such that 
    \[ |\xilfphia(s;\delta)|\leq C_\xi A_\xi^{|\Re(s)|}, \]
    as this function is bounded when $|\Re(s)|$ is bounded as is the case with vertical strips. 
    
    To that end, let $\sigma=\Re(s)$ and define $\Lambda_+=\max(\set{\lambda_\ph^\alpha}_{\ph\in\Phi})$ and $\Lambda_-=\min(\set{\lambda_\ph^\alpha}_{\ph\in\Phi})$. Then we have that 
    \begin{align*}
        |\xilfphia(s;\delta)| 
            &\leq \sum_{\ph\in\Phi} \lambda_\ph^{\alpha\sigma}
                \left|\Mellin_\delta^{\delta/\lambda_\ph^\alpha}[f](s)\right| \\
            &\leq \max(\Lambda_+^\sigma,\Lambda_-^\sigma) \sum_{\ph\in\Phi}
                \left|\Mellin_\delta^{\delta/\lambda_\ph^\alpha}[f](s)\right|.
    \end{align*}
    By Corollary~\ref{cor:MellinBounds}, we have that each Mellin transform $\Mellin_\delta^{\delta/\lambda_\ph^\alpha}[f]$ is bounded on any vertical strip of the form $\HH_{a}^{b}$. It follows that a finite sum of such functions is bounded on any such vertical strip by the sum of these constants, which we denote by $C_\xi$. Choosing $A_\xi=\max(\Lambda_-,\Lambda_+,\Lambda_-^{-1},\Lambda_+^{-1})$ allows us to write that $A_\xi^{|\sigma|}\geq \max(\Lambda_+^\sigma,\Lambda_-^\sigma)$, and thus obtaining the desired bound. It follows that $|\xilfphia|$ is bounded in any vertical strip. 

    Now let $\set{\tau_n}_{n\in\ZZ}$ be a sequence of admissible heights shared jointly by $\zeta_\Phi(\alpha s)$ and $\zeta_R(s;\delta)$, which exists by fiat. We have that each of these respective functions is uniformly bounded on all of the horizontal contours $I_n=[S(\tau)+i\tau_n,c+i\tau_n]$, for some particular $c$ sufficiently large so that $\zeta_\Phi(\alpha s)$ and $\zeta_R(s;\delta)$ are holomorphic in $\HH_c$. Since $\xilfphia$ is bounded on the strip $\HH_{\inf(S)}^c$, it follows that it is uniformly bounded on all of these intervals $I_n\subset\HH_{\inf(S)}^c$. It follows that $\zeta_f(s;\delta)=\zeta_\Phi(\alpha s)(\xilfphi(s;\delta)+\zeta_R(s;\delta))$ is uniformly bounded on these intervals. This establishes hypothesis \textbf{L1} for the screen $S$. 

    For hypothesis \textbf{L2}, we note that $\zeta_R(s;\delta)$ and $\zeta_\Phi(\alpha s)$ are bounded on $S$ by assumption of their joint languidity. Because $S$ is contained in the strip $\HH_{\inf(S)}^{\sup(S)}$ (noting that screens are by definition bounded), we have that $\xilfphia$ is bounded on $S$. The boundedness of $\zeta_f$ on this screen thus follows. Thus, $\zeta_f$ is languid with respect to the same screen $S$ (with exponent $\kappa=0$). 

\end{proof}

\subsubsection{Proof of the Explicit Formulae}

With the languid growth conditions of $\zeta_f$ having been established, we may now complete the proofs of the main theorems, Theorems~\ref{thm:pointwiseFormula} and \ref{thm:distFormula}. We shall treat the proofs of these theorems together as they require similar setup and estimates. The main differences come in the application of different theorems to provide the relevant result. Starting from the result of Proposition~\ref{prop:MellinInversion}, we will compute the resulting contour integral by the methods in \cite[Theorem~5.1.11]{LRZ17_FZF} and \cite[Theorem~5.2.2]{LRZ17_FZF} for the pointwise and distributional explicit formula, respectively. 

\begin{proof}
    Fix $c$ such that $\beta/\alpha+1>c>\max(D/\alpha,\sigma_R)$. By Proposition~\ref{prop:MellinInversion}, we have that for any $t\in(0,\delta)$,
    \begin{align*}
        f(t) &= \frac{1}{2\pi i} \int_{c-i\infty}^{c+i\infty} 
            t^{-s} \zeta_\Phi(\alpha s)(\xilfphia(s;\delta)+\zeta_R(s;\delta))\,ds,
    \end{align*}
    where $\zeta_f$, $\xilfphia$, and $\zeta_R$ are as in Corollary~\ref{cor:structureOfPoles}. By Theorem~\ref{thm:languidityFromSFEs}, we have that $\zeta_f$ is languid with exponent $\kappa=0$ on the admissible screen $S$.

    Noting that $F(t)=t^{\beta/\alpha}f(t)$, we have that 
    \begin{align*}
        F(t) &= \frac{1}{2\pi i} \int_{c-i\infty}^{c+i\infty} 
            t^{\beta/\alpha-s} \zeta_\Phi(\alpha s)(\xilfphia(s;\delta)+\zeta_R(s;\delta))\,ds.
    \end{align*}
    It is this expression that is analogous to \cite[Equation~(5.1.18)]{LRZ17_FZF}, but with $\beta$ in place of $\dimension$, $\alpha=1$, and $\zeta_f(s;\delta)$ in place of the relative tube zeta function $\tubezeta_{A,\Omega}(s;\delta)$. Under these conditions, \cite[Theorem~5.1.11]{LRZ17_FZF} and \cite[Theorem~5.2.2]{LRZ17_FZF} are essentially applicable with only minor modifications and justifications required in the general case. 
    
    Firstly, \cite[Proposition~5.1.8]{LRZ17_FZF} is applicable, where the interchange of the order of integration may be justified by the explicit formula for $\zeta_f$ (cf. Corollary~\ref{cor:structureOfPoles}) and explicit bounds for the constituent functions. The uniform estimates for $\zeta_R$ and $\xilfphia$ follow from Corollary~\ref{cor:MellinBounds}. 
    
    For an estimate of $\zeta_\Phi$, note that when $\sigma=\Re(s)>D/\alpha$,
    \begin{align*}
        \left| \sum_{\ph\in\Phi} \lambda_\ph^{\alpha s}  \right| \leq p(\alpha \sigma):=\sum_{\ph\in\Phi} \lambda_\ph^{\alpha\sigma} < p(D) =1,
    \end{align*}
    using the fact that $p$ is strictly decreasing. It follows that there is a uniform bound (based on $c>D/\alpha$) for $\zeta_\Phi(\alpha s)$ when $\Re(s)=c$. Put together, we obtain that the function $t^{\beta/\alpha-s}\zeta_f(s;\delta)$, viewed as a multivariate function with variables $t$ and $s$, is integrable for $(t,s)\in [0,\delta]\times (c-i\infty,c+i\infty)$ provided that $\beta/\alpha+1>c>D/\alpha$. The assumption that $\beta/\alpha+1>c$ ensures that $t^{\beta/\alpha-s}$ is integrable as $t\to0^+$. Next, \cite[Lemma~5.1.10]{LRZ17_FZF} (with $\kappa=0$) is directly applicable based on the languidity of the function $\zeta_f$ (see Theorem~\ref{thm:languidityFromSFEs}) followed by \cite[Theorem~5.1.11]{LRZ17_FZF}. This establishes the proof of Theorem~\ref{thm:pointwiseFormula}. 

    Next, this integrability argument is also necessary to prove Theorem~\ref{thm:distFormula}, namely to justify the use of the Fubini--Tonelli theorem in \cite[Equation~(5.2.10)]{LRZ17_FZF} (in the proof of \cite[Theorem~5.5.5]{LRZ17_FZF}). In this case, because $\testfn\in\Ss(0,\delta)$, we may write the corresponding integral in question as  
    \begin{align*}
        &\int_{c-i\infty}^{c+i\infty}\int_0^\infty 
            \testfn(t)\frac{t^{\beta/\alpha-s+k}}{(\beta/\alpha-s+1)_k}\zeta_f(s;\delta)  \,dt \\
        &= \int_{c-i\infty}^{c+i\infty}\int_0^\delta 
            \testfn(t)\frac{t^{\beta/\alpha-s+k}}{(\beta/\alpha-s+1)_k}\zeta_f(s;\delta)  \,dt.
    \end{align*}

    We claim that once again, the integrand is integrable over the product space, that is, for $(t,s)\in[0,\delta]\times(c-i\infty,c+i\infty)$. Firstly, we have the same estimates for $\zeta_f$ as before. For the Pochhammer symbol, we may write that 
    \begin{align*}
        |(\beta/\alpha-s+1)_k| 
            &\geq |\beta/\alpha-c+1|\cdot |\beta/\alpha-c+2|\cdots |\beta/\alpha-c+k-1| \\
            &\geq |\beta/\alpha-c+1|^k,
    \end{align*}
    noting that on the vertical lines $z=\beta/\alpha-c+k+i\RR$, the closest point to the origin occurs exactly on the real axis. The second estimate follows from using the fact that $\beta/\alpha-c+1>0$ is the smallest term in the product. This gives a uniform upper bound for $|(\beta/\alpha-s+1)_k^{-1}|$. Lastly, the power of $t$ may be arbitrary since $\testfn$ is a function of rapid decrease: for any $\gamma\in\RR$, $t^\gamma\testfn(t)\to0$ as $t\to0^+$.  
    
    The final step is a minor change of variables. Rather than indexing the sum by the poles of $\zeta_f$ in $W_S$, we note that as long as $\Re(s)>\sigma_R$, the abscissa of absolute convergence of $\zeta_R$, a pole of $\zeta_f$ only occurs at $s$ if $\omega=\alpha s$ is a pole of $\zeta_\Phi$. In other words, we have that 
    \[
        \sum_{s\in\Dd_f(\HH_{\sigma_R})} g(s) = \sum_{\omega/\alpha\in\Dd_f(\HH_{\sigma_R})} g(\omega/\alpha) = \sum_{\omega\in\Dd_\Phi(\HH_{\alpha \sigma_R})} g(\omega/\alpha).
    \]
    Since we assume that $S$ lies in the half-plane $\HH_{\sigma_R}$, this same argument applies with respect to the window $W_S$ and its scaled transformation $\alpha W_S$. This establishes the proof of Theorem~\ref{thm:distFormula}.
    
\end{proof}

\section{Heat Content Explicit Formulae}
\label{sec:heatAnalysis}

\subsection{Heat Zeta Functions}
\label{ssc:heatZF}

Suppose that $\Region\subset\RR^\dimension$ is a bounded, self-similar set and let $\Phi$ be a self-similar system with $\Region$ as its attractor. Suppose that $\Omega\subset\RR^\dimension$ is an osculating set for $\Phi$ so that $(X,\Omega)$ is an osculant fractal drum (see Definition~\ref{def:oscRFD}). Let 
\begin{equation}
    \label{eqn:partition}
    X = \enclose{\biguplus_{\ph\in\Phi} \ph[\Omega]} \uplus \remset
\end{equation}
be a partition of $X$ where $\remset$ is defined as $\remset:=X\setminus (\cup_{\ph\in\Phi}\ph[\Omega])$. Here, $\uplus$ denotes a disjoint union, and that this union is disjoint follows from the definition of an osculating set.  

Our goal is to use the methods of Section~\ref{sec:SFE} to study Problem~\ref{prob:specificHeatProblem} on the region $\Region$. If the self-similar system $\Phi$ induces a decomposition of the total heat contents (in the sense of Definition~\ref{def:inducedDecomp}), we can use Corollary~\ref{cor:heatScalingLaw} to obtain a scaling functional equation. To this end, we now establish some preliminaries and notation for the application of this theory. 

We begin with the notion of a \textit{heat zeta function}, defined in analogy with the definition for tube zeta functions and their normalized quantities which satisfy a scaling law. We define it for the general problem on a cylindrical set and in the case of Problem~\ref{prob:specificHeatProblem}, we say that the heat zeta function is a parabolic (heat) zeta function owing to the nature of the solution $u_\Region$ as a parabolic average at the points of the given set $\Region$. 
\begin{Definition}[Heat Zeta Function]
    \label{def:heatZetaFn}
    Let $u_\Region(x,t)$ be the PWB solution to Problem~\ref{prob:generalHeatProblem} on the open set $\Region\subset\RRNplus$ with a resolutive boundary function $F=f\1_{t=0}+g\1_{x\in\partial\Region}$. \medskip

    Then we define the \textbf{heat zeta function} of $\Region$ with boundary conditions $F$ to be
    \begin{equation}
        \label{eqn:defHeatZeta}
        \genheatzeta(s;\delta) := \Mellin^\delta[t^{-\dimension/2}\heatContent(t)](s)
    \end{equation}
    for all $s\in\CC$ for which the integral converges and extended by analytic continuation to some open connected subset of $\CC$. If $F=\1_{\partial\Region}$ as in Problem~\ref{prob:specificHeatProblem}, we say that $\heatzeta:=\genheatzeta$ is a \textbf{parabolic} (heat) zeta function. 
\end{Definition}
Note that if $\heatContent(t)$ is integrable, bounded, and satisfies $\heatContent(t)=O(t^{-\sigmaRem})$ as $t\to0^+$ for some $\sigmaRem\in\RR$, then by Lemma~\ref{lem:MellinHolo} we have that $\genheatzeta$ is holomorphic in $\HH_{\sigmaRem}$. When we establish functional equations for the heat content induced by a self-similar system $\Phi$, we also note that the relevant scaling operator will be $L_\Phi^2:=\sum_{\ph\in\Phi}M_{\lambda_\ph^2}$, but the associated scaling zeta function $\zeta_\Phi$ is still the primary function of interest since $\zeta_\Phi(2s)=\zeta_{L_\Phi^2}(s)$. Owing to the effect of this quadratic scaling, viz. $\alpha=2$ as in Section~\ref{sec:SFE}, we will also state our estimates for the heat content remainders in the form $R(t)=O(t^{-\sigmaRem/2})$ (as $t\to0^+$). We state our results provided some control over the decomposition remainder, as in Definition~\ref{def:heatDecompRem}. 

\begin{Theorem}[Formula for the Heat Zeta Function]
    \label{thm:heatZetaFormula}
    \index{Heat zeta function!Explicit formula}
    Let $\heatContent$ be the heat content of the PWB solution to Problem~\ref{prob:specificHeatProblem}. Let $\Phi$ be a self-similar system and suppose that $\heatContent$ decomposes according to $\Phi$ with decomposition remainder $\decompRem$ (as in Definition~\ref{def:heatDecompRem}) which satisfies $\decompRem(t)=O(t^{\rempow/2})$ as $t\to0^+$ for some $\sigmaRem\in\RR$.
    \medskip 

    Define $\sigma_R:=\sigmaRem/2$ and let $\zeta_\Phi$ be the scaling zeta function associated to $\Phi$. Then for any $\delta>0$ and for all $s\in\HH_{\sigma_R}\setminus\Dd_\Phi$,
    \begin{equation}
        \label{eqn:heatZetaFormula}
        \heatzeta(s;\delta) = \zeta_\Phi(2s)(\partialheatzeta(s;\delta)+\zeta_R(s;\delta)),
    \end{equation}
    with $\heatzeta(s;\delta)$ meromorphic in $\HH_{\sigma_R}$, having poles contained in a subset of $\Dd_\Phi$, the set of poles of $\zeta_\Phi$. Here, $\zeta_R(s;\delta):= \Mellin^\delta[t^{-\dimension/2}\decompRem(t)](s)$ is a holomorphic function in $\HH_{\sigma_R}$ and 
    \begin{equation}
        \label{eqn:partialZetaDef}
        \partialheatzeta(s;\delta):= \sum_{\ph\in\Phi}\lambda_\ph^{2s}\Mellin_\delta^{\delta/\lambda_\ph^2}[t^{-\dimension/2}\heatContent(t)](s)
    \end{equation}
    is an entire function. 
\end{Theorem}

\begin{proof}
    By Corollary~\ref{cor:heatScalingLaw}, we have that $\heatContent$ satisfies a $2$-scaling law. Together with the presupposed induced decomposition, by Proposition~\ref{prop:inducedSFE} we have that the normalized $\heatContent$ satisfies the scaling functional equation 
    \begin{equation}
        \label{eqn:heatSFE}
        t^{-\dimension/2}\heatContent(t) 
            = L_\Phi^2[t^{-\dimension/2}\heatContent(t)] + t^{-\dimension/2}\decompRem(t)
    \end{equation}
    for any $t\geq 0$. The result is then a corollary of Theorem~\ref{thm:zetaFormula}, noting that the normalized remainder satisfies $R(t)=t^{-\dimension/2}\decompRem(t)=O(t^{-\sigma_R})$ at $t\to0^+$, with $\sigma_R=\sigmaRem/2$.
\end{proof}

Note that this formula implies that the poles of the heat zeta function $\heatzeta$ in the half-plane $\HH_{\sigma_R}=\HH_{\sigmaRem/2}$ occur exactly at the points $\omega/2$ where $\omega\in\Dd_\Phi(\HH_{\sigmaRem})$. Here, we see that the scaling zeta function $\zeta_\Phi$, determined from the scaling ratios and their multiplicities alone, also governs the nature of the zeta functions for self-similar fractals.

As we will use this scaling functional equation in order to obtain explicit formulae in what follows, we record this result as a proposition to which we will refer later. 
\begin{Proposition}[Heat Scaling Functional Equation]
    \label{prop:heatSFE}
    \index{Scaling functional equation (SFE)!Induced SFE of heat content}
    Let $\heatContent$ be the heat content of the PWB solution to Problem~\ref{prob:specificHeatProblem}. Let $\Phi$ be a self-similar system and suppose that $\heatContent$ decomposes according to $\Phi$ with decomposition remainder $\decompRem$ (as in Definition~\ref{def:heatDecompRem}) which satisfies $\decompRem(t)=O(t^{\rempow/2})$ as $t\to0^+$ for some $\sigmaRem\in\RR$.
    \medskip 

    Then for all $t\geq 0$, the heat content $\heatContent$ satisfies the scaling functional equation given by \eqref{eqn:heatSFE}. Note that the normalized remainder $R(t)=t^{-\dimension/2}\decompRem(t)$ satisfies the corresponding estimate $R(t)=O(t^{-\sigma_R})$ as $t\to0^+$, where $\sigma_R:=\sigmaRem/2$.  
\end{Proposition}

\subsection{General Heat Content Results for Self-Similar Sets}
\label{ssc:genHeatFormulae}

We now state some general results regarding pointwise and distributional explicit formulae for the heat contents of open bounded regions $\Region\subset\RRN$ whose boundary $\Regionbd$ is a self-similar set and for which $(\Regionbd,\Region)$ is an osculant fractal drum. The pointwise expansions require integrating the heat content sufficiently many times to improve regularity of the function to be obtained. Meanwhile, the distributional identities (while less direct) remove this technical hurdle.

First, we state the pointwise identity, given in terms of antiderivatives of the heat content. We denote by  $\heatContent^{[k]}$ the $k^{\text{th}}$ antiderivative of the heat content defined so that $\heatContent^{[k]}(0)=0$. Explicitly, it may be defined by recursion with $\heatContent^{[0]}:=\heatContent$ and for $k>0$ by
\[
    \heatContent^{[k]}(t) := \int_0^t \heatContent^{[k-1]}(\tau)\,d\tau.
\] 
As an additional preliminary to stating the result, we define the Pochhammer symbol $(z)_w:=\Gamma(z+w)/\Gamma(w)$ for $z,w\in\CC$. Note that when $w=k$ is a positive integer, this simplifies to $(z)_k=z(z+1)\cdots(z+k-1)$ or, when $w=0$, to $(z)_0=1$. Lastly, we note that the summations to follow are defined as symmetric limits of the sums taken over values of $\omega$ with bounded imaginary parts for increasingly large upper bounds. 

\begin{Theorem}[Heat Content, Pointwise Expansion]
    \label{thm:heatFormulaPtw}
    \index{Heat content!Pointwise explicit formulae}
    Let $\heatContent$ be the heat content of the PWB solution to Dirichlet Problem~\ref{prob:specificHeatProblem}. Let $\Phi$ be a self-similar system and suppose that $\heatContent$ decomposes according to $\Phi$ with decomposition remainder $\decompRem$ satisfying $\decompRem(t)=O(t^{\rempow/2})$ as $t\to0^+$, for some $\sigma_0\in\RR$. Suppose either that $\sigmaRem<D_\ell\leq\lowersimdim(\Phi)$, the bound for $\lowersimdim(\Phi)$ from Proposition~\ref{prop:simDimBounds}, or that the (distinct) scaling ratios of $\Phi$ are arithmetically related (see Definition~\ref{def:latticeDichotomy}).
    \medskip

    Let $k\in\ZZ$ with $k\geq 2$ and let $\delta>0$. Then for all $t\in(0,\delta)$, we have that
    \[ 
        \heatContent^{[k]}(t) = \sum_{\omega\in\Dd_\Phi(\HH_{\sigmaRem})} 
            \Res\Bigg(\cfrac{t^{(\dimension-s)/2+k}}{((\dimension-s)/2+1)_k}\heatzeta(s/2;\delta);\omega\Bigg)
            + \Rr^k(t),
    \]
    where $\heatzeta$ is as in Theorem~\ref{thm:heatZetaFormula}. For any $\e>0$ sufficiently small, the remainder satisfies $\Rr^k(t)=O(t^{\rempow/2-\e+k})$ as $t\to0^+$. 
\end{Theorem}

\begin{proof}
    By Proposition~\ref{prop:heatSFE}, we have that the normalized heat content satisfies a scaling functional equation with respect to the operator $L_\Phi^2$ and remainder $R(t):=t^{-\dimension/2}\decompRem(t)$. Letting $\sigma_R=\sigmaRem/2$, we note that $R(t)=O(t^{-\sigma_R})$ as $t\to0^+$ and that $\sigmaRem<D_\ell\leq\lowersimdim(\Phi)$ implies that $\sigma_R<D_\ell/2$. So in the first case, we apply Corollary~\ref{cor:lowerDimAdmissibilityRescaled} to obtain that $R$ is an admissible remainder with respect to screens of the form $S_\e(\tau)\equiv \sigma_R+\e$ when $0<\e<D_\ell/2-\sigma_R$. Meanwhile, in the lattice case, we apply Theorem~\ref{thm:latticeCaseAdmissibility} to obtain that any screen of the form $S_\e(\tau)=\sigma_R+\e$ are admissible for all but finitely many $\e>0$. In either case, $S_\e$ is admissible for any sufficiently small $\e$. 
    
    We may now apply Theorem~\ref{thm:pointwiseFormula}. Here, $\beta=\dimension$, $\alpha=2$, and $\delta>0$ is arbitrary. Note that $\dimension/2\leq\simdim(\Phi)/2$ as a corollary of Moran's theorem and the open set condition and that $\dimension/2\geq \sigma_R$ since Proposition~\ref{prop:heatContentBounded} implies that $\zeta_R$ is holomorphic in the half-plane $\HH_{\dimension/2}$ as noted in the discussion following Definition~\ref{def:heatDecompRem}. Here, with the estimate $\decompRem(t)=O(t^{\rempow/2})$ as $t\to0^+$, we obtain by Lemma~\ref{lem:MellinHolo} that $\zeta_R$ is holomorphic in $\HH_{\sigma_R}$. Note also that $S_\e$ is chosen with $S_\e(\tau)\equiv\sigma_R+\e>\sigma_R$, so $W_{S_\e}=\HH_{\sigma_R+\e}$ and is contained in $\HH_{\sigma_R}$. Further, we note that for the estimate of the remainder, $\beta/\alpha-\sup(S)+k$ in the theorem is $\rempow/2-\e+k$ since $\sup(S)=\sigma_R+\e=\sigmaRem/2+\e$ and $\beta/\alpha=\dimension/2$. 

    Lastly, we simplify the sum over $\Dd_\Phi(\alpha W_S)=\Dd_\Phi(2\HH_{\sigma_R+\e})$. Firstly, note that $2\HH_{\sigma_R+\e}=\HH_{2\sigma_R+2\e}$. Since $\zeta_R$ is holomorphic in $\HH_{\sigma_R}$, $\zeta_R(s/2)$ is holomorphic when $s\in\HH_{2\sigma_R}=\HH_{\sigmaRem}$. In both cases, we may choose $\e$ sufficiently small so that $\zeta_\Phi(2s)$ has no poles of the form $s=\omega$ with $\Re(\omega)\in(\sigmaRem,\sigmaRem+2\e)$. In the case when $\sigmaRem<D_\ell$, take $\e<(D_\ell-\sigmaRem)/2$ and in the lattice case choose $2\e$ smaller than the distance between $\sigmaRem$ and the closest exceptional point. Then, with the change of variables, we see that when $\Re(s)\in(\sigmaRem,\sigmaRem+2\e)$, the function $\heatzeta(s/2)$ has no poles because $\zeta_\Phi(s)$ has no poles with $\Re(s)\in(\sigmaRem,\sigmaRem+2\e)$ (using Theorem~\ref{thm:heatZetaFormula}). Thus the sum over $\Dd_\Phi(\HH_{\sigmaRem+2\e})$ is the same as over $\Dd_\Phi(\HH_{\sigmaRem})$. 
\end{proof}

The leading order term occurs at the pole $\omega=D:=\simdim(\Phi)$, as this is a pole of $\zeta_\Phi$ (and hence of $\heatzeta$) having the largest real part, provided that the residue is nonvanishing. If $D$ is a simple pole and the only pole with real part $D$, then the leading order term is explicitly 
\[ \frac1{((\dimension-D)/2+1)_k}\Res(\heatzeta(s/2;\delta);D)\,t^{(\dimension-D)/2+k}. \]
When the scaling ratios are non-arithmetically related, $D$ will be the unique pole with real part $D$; when the scaling ratios are arithmetically related, this will not be true. Instead, the leading order terms will form the Fourier expansion of a periodic function (when viewed additively after a change of variable) owing to the nature of the structure of the poles of $\zeta_\Phi$ in this case: they lie on finitely many vertical lines and are periodically spaced. See for instance the proof of Theorem~\ref{thm:latticeCaseAdmissibility} in this work for an explicit proof of this and more generally we refer the reader to the discussion in \cite[Chapters~2 and 3]{LapvFr13_FGCD} regarding the structure of the complex dimensions of self-similar fractal harps (and their natural generalizations), noting that these zeta functions have the same structure as $\zeta_\Phi$, and thus controlling the poles of $\heatzeta$.

For the distributional setting, the restriction of $k\geq 2$ may be relaxed. For these formulae, recall from Section~\ref{sec:SFE} that (just as in \cite[Chapter~5]{LapvFr13_FGCD} and \cite[Chapter~5]{LRZ17_FZF}) we will use as the space of test functions the set of Schwartz functions. These are the functions of rapid decrease near the boundary, given explicitly by
\begin{equation}
    \label{eqn:defSchwartzFns}
    \Ss(0,\delta):=\Bigg\{\testfn\in C^\infty(0,\delta)\,\Bigg|\, 
    \begin{aligned}
        &\forall m\in\ZZ,\,\forall q\in\NN,\,t^m\testfn^{(q)}(t)\to0\\
        &\text{and }(t-\delta)^m\testfn^{(q)}(t)\to0\text{ as }t\to0^+
    \end{aligned}
    \Bigg\}.
\end{equation}
The tempered distributions are elements of the dual space, $\Ss'(0,\delta)$. 

\begin{Theorem}[Heat Content, Distributional Expansion]
    \label{thm:heatFormulaDist}
    \index{Heat content!Distributional explicit formulae}
    Let $\heatContent$ be the heat content of the PWB solution to Dirichlet Problem~\ref{prob:specificHeatProblem}. Let $\Phi$ be a self-similar system and suppose that $\heatContent$ decomposes according to $\Phi$ with decomposition remainder $\decompRem$ satisfying $\decompRem(t)=O(t^{\rempow/2})$ as $t\to0^+$, for some $\sigmaRem\in\RR$. Suppose either that $\sigmaRem<D_\ell\leq\lowersimdim(\Phi)$, the bound for $\lowersimdim(\Phi)$ from Proposition~\ref{prop:simDimBounds}, or that the (distinct) scaling ratios of $\Phi$ are arithmetically related (see Definition~\ref{def:latticeDichotomy}).
    \medskip

    Let $k\in\ZZ$ and let $\delta>0$. Then the heat content, viewed as a tempered distribution, satisfies the identity
    \begin{equation}
        \label{eqn:heatFormulaDist}
        \begin{split}
            \heatContent^{[k]}(t) &= \sum_{\omega\in\Dd_\Phi(\HH_{\sigmaRem})} 
                \Res\Bigg(\cfrac{t^{(\dimension-s)/2+k}}{((\dimension-s)/2+1)_k}\heatzeta(s/2;\delta);\omega\Bigg)
                + \Rr^{[k]}(t),
        \end{split} 
    \end{equation}
    as $t\to0^+$, where two distributions are equal if they action on an arbitrary test function agrees. See \eqref{eqn:bracketIdentityHeat} for the explicit identity of action on test functions. Here, $\heatzeta$ is as in Theorem~\ref{thm:heatZetaFormula}. The remainder term, as a distribution, satisfies the estimate that for any $\e>0$ sufficiently small, $\Rr(t)=O(t^{\rempow/2-\e+k})$ as $t\to0^+$, in the sense of \eqref{eqn:distRemEstHeat}.
\end{Theorem}
\begin{proof}
    The proof is the same as Theorem~\ref{thm:heatFormulaPtw} but through application of Theorem~\ref{thm:distFormula}. 
\end{proof}

To say that \eqref{eqn:heatFormulaDist} is an identity of distributions means the following. For any test function $\testfn\in\Ss(0,\delta)$, 
\begin{equation}
    \label{eqn:bracketIdentityHeat}
    \begin{split}
        \bracket{\heatContent^{[k]},\testfn} &= \sum_{\omega\in\Dd_\Phi(\HH_{\sigmaRem})} 
        \Res\Bigg(\cfrac{\Mellin[\testfn]({(\dimension-s)/2+k+1})}{((\dimension-s)/2+1)_k}\heatzeta(s/2);\omega\Bigg)
        + \bracket{\Rr^{[k]},\testfn}.
    \end{split}
\end{equation}
The distributional remainder estimate is equivalent to the statement that for all $a>0$ and for all $\testfn\in\Ss(0,\delta)$,
\begin{equation}
    \label{eqn:distRemEstHeat}
    \Big\langle\Rr^{[k]}(t),\frac1a\testfn(t/a)\Big\rangle = O(a^{\rempow/2-\e+k}),
\end{equation}
as $t\to0^+$. While this formulation is less direct compared to the pointwise expansion, it does have the advantage of requiring less regularity to leverage the expansion. Explicitly, one may set $k=0$ and obtain a distributional identity for the heat content itself. Note that in this case, the Pochhammer symbol simplifies to $((\dimension-s)/2+1)_0=1$.

\subsection{Heat Content Results for Generalized von Koch Snowflakes}
\label{ssc:heatAppGKF}

We apply here the general theory developed in Section~\ref{ssc:genHeatFormulae} to the family of generalized von Koch snowflake domains, a family of self-similar fractal drums including the classic von Koch snowflake and generalizations of its construction. In particular, we provide explicit formulae for the heat content of these domains with self-similar boundary, which are expressed in terms of the complex dimensions of the generalized von Koch fractals. The analysis of the heat content of fractal snowflakes originates with the work of Fleckinger, Levitin, and Vassiliev \cite{FLV95a,FLV95b} on the von Koch snowflake and then the later work of van den Berg and collaborators \cite{vdB00_generalGKF,vdB00_squareGKF,vdBGil98,vdBHol99} on generalized von Koch fractals. The main tool in their analysis is the renewal theorem, developed by Feller \cite{Fel50} in the field of probability. Here, we recover and extend these results as a corollary of the results of Section~\ref{sec:SFE} in the case of heat content (as in Section~\ref{ssc:genHeatFormulae}).

\subsubsection{Generalized von Koch Fractals}

\begin{figure}[t]
    \centering
    \subfloat{\includegraphics[width=4cm]{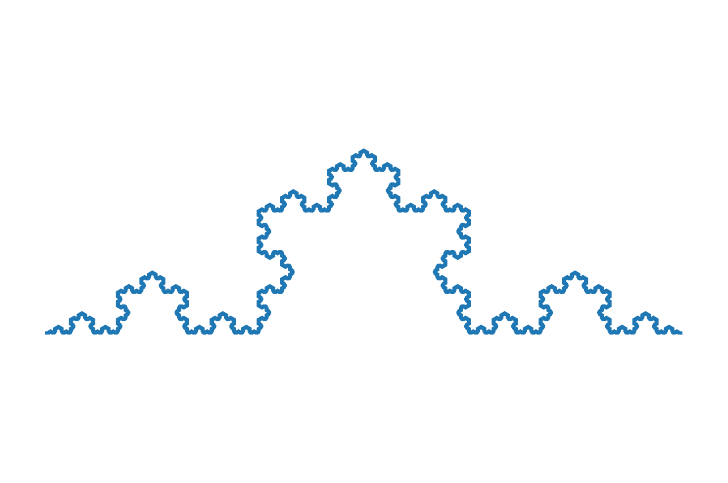}}
    \qquad
    \subfloat{\includegraphics[width=4cm]{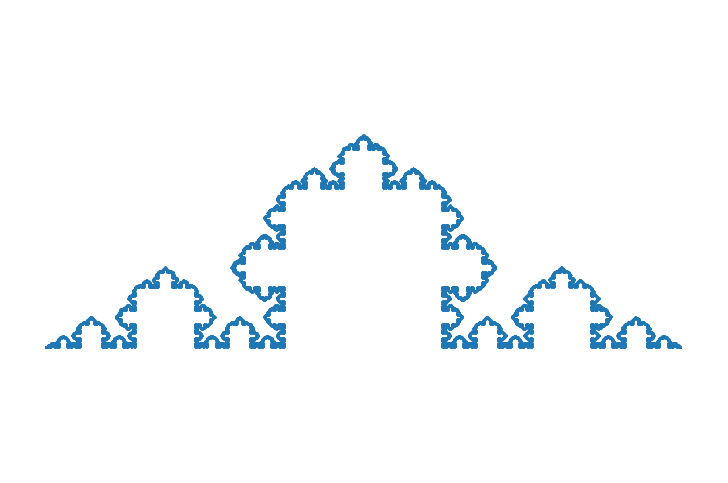}}
    \caption[The von Koch curve and a generalization thereof]{Two fractal curves: a planar curve having nowhere-defined tangent introduced by von Koch (left) \cite{Koch1904} and a generalized von Koch curve (right).}
    \label{fig:vKCurves}
\end{figure}

Generalized von Koch fractals are a class of domains with self-similar fractal boundary obtained by generalizing the construction of its namesake, the von Koch snowflake. The von Koch curve (depicted in Figure~\ref{fig:vKCurves}) was introduced and studied by von Koch as a more geometric/elementary alternative to Weierstrass' nowhere differentiable function, being a planar curve with nowhere-defined tangents \cite{Koch1904,Koch1906}. 

In general, an $(n,r)$-von Koch curve may be defined as the invariant set of an explicit self-similar system $\Phinr$ consisting of two mappings with scaling ratio $\ell=(1-r)/2$ and $n-1$ mappings of scaling ratio $r$; see \eqref{eqn:defGKCsystem} below for an explicit definition. Here, $n\geq3$ refers to the type of regular polygon used in the construction and $r$ refers to the proportion of the line segments removed from the middle at each step (where a regular $n$-gon of that same length is attached, with the bottom edge removed). When $r$ is sufficiently small, the resulting curve is topologically simple \cite{KP10}. The following bound is a sufficient condition.
\begin{Proposition}[Self-Avoidance of GKFs \cite{KP10}]
    \label{prop:selfAvoid}
    An $(n,r)$-von Koch curve has no self-intersections if the scaling ratio $r>0$ satisfies
    \begin{align*} 
        r &< \frac{\sin^2(\pi/n)}{\cos^2(\pi/n)+1}, &&\text{ if }n\text{ is even, and }\\
        r &< 1-\cos(\pi/n),                         &&\text{ if }n\text{ is odd.}
    \end{align*}
    The boundary of a corresponding $(n,r)$-von Koch snowflake is topologically simple under these conditions.
\end{Proposition}

\begin{figure}[t]
    \centering
    \subfloat{\includegraphics[trim=60 0 60 0,clip,width=3cm]{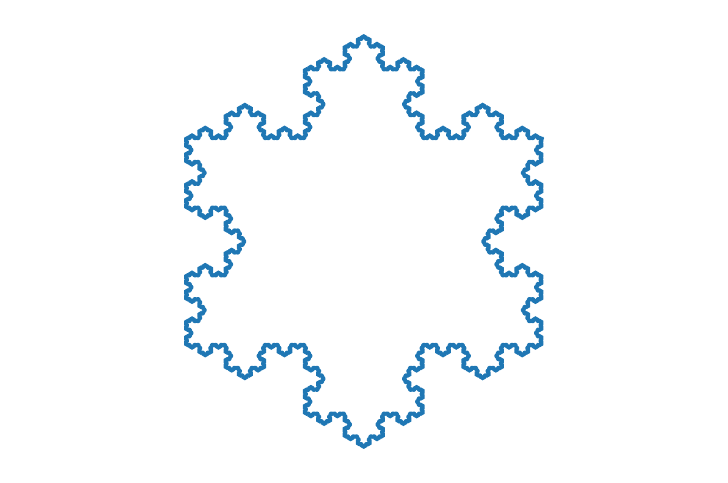}}
    \qquad 
    \subfloat{\includegraphics[trim=60 0 60 0,clip,width=3cm]{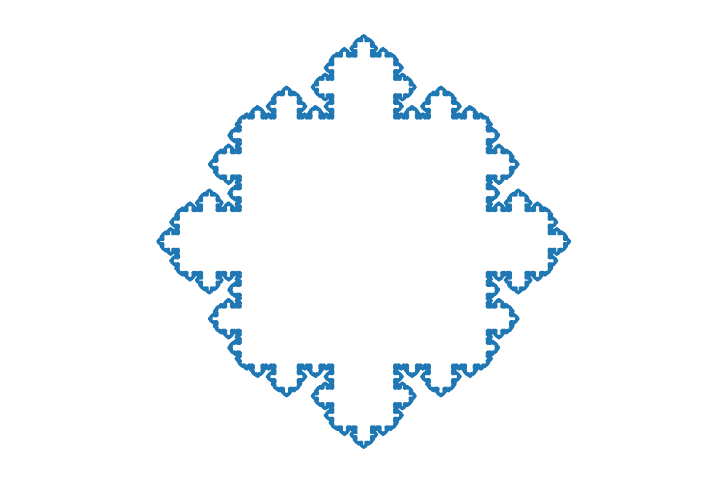}}
    \qquad 
    \subfloat{\includegraphics[trim=60 0 60 0,clip,width=3cm]{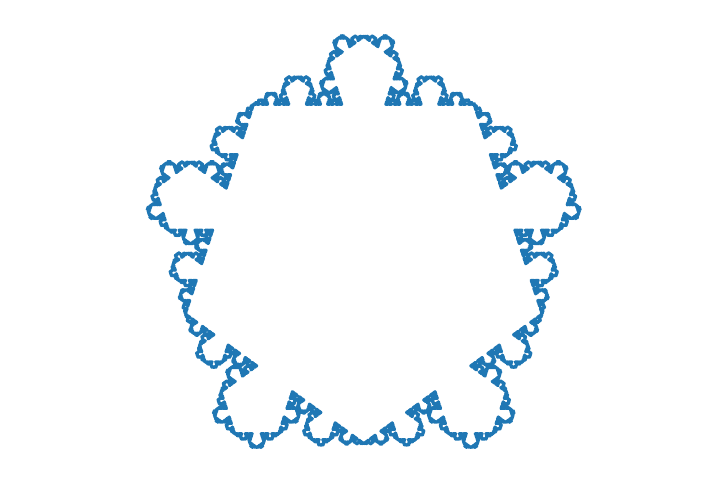}}
    \caption[The von Koch snowflake and two generalizations thereof]{The von Koch snowflake (left) and two of its generalizations, the ``squareflake'' (middle) and the ``pentaflake'' (right).}
    \label{fig:threeGKFs}
\end{figure}

An $(n,r)$-von Koch snowflake is a union of $n$ copies of an $(n,r)$-von Koch curve arranged along the edges of a regular $n$-gon. Figure~\ref{fig:threeGKFs} depicts the von Koch snowflake (with $n=3$, $r=\frac13$) as well as two generalizations both corresponding to simple curves, with $n=4$ and $r=\frac14$ and with $n=5$ and $r=\frac15$, respectively. 

To be precise, the union of curves (when $r$ is sufficiently small) defines a topologically simple, closed curve which we call an $(n,r)$-von Koch snowflake boundary. By the Jordan curve theorem, such a curve partitions its complement into two domains; we call the bounded component an $(n,r)$-von Koch snowflake domain or snowflake interior. The Dirichlet problem we will consider is for the domain $\Region$ which is this interior, and $\Regionbd$ is its boundary, which is a union of $(n,r)$-von Koch curves.

\subsubsection{Preliminaries}

Using the results of Section~\ref{sec:SFE}, we will recover the results of \cite{vdB00_generalGKF,vdB00_squareGKF,FLV95a} when either $n\geq5$ or in the lattice case. Notably, this extends the results for generalized von Koch fractals to the nonlattice case when $n\geq5$. Furthermore, we explicitly show the role of the complex dimensions in the explicit formulae for the heat content. Results in the nonlattice case for $n=3$ or $4$ may be deduced provided a priori knowledge of pole-free regions and estimates for the explicit function $\zeta_\Phinr$ with the methods established in this work as well.  

Given an $(n,r)$-von Koch snowflake, let $\Knr=\Regionbd$ denote its boundary. Assume that $r$ is sufficiently small (viz. Proposition~\ref{prop:selfAvoid}) so that $\Knr$ is a simple closed curve with a well-defined interior domain, $\Region$.

In what follows, we will consider Problem~\ref{prob:specificHeatProblem}. We let $u_\Region$ be the PWB solution and $\heatContent$ be the associated heat content. Our goal will be to explicitly describe this heat content in the limit as $t\to0^+$ by applying the explicit formulae results. We note that by symmetry, we need only consider the heat content in one portion of $\Omega$ contained in a sector of angle $2\pi/n$ and that the heat content of the total set is $n$ times the heat content in this restricted region. 

To define an $(n,r)$-von Koch curve, where $n\geq3$ is an integer and $r\in(0,1)$, we explicitly specify the self-similar system $\Phinr$ whose invariant set is the curve. To that end, we define the mappings $T_{(a,b)}:\RR^2\to\RR^2$ which translates by the point $(a,b)\in\RR^2$, $R_\theta:\RR^2\to\RR^2$ which rotates a point counterclockwise about the origin by the angle $\theta$, and $S_\lambda:\RR^2\to\RR^2$ which uniformly scales in both coordinates by the factor of $\lambda$. Next, let $\ell=(1-r)/2$ be the conjugate scaling ratio, let $\theta_n=2\pi/n$ be the central angle of a regular $n$-gon, and let $\alpha_n=\pi-2\pi/n$ be the interior angle of a regular $n$-gon. Then $\Phi_{n,r}$ may be defined by
\begin{align}
    \label{eqn:defGKCsystem}
    \begin{split}
        \Phi_{n,r}  &:= \set{ \ph_L, \ph_R, \psi_k:\RR^2\to\RR^2, k=1,...,n-1 }, \\
        \ph_L       &:= S_\ell, \\
        \ph_R       &:= T_{(\ell+r,0)}\circ S_\ell, \\
        \psi_1      &:= T_{(\ell,0)}\circ R_{\alpha_n} \circ S_r, \\
        \psi_k      &:= T_{\psi_{k-1}(1,0)}\circ R_{\alpha_n-(k-1)\theta_n}\circ S_r,\quad k>1. 
    \end{split}
\end{align} 
The maps $\ph_L$ and $\ph_R$ correspond to the left and right pieces of the curve and the mappings $\psi_k$, for $k=1,...,n-1$, correspond to the $n-1$ pieces attached to the edges of a regular $n$-gon adjoined about the middle gap. 

Next, we impose that $r\in(0,\frac13]$ and satisfies the bound in Proposition~\ref{prop:selfAvoid} based on $n$. When $r=\ell=\frac13$ and $n=3$, this is exactly the standard von Koch snowflake (see Figure~\ref{fig:threeGKFs}). The scaling zeta function associated to $\Phinr$ is given explicitly by 
\begin{equation}
    \label{eqn:defKochSZF}
    \zeta_{\Phinr}(s) = \cfrac1{1-(n-1)r^{s}-2\ell^{s}},
\end{equation}
and the set of its poles is precisely the set 
\begin{equation}
    \label{eqn:defPolesKochSZF}
    \Dd_\Phinr(\CC) := \set{\omega\in \CC\suchthat 1 = (n-1)\,r^\omega+2\,\ell^\omega }.
\end{equation}
In the lattice case (Definition~\ref{def:latticeDichotomy}), we have that the poles of $\zeta_\Phinr$ are simple. This is essentially the content of \cite[Proposition 1.3]{vdB00_generalGKF}, but without reference to fractal zeta functions. The Dirichlet polynomial therein, under a change of variables made possible by the lattice constraint that $\log\ell/\log r=p/q$, is the denominator of $\zeta_\Phinr$. Explicitly, $z=r_0^s$, where $r_0$ is the multiplicative generator such that $r=r_0^q$ and $\ell=r_0^p$. 

In order to determine the admissibility of the remainder terms in the nonlattice case, we establish the following bound regarding the lower similarity dimension of $\Phinr$.
\begin{Lemma}[Sufficient Lower Dimension Bounds]
    \label{lem:lowerDimBoundsGKF}
    Let $n\geq3$, $r\in(0,\frac13]$, and define $\ell=(1-r)/2$. Let $\Phi$ a self similar system having scaling ratios $r$ with multiplicity $n-1$ and $\ell$ with multiplicity $2$. Denote by $D_\ell\leq\lowersimdim(\Phi)$ the bound from Proposition~\ref{prop:simDimBounds}. Then we have the following.
    \begin{align*}
        \text{If }n=3,\,D_\ell<0. && \text{If }n=4,\,D_\ell=0. && \text{If }n\geq5,\,D_\ell>0.
    \end{align*}
\end{Lemma}
\begin{proof}
    Let $\zeta_\Phi$ the zeta function of $\Phi$ and let $P(s)$ denote its denominator. Because $r\leq\frac13$, we have that $r\leq \ell$. Suppose first that $r=\ell=\frac13$. Then we have that 
    \[ P(s) = 1 - 2\ell^s-(n-1)r^s = 1-(n+1)3^{-s}. \]
    We may explicitly solve to find that $P(\omega)=0$ when $(n+1)=3^\omega$, or $\omega = \log_3(n+1)+2\pi i k/\log 3$ for $k\in\ZZ$. Since $n\geq3$, the real part of these poles are always positive and exactly equal to $\log_3(n+1)$. Thus, $0<D_\ell=\lowersimdim(\Phi)=\log_3(n+1)$, noting that $D_\ell=\log_3(n+1)$ solves \eqref{eqn:defLowerDimBound}.

    Now suppose that $r<\frac13$, in which case $r<\ell$. Consider the polynomial 
    \[ p(t) = \frac1{n-1} \enclose{r^{-1}}^t + \frac2{n-1} \enclose{\frac{\ell}{r}}^t. \]
    By definition, $D_\ell$ is the unique real solution to $p(t)=1$. Note that $p$ is strictly increasing, so if $p(t)<1$, then $t<D_\ell$. We have that $p(0)=\frac{3}{n-1}$. If $n\geq5$, then $p(0)<1$. If $n=4$, $p(0)=1$ and thus $D_\ell=0$. If $n=3$, $p(0)>1$ whence $D_\ell<0$.  
\end{proof}
It is for this reason that our results for nonlattice $(n,r)$-von Koch snowflakes will be restricted to $n\geq5$ since we need this condition to apply the criterion of Theorem~\ref{thm:lowerDimAdmissibility} with the available estimates.

The last major preliminary is an induced decomposition (in the sense of Definition~\ref{def:inducedDecomp}) and an estimate of the decomposition remainder term, $\decompRem$. This is needed to find a scaling functional equation for the heat content $\heatContent$. A scaling functional equation for $\heatContent$ in the case of $(n,r)$-von Koch fractals was established in \cite{vdB00_generalGKF} (without this terminology). The decomposition remainder, which we will denote by $\decompRem$ is given explicitly by \cite[Equation~(1.12)]{vdB00_generalGKF} and \cite[Proposition~1.2]{vdB00_generalGKF} gives an explicit estimate. Namely, $\decompRem(t)=O(t)$ as $t\to0^+$, in which case we have that $\sigmaRem=0$ since $\dimension/2=1$ in $\RR^2$.

\subsubsection{Heat Content Explicit Formulae}

We will give two types of explicit formulae. The first result is regarding the pointwise-valid formulae for $k\geq2$, and the second will be the distributional formulae specialized to the case when $k=0$. Note that the general pointwise formulae may be deduced in the same way using the general results for heat contents of self-similar sets. 

For the preliminaries regarding the notation and conventions defining the antiderivatives $\heatContent^{[k]}$, the Pochhammer symbol $(z)_w$, $z,w\in\CC$, and the definition of the sums over complex dimensions as symmetric limits, we refer the reader to Section~\ref{ssc:genHeatFormulae}.

\begin{Theorem}[Heat Content of GKFs (Pointwise)]
    \label{thm:pointwiseHeatFormulaGKF}
Let $\Knr$ be an $(n,r)$-von Koch Snowflake boundary satisfying the self-avoidance criterion in Proposition~\ref{prop:selfAvoid} and let $\Omega$ be the interior region defined by $\Knr$ (i.e. the bounded component of $\RR^2\setminus\Knr$). Suppose that either $n\geq5$ or that the scaling ratios $r$ and $\ell=(1-r)/2$ are arithmetically related (i.e. the ratio of their logarithms is rational). 
    \medskip

    Then for every integer $k\geq 2$, any $\delta>0$, and every $t\in(0,\delta)$, we have that the antiderivatives $\heatContent^{[k]}$ of the heat content $\heatContent$ for Problem~\ref{prob:specificHeatProblem} on $\Region$ satisfy
    \[ 
        \heatContent^{[k]}(t) = \sum_{\omega\in\Dd_\Phinr(\HH_{0})} 
            \Res\Bigg(\cfrac{t^{(2-s)/2+k}}{\left(\frac{2-s}{2}+1\right)_k}\heatzeta(s/2;\delta);\omega\Bigg)
            + \Rr^k(t),
    \]
    where $\heatzeta$ is as in Theorem~\ref{thm:heatZetaFormula} and where $\Dd_\Phinr(\HH_{0})$ is the set of possible complex dimensions of the generalized von Koch snowflake domain as given in \eqref{eqn:defPolesKochSZF} belonging to the open right half-plane $\HH_0=\set{s\in\CC\suchthat \Re(s)>0}$. For any $\e>0$ sufficiently small, the error term satisfies $\Rr^k(t)=O(t^{1-\e+k})$ as $t\to0^+$. 
\end{Theorem}
\begin{proof}
    We have that $\heatContent$ satisfies a scaling functional equation induced by $\Phinr$ with decomposition remainder $\decompRem(t)=O(t)$ as $t\to0^+$ by Proposition~1.2 of \cite{vdB00_generalGKF}, where $\Phinr$ is as in \eqref{eqn:defGKCsystem}. When normalized (noting that $\dimension/2=1$ in $\RR^2$), the heat content scaling functional equation takes the form 
    \[ t^{-1}\heatContent(t) = L_\Phinr[t^{-1}\heatContent(t)](t) + t^{-1}\decompRem(t). \]
    Further, the remainder $R(t):=t^{-1}\decompRem(t)=O(t^{-\sigma_R})$, with $\sigma_R=\sigmaRem/2=0$, as $t\to0^+$ and is continuous on $\RR^+$. Note that we have also multiplied the expression in \cite{vdB00_generalGKF} by $n$ (and distributed by linearity) to deduce a scaling functional equation for the total heat content, rather than the amount of heat content contained in a single symmetric sector. 

    The result then follows from application of Theorem~\ref{thm:heatFormulaPtw} with the following notes. If we assume that $n\geq5$, then we have that $\lowersimdim(\Phinr)\geq D_\ell>0=\sigmaRem$ by Lemma~\ref{lem:lowerDimBoundsGKF}. If $n=3$ or $n=4$, we have assumed that $r$ and $\ell$ are arithmetically related. 
\end{proof}

In the lattice case, as can be verified in light of \eqref{eqn:defKochSZF}, the poles of $\zeta_\Phinr$ are simple. Thus, we may simplify the residues to obtain an expansion in powers of $t^{(2-\omega)/2+k}$. 
\begin{Corollary}[Heat Content of Lattice GKFs (Pointwise)]
    \label{cor:pointwiseheatFormulaGKFLattice}
Let $\Knr$ be an $(n,r)$-von Koch Snowflake boundary satisfying the self-avoidance criterion in Proposition~\ref{prop:selfAvoid} and let $\Omega$ be the interior region defined by $\Knr$ (i.e. the bounded component of $\RR^2\setminus\Knr$). Suppose that its scaling ratios $r$ and $\ell=(1-r)/2$ are arithmetically related (i.e. the ratio of their logarithms is rational).
    \medskip 

    Then for every integer $k\geq 2$, any $\delta>0$, and every $t\in(0,\delta)$, we have that the antiderivatives $\heatContent^{[k]}$ of the heat content $\heatContent$ for Problem~\ref{prob:specificHeatProblem} on $\Region$ satisfy
    \[ 
        \heatContent^{[k]}(t) = \sum_{\omega\in\Dd_\Phinr(\HH_{0})} 
            \cfrac{r_\omega}{\left(\frac{2-\omega}{2}+1\right)_k}\,t^{(2-\omega)/2+k}
            + \Rr^k(t),
    \]
    where $r_\omega$ is a constant given by $r_\omega:=\Res\enclose{\heatzeta(s/2;\delta);\omega}$, with $\heatzeta$ is as in Theorem~\ref{thm:heatZetaFormula}, and where $\Dd_\Phinr(\HH_{0})$ is the set of possible complex dimensions of the generalized von Koch snowflake domain as given in \eqref{eqn:defPolesKochSZF} belonging to the open right half-plane $\HH_0=\set{s\in\CC\suchthat \Re(s)>0}$. For any $\e>0$ sufficiently small, the error term satisfies $\Rr^k(t)=O(t^{1-\e+k})$ as $t\to0^+$.
\end{Corollary}

In order to obtain expansions for the heat content itself when $k=0$, we now move to the distributional formulation of the explicit formulae. We refer the reader to Section~\ref{ssc:genHeatFormulae} for the relevant preliminaries and more information about the action of distributions or the estimates of their remainder terms. The space of test functions is the class of Schwartz functions on $(0,\delta)$, $\Ss(0,\delta)$, and the tempered distributions are the elements of its dual space, $\Ss'(0,\delta)$. 

\begin{Theorem}[Heat Content of GKFs (Distributionally, $k=0$)]
    \label{thm:distHeatFormulaGKF}
Let $\Knr$ be an $(n,r)$-von Koch Snowflake boundary satisfying the self-avoidance criterion in Proposition~\ref{prop:selfAvoid} and let $\Omega$ be the interior region defined by $\Knr$ (i.e. the bounded component of $\RR^2\setminus\Knr$). Suppose that either $n\geq5$ or that the scaling ratios $r$ and $\ell=(1-r)/2$ are arithmetically related (i.e. the ratio of their logarithms is rational). 
    \medskip

    Then for any $\delta>0$ and every $t\in(0,\delta)$, we have that as an equality of distributions in the Schwartz space $\Ss'(0,\delta)$ (the dual of the space defined in \eqref{eqn:defSchwartzFns}),
    \[ 
        \heatContent(t) = \sum_{\omega\in\Dd_\Phinr(\HH_{0})} 
            \Res\Bigg(t^{(2-s)/2}\heatzeta(s/2;\delta);\omega\Bigg)
            + \Rr(t),
    \]
    where $\heatzeta$ is as in Theorem~\ref{thm:heatZetaFormula} and where $\Dd_\Phinr(\HH_{0})$ is the set of possible complex dimensions of the generalized von Koch snowflake domain as given in \eqref{eqn:defPolesKochSZF} belonging to the open right half-plane $\HH_0=\set{s\in\CC\suchthat \Re(s)>0}$. For any $\e>0$ sufficiently small, the error term satisfies $\Rr(t)=O(t^{1-\e})$ as $t\to0^+$ in the sense of \eqref{eqn:distRemEstHeat}. The action of $\heatContent$ on a test function is given explicitly by \eqref{eqn:bracketIdentityHeat} with $k=0$.
\end{Theorem}
\begin{proof}
    The proof is the same as that of Theorem~\ref{thm:pointwiseHeatFormulaGKF} but with application of Theorem~\ref{thm:heatFormulaDist}. Note that we specialize to the case of $k=0$. 
\end{proof}

When we assume that the scaling ratios are arithmetically related, we may simplify this expression regarding the residues. Namely, we obtain an expansion in powers of $t^{(2-\omega)/2}$ with coefficients determined by the residues of $\heatzeta$ at the point $\omega/2$, where $\omega$ is a pole of the scaling zeta function $\zeta_\Phinr$ given by \eqref{eqn:defKochSZF}. 
\begin{Corollary}[Heat Content of Lattice GKFs (Distributionally)]
    \label{cor:distHeatFormulaGKFLattice}
Let $\Knr$ be an $(n,r)$-von Koch Snowflake boundary satisfying the self-avoidance criterion in Proposition~\ref{prop:selfAvoid} and let $\Omega$ be the interior region defined by $\Knr$ (i.e. the bounded component of $\RR^2\setminus\Knr$). Suppose that the scaling ratios $r$ and $\ell=(1-r)/2$ are arithmetically related (viz. the ratio of their logarithms is rational). 
    \medskip

    Then for any $\delta>0$ and every $t\in(0,\delta)$, we have that as an equality of distributions in the Schwartz space $\Ss'(0,\delta)$ (the dual of the space defined in \eqref{eqn:defSchwartzFns}),
    \[ 
        \heatContent(t) = \sum_{\omega\in\Dd_\Phinr(\HH_{0})} 
            r_\omega\,t^{(2-\omega)/2}
            + \Rr(t),
    \]
    where $r_\omega$ is a constant given by $r_\omega:=\Res\enclose{\heatzeta(s/2;\delta);\omega}$, with $\heatzeta$ is as in Theorem~\ref{thm:heatZetaFormula}, and where $\Dd_\Phinr(\HH_{0})$ is the set of possible complex dimensions of the generalized von Koch snowflake domain as given in \eqref{eqn:defPolesKochSZF} belonging to the open right half-plane $\HH_0=\set{s\in\CC\suchthat \Re(s)>0}$. For any $\e>0$ sufficiently small, the error term satisfies $\Rr(t)=O(t^{1-\e})$ as $t\to0^+$ in the sense of \eqref{eqn:distRemEstHeat}. The action of $V_{\Knr,\Omega}$ on a test function is given explicitly by \eqref{eqn:bracketIdentityHeat} with $k=0$.
\end{Corollary}

\subsection{Connection to Complex Dimensions}
\label{ssc:cDims}

Let $\Omega\subset\RR^\dimension$ be a bounded open set with boundary $\partial\Omega$. Suppose that $\partial\Omega$ is a self-similar set, arising as the attractor of the self-similar system $\Phi$, and suppose that $(\partial\Omega,\Omega)$ is an osculant fractal drum with respect to $\Phi$. (Per Definition~\ref{def:oscRFD}, this means that $\Phi$ satisfies the open set condition with $\Omega$ as a feasible open set and that the osculating condition holds for $\Omega$.) For explicit examples of such fractals, one may consider when $\Omega$ is the interior of an $(n,r)$-von Koch snowflake and $\partial\Omega$ is the snowflake boundary. 

Letting $R = \Region\setminus(\cup_{\ph\in\Phi}\ph[\Region])$, denote by $V_{\Regionbd,R}$ the tube function of $\Regionbd$ relative to the residual set $R$. By \cite[Theorem 5.3]{Hof25}, it follows that the normalized tube function $t^{-\dimension}V_{\Regionbd,\Region}(t)$ satisfies the scaling function equation 
\begin{equation}
    \label{eqn:tubeSFE}
    t^{-\dimension}V_{\Regionbd,\Region}(t)=L_\Phi[t^{-\dimension}V_{\Regionbd,\Region}(t)](t)+t^{-\dimension}V_{\Regionbd,R}(t).
\end{equation}
Further, let $\sigmaRem\in\RR$ be such that as $t\to0^+$,
\begin{align*}
    V_{\partial\Omega,R}(t) &= O(t^{\dimension-\sigmaRem}), \\
    R_\Omega(t)             &= O(t^{(\dimension-\sigmaRem)/2}).
\end{align*}
Lastly, we suppose that either $\sigmaRem<D_\ell\leq\lowersimdim(\Phi)$ (as in Proposition~\ref{prop:simDimBounds}) or that the distinct scaling ratios of the similitudes in $\Phi$ are arithmetically related, which are sufficient criteria for the admissibility of remainder terms.

Under these assumptions, explicit formulae for $V_{\Regionbd,\Region}$ and $\heatContent$ may both be written in terms of the possible complex dimensions, as governed by the poles of $\zeta_\Phi$. For simplicity, suppose that the poles of $\zeta_\Phi$ are simple, such as in the case of lattice GKFs, and consider the distributional identities specialized to $k=0$. By \cite[Theorem 5.2.2]{LRZ17_FZF} and Theorem~\ref{thm:heatFormulaDist}, respectively, we have that:
\begin{align}
    \label{eqn:volumeEx}
    V_{\partial\Omega,\Omega}(t) 
        &= \sum_{\omega\in\Dd_\Phi(\HH_{\sigmaRem})} a_\omega\, t^{\dimension-\omega} + \Rr_V(t); \\
    \label{eqn:heatEx}
    \heatContent(t) 
        &= \sum_{\omega\in\Dd_\Phi(\HH_{\sigmaRem})} b_\omega\, t^{(\dimension-\omega)/2} + \Rr_E(t), 
\end{align}
where $\Rr_V(t)=O(t^{\dimension-\sigmaRem-\e})$ and $\Rr_E(t^{(\dimension-\sigmaRem)/2-\e})$ as $t\to0^+$ for sufficiently small $\e>0$. Here, $a_\omega =\Res(\tubezeta_{\partial\Omega,\Omega}(s;\delta);\omega)$ and $b_\omega=\Res(\heatzeta(s;\delta);\omega)$ are constants determined by residues of the tube or heat zeta functions respectively, both over the same set of (possible) complex dimensions. The heat content, just as with the tube function, can therefore be seen to have an expansion governed by the complex dimensions of the boundary $\Regionbd$ relative to $\Region$.

\appendix

\section{Perron-Wiener-Brelot Solutions}
\label{app:PWBsolutions}

\subsection{History of the PWB Method for the Heat Equation}
\label{ssc:PWBhistory}

Perron introduced his famous method for solving the Dirichlet problem for Laplace's equation \cite{Per23}. The basic premise is to obtain the solution, a harmonic function, as a supremum over a family of subharmonic functions. This method was studied by Wiener, who in particular defined a generalized solution as one obtained in this way with limits converging to the prescribed boundary values at regular boundary points and proved a criterion for the convergence of such solutions to the prescribed boundary value conditions \cite{Wie24_Dirichlet,Wie24_Potential}. 

Wiener's criterion established existence of such generalized solutions, and the uniqueness of such solutions (namely, that such a solution is identically zero when all its limits approaching regular boundary points are zero) was established by Kellogg \cite{Kel28} and Evans \cite{Eva33}. Brelot's contributions include simplifying this proof \cite{Bre55} as well as developing a general theory and extending the method (cf. \cite{Bre67}). His general study in potential theory was already inspired by others, notably including the work of Doob \cite{Doo66}. The potential theory was further developed and axiomatized by Bauer \cite{Bau66} and Constantinescu and Cornea \cite{CC72}.

The application of the PWB method to the heat equation relates to the probabilistic approach, owing to the connection of this equation to Brownian motion. This connection was discovered as early as the work of Bachelier in mathematical finance \cite{Bac00}, and it was independently analyzed by Einstein \cite{Ein56} and Smoluchowski \cite{Smo06}. Einstein's work especially was a large inspiration for Wiener's mathematical development of the theory of Brownian motion \cite{Wie23,Wie24_Average,Wie76}. It is for this reason that Brownian motion may be instead called a Wiener process. 

The analysis of the heat equation on an arbitrary bounded open set in Euclidean space, using the probabilistic approach, is due to Doob \cite{Doo55}. He obtained the solution in the sense of Perron and Wiener, using the martingale analogues of subharmonic functions, though he did not provide details for the classification of regular and irregular boundary points. Evans and Gariepy proved an analogue of Wiener's criterion for the classification of regular and irregular boundary points for solutions of the heat equation \cite{EG82}. Watson would later show that Doob's approach is equivalent to the later developments in the theory, including filling in details not originally provided by Doob \cite{Wat15}. For a modern reference, we recommend Watson's book on the subject \cite{Wat12}. 

Lastly, we note that the Dirichlet problem for the heat equation requires some extra assumptions in order to be well posed. (A Dirichlet problem is well posed if there exists a unique solution and, further, this solution depends continuously on the initial conditions.) As was discovered by Tychonoff, functions with growth larger than quadratic exponentials with certain initial conditions can yield nonunique solutions, but under the restriction to this growth condition, solutions are unique \cite{Tyc35}. Widder further analyzed the problem and showed that nonnegative solutions are unique \cite{Wid44}. For more information about the (non)uniqueness, we recommend Widder's paper.

\subsection{Classes of Temperature Functions}
\label{ssc:temperatures}

In order to define PWB solutions, we must define a class of \textit{hypertemperatures} (and respectively \textit{hypotemperatures}) to be considered. Let $\Regionplus\subseteq\RRNplus$ be an arbitrary open set and we write $p=(x,t)\in\RR^\dimension\times\RR$ to distinguish the \textit{spatial coordinates} $x$ and the \textit{time coordinate} $t$. Let 
\[ 
    C^{2,1}(\Regionplus) = \set{f:\Regionplus\to\RR\suchthat \partial_{x_k}\partial_{x_j}f\in C^0(\Regionplus),\,k,j\in\set{1,...,\dimension},\text{ and }\partial_t f\in C^0(\Regionplus)} 
\]
denote the class of continuously differentiable functions on $\Regionplus$ with continuous second partial derivatives with respect to the spatial coordinates (each $x_k$, $k=1,...,\dimension$) and one continuous derivative with respect to the time coordinate $t$. 

In the simplest case, if we consider $u\in C^{2,1}(\Regionplus)$, then we may define the heat operator\footnote{Note that we take the opposite convention as that in \cite{Wat12}, where the heat operator therein is $\Theta:=-\heatop$.} $\heatop$ by $\heatop u:= \partial_t u -\Delta u$, in which case the heat equation is simply $\heatop u =0$. A \index{Temperature}\textbf{temperature} is a function which is in $C^{2,1}(\Regionplus)$ and which solves the heat equation, i.e. $\heatop u=0$. 

A \textbf{subtemperature} is then a function $u\in C^{2,1}(\Regionplus)$ such that $\heatop u\leq 0$. A \textbf{supertemperature} is a function such that $\heatop u\geq 0$. Note that a function $u$ is a supertemperature if and only if $-u$ is a subtemperature, and in general we need only define one class of functions and we may use this relation to define the other. We see already the theme of the PWB approach: a temperature function is both a subtemperature and a supertemperature. 

However, for the study of the heat problem on an arbitrary open set, we will need to consider functions which are not necessarily continuously differentiable. Thus, we will need an analogue of the heat operator that does not require differentiability, as we cannot define $\heatop$ for such functions to consider sub/supertemperatures. There are several equivalent approaches. For instance, one may consider a mean value operator over an appropriate type of set (so-called \textit{heat balls}) and look for functions which agree with their mean values. This is a fundamental property of solutions to the classical heat equation, and can be used to show that such functions obey the heat equation. One may also use a classification in terms of a type of integral called a \textit{Poisson integral} which involves types of measures called \textit{parabolic measures}. 

Let us presently take the mean value approach. Let $p_0=(x_0,t_0)$ and $r>0$. A \textbf{heat ball} is a set of the form 
\[ 
    E_{p_0,r} := \set{p=(x,t)\in\Regionplus \suchthat 
        |x-x_0| \leq \sqrt{2\dimension(t_0-t)\log(r/(t_0-t))},\, 0<t_0-r<t<t_0}.
\]
(One may represent this set in terms of a lower bound for the \textit{fundamental solution} to the heat equation, hence the name.) We say that a \textbf{heat sphere} is the boundary $\partial E_{p_0,r}$ of a heat ball.

In order to avoid the singular behavior which occurs near the center $p_0$ of a heat ball, we will introduce the function 
\begin{equation}
    \label{eqn:cutoffKernel}
    Q(p) = Q(x,t) := \begin{cases}
        \cfrac{|x|^2}{\strut \sqrt{ 4|x|^2|t|^2 + (|x|^2 - 2\dimension t)^2 } } & (x,t)\neq 0, \\
        1 & (x,t)=(0,0).
    \end{cases}
\end{equation}
Note that $Q$ has the property that for any sequence of points $p\in\partial E_{p_0,r}$ converging to $p_0$, 
$Q(p) \to 1 $, i.e. it is continuous on $\partial E_{p_0,r}$. 

\begin{Definition}[Heat Sphere Mean Value Operator]
    \label{def:heatMeanValueOp}
    \index{Mean value operator!on a heat sphere}
    Letting $p_0=(x_0,t_0)\in\RR^\dimension\times\RR$ and $r>0$, consider the heat ball $E_{p_0,r}$. Let $dS$ denote the surface area measure on $\partial E_{p_0,r}$. Then for any function $u$ for which the integral converges, we define the \textbf{mean value operator} $\heatmean_{p_0,r}$ on $\partial E_{p_0,r}$ to be 
    \begin{equation}
        \label{eqn:defHeatMean}
        \heatmean_{p_0,r}[u]=\heatmean[u;\partial E_{p_0,r}]:= \frac1{(4\pi)^{\dimension/2}} \int_{\partial E_{p_0,r}} Q(p-p_0) u(p) dS(p).
    \end{equation}
\end{Definition}

The last preliminary is to define the properties that our functions must satisfy in lieu of continuous differentiability. We first define the notions of upper and lower \textbf{semicontinuity}: a function $f:\RRNplus\to[-\infty,\infty]$ is \textbf{upper semicontinuous} on a set $\Regionplus\subseteq \RRNplus$ if for any $r\in\RR$, the set $\set{p\in\Regionplus \suchthat f(p)<a}$ is a relatively open subset of $S$ (i.e. open in the subspace topology it inherits from $\RRNplus$). We say that an extended real-valued function $f$ is \textbf{lower semicontinuous} if $-f$ is upper semicontinuous. 

Next, we say that an extended real-valued function $f:\RRNplus\to[-\infty,\infty]$ is \textbf{upper finite} on a set $\Regionplus\subseteq\RRNplus$ if $f[\Regionplus]\subseteq [-\infty,\infty)$ and \textbf{lower finite} on $\Regionplus$ if $f[\Regionplus]\subset(-\infty,\infty]$. Note that when $f$ is both upper and lower finite, it is simply a real-valued function. 

We may now define hypertemperatures and hypotemperatures. 
\begin{Definition}[Hypertemperatures]
    \label{def:hypertemperature}
    \index{Temperature!Hypertemperature}\index{Temperature!Hypotemperature}
    Let $\Regionplus\subseteq\RRNplus$. We say that an extended real-valued function $f$ is a \textbf{hypertemperature} on $\Regionplus$ if $f$ is \textit{lower finite}, \textit{lower semicontinuous}, and with the following property. For any $p\in\Regionplus$ and any $\e>0$, there exists $r<\e$ such that 
    \[ f(p)\geq \heatmean[f;\partial E_{p,r}]. \]
    Let $\hyperspace(\Regionplus)$ denote the set of all hypertemperatures on $\Regionplus$.
\end{Definition}

The efficient way to define when a function $f$ is a hypotemperature is to require that $-f$ be a hypertemperature. However, for the sake of completeness and establishing notation, we will give an independent definition and let this fact be an immediate corollary. 
\begin{Definition}[Hypotemperatures]
    \label{def:hypotemperature}
    \index{Temperature!Hypertemperature}\index{Temperature!Hypotemperature}
    Let $\Regionplus\subseteq\RRNplus$. We say that an extended real-valued function $f$ is a \textbf{hypotemperature} on $\Regionplus$ if $f$ is \textit{upper finite}, \textit{upper semicontinuous}, and with the following property. For any $p\in\Regionplus$ and any $\e>0$, there exists $r<\e$ such that 
    \[ f(p)\leq \heatmean[f;\partial E_{p,r}]. \]
    Let $\hypospace(\Regionplus)$ denote the space of all hypotemperatures on $\Regionplus$. 
\end{Definition}

Ultimately, when we look for solutions of the heat equation, we will consider the collection of all hypertemperatures and hypotemperatures that satisfy the prescribed boundary conditions. By taking an infimum of hypertemperatures or a supremum of hypotemperatures, then ideally we arrive at the same function which now satisfies the boundary conditions and will be a \text{temperature}, satisfying the heat equation proper.

\subsection{Classification of Boundary Points}
\label{ssc:PWBboundary}

In what follows, let $\Regionplus\subseteq\RRNplus$ be an arbitrary open set. We consider the boundary of $\Regionplus$ relative to the one-point compactification of $\RRNplus$, so that $\infty\in\partial\Regionplus$ exactly when $\Regionplus$ is unbounded. With this proviso, $\partial\Regionplus$ is otherwise the standard Euclidean boundary of $\Regionplus$. For an arbitrary open set $\Regionplus$, it turns out that we cannot prescribe boundary conditions on all of $\partial\Regionplus$. This is to be \textit{reasonably expected}, however. 

Consider the simple problem of a one-dimensional rod given by the interval $[a,b]$ and suppose that we wish to consider the problem for time $t\in[0,1]$. Suppose now to look for solutions of the heat equation $\partial_x^2u=\partial_tu$ in the set $(a,b)\times(0,1)$. Prescribing conditions on the boundary of $[a,b]\times[0,1]$ amounts to three types of constraints. On the set $(a,b)\times\set{0}$, we prescribe the initial amount of heat on the inside of the rod at time $t=0$. On each of the sets $\set{a}\times[0,1]$ and $\set{b}\times[0,1]$, we are prescribing the temperature values of the endpoints of the rod. Lastly, on the boundary piece $(a,b)\times\set{1}$, we are prescribing the {future} temperature of the inside of rod at time one. However, in this deterministic model, the conditions at time $t=0$ and on the boundary for all time $t\in[0,1]$ will determine the future temperature; it cannot be arbitrarily chosen. 

This leads us to a general classification of the types of boundary points of $\partial\Regionplus$. In order to do so, let us establish some notation regarding half-balls with respect to the time coordinate. Letting $p_0=(x_0,t_0)\in\RRNplus$, we write 
\[ H_*(p_0,r) = \set{(x,t)\suchthat |x-x_0|^2+|t-t_0|^2<r^2,\, t<t_0} \] 
for the open \textbf{lower half-ball}. Similarly, we write $H^*(p_0,r)$ for the open \textbf{upper half-ball}, defined analogously but with $t>t_0$. Note that these are standard balls cut in half, not heat balls. 

The boundary $\partial\Regionplus$ is classified into three types of points: \textit{normal}, \textit{semi-singular} (or \textit{semi-normal}), and (\textit{fully}) \textit{singular points}. If $q\in\partial\Regionplus$, then $q$ is said to be \textbf{normal} if either $q=\infty$ is the point at infinity or if every lower half-ball $H_*(q,r)$ centered at $q$ intersects the complement of $\Regionplus$, viz. $H_*(q,r)\setminus\Regionplus\neq\emptyset$ for all $r>0$, namely as $r\to0^+$. For abnormal boundary points $q$, eventually there exists an $r_0>0$ so that $H_*(q,r)\subseteq\Regionplus$. If there exists $r_1<r_0$ such that $H^*(q,r_1)\cap \Regionplus=\emptyset$, then $q$ is a {(fully) singular} boundary point. (Note that this implies that any smaller upper half-ball $H^*(p,r_2)$, $r_2<r_1$, will also be contained in the complement of $Y$.) Otherwise, for every $r<r_0$ we have that $H^*(q,r_1)\cap \Regionplus\neq\emptyset$ and $q$ is said to be \textbf{semi-singular}.

Returning to the example from earlier, we can see now how these criteria describe the different parts of the boundary. The initial condition set, $(a,b)\times\set{0}$, clearly contains only normal points since the lower half-balls $H_*(p,r)$ are contained in the complement for any $r$, let alone merely intersecting the complement. On the edges of the rod, we also have that arbitrarily small lower half-balls always intersect the complement to the left or right, respectively. Thus this part of the boundary is also normal. 

The top part of the boundary, however, is fully singular: the lower half-balls are strictly contained in the set $\Regionplus$ and the upper half-balls are strictly contained in the complement. For these fully singular points, they can only be approached from within the set $\Regionplus$ and only from past times, not future times. The conditions above amount to requiring that fully singular boundary points look locally like such points. 

This example does not illustrate a semi-singular point, but for a simple example puncture the square $(a,b)\times(0,1)$ at an interior point, say $(x_0,t_0)$. This new point in the boundary is then semi-singular. Imposing a boundary condition here amounts to instantaneously changing the temperature on the rod at $x_0$ at the time $t_0$. In this case, one could expect a solution for times $t>t_0$ which are compatible with the current values of the temperature at $t_0$ for $x\neq x_0$ and the new prescribed temperature at $x_0$. However, the solution in the larger set will not approach this arbitrarily changed value from times $t<t_0$. 

So, for the Perron-Wiener-Brelot solutions to follow, we will impose boundary conditions only on the set of normal and semi-singular boundary points. For this reason, we say that the \textbf{essential boundary} $\partial_e\Regionplus$ of a set $\Regionplus\subseteq\RRNplus$ is the union of the set of all \textit{normal} boundary points, the \textbf{normal boundary} $\partial_n\Regionplus$, and the set of all \textit{semi-singular} boundary points, the \textbf{semi-singular boundary} $\partial_{ss}\Regionplus$. If we denote the set of \textit{fully singular} boundary points by $\partial_s\Regionplus$, called the \textbf{singular boundary}, then we have that $\partial_e\Regionplus=\partial\Regionplus\setminus\partial_s\Regionplus$. Note also that the normal boundary and the semi-singular boundary are disjoint by the definitions of normal and semi-singular boundary points, and thus $\partial_e\Regionplus=\partial_n\Regionplus\uplus\partial_{ss}\Regionplus$. Recall that $\partial\Regionplus$ is considered with respect to the one-point compactification of $\RRNplus$ and that $\infty\in\partial_n\Regionplus$ is always a normal boundary point when $\Regionplus$ is unbounded.

We now may define the sense in which boundary conditions will be satisfied. We will require convergence of the function to the boundary function at normal points to be for any sequence in $\Regionplus$ converging to the boundary points. Meanwhile, for convergence to the value of the function at semi-singular boundary points, we will only impose convergence when the sequence approaches from future times, not past times.
\begin{Definition}[Regularity and Boundary Convergence]
    \label{def:convergenceOnEssenBd}
    Let $\Regionplus\subseteq\RRNplus$ be an open set and let $\partial_e\Regionplus=\partial_n\Regionplus\uplus\partial_{ss}\Regionplus$ be its essential boundary. Let $f$ be a function on $\partial\Regionplus$ and $u$ a function on $\Regionplus$ itself. 
    \medskip 

    We say that $u$ \textbf{converges to} $f$ at a point $q\in\partial_e\Regionplus$ if we have the following.
    \begin{itemize}
        \item If $q\in\partial_n\Regionplus$, then we have that for any sequence $\set{p_n}_{n\in\NN}\subset\Regionplus$ with $p_n\to q$, $u(p_n)\to f(q)$. 
        \item If $q=(y,s)\in\partial_{ss}\Regionplus$, then we have that for any sequence $\set{p_n=(x_n,t_n)}_{n\in\NN}\subset\Regionplus$ with $x_n\to y$ and $t_n\to s^+$, $u(p_n)\to f(q)$.
    \end{itemize}
    When $q\in \partial_e\Regionplus$ is such that these limits exist, $q$ is called a \textbf{regular boundary point}. We say that $u$ \textbf{converges to} $f$ \textbf{on the regular (essential) boundary} if $u$ converges to $f$ at every regular point $q$ in the essential boundary.
\end{Definition}

The last caveat is that given a set $\Regionplus$, it can have both \textit{regular} and \textit{irregular} boundary points in its essential boundary. (A boundary point is called \index{Boundary point!Irregular boundary point}\textbf{irregular} if it is not regular in the sense above.) So, the final task is to determine which points in the essential boundary are regular. In the case that the entire essential boundary is not regular, i.e. there exist irregular boundary points, then one must seek to understand which points are regular and which are irregular. We will not explore this issue further here, but refer the reader to the work of Evans and Gariepy \cite{EG82} or to \cite[Section~8.6]{Wat12} for more information, including sufficient conditions to ensure that an entire boundary or a specific point in the boundary is regular.

\subsection{Perron-Wiener-Brelot Solutions}

We now have the means to define a Perron-Wiener-Brelot solution to the heat equation on an arbitrary open set in $\RR^{\dimension+1}$, identified with $\RR^\dimension\times\RR$. Consider the following Dirichlet problem for the heat equation on an arbitrary open set $\Regionplus\subseteq\RRNplus$, given by 
\begin{equation}
    \label{prob:arbitraryHeatProb}
    \left\{
    \begin{aligned}
        \partial_t u - C\Delta u &= 0   &&\text{in }\Regionplus, \\
            u &= f                      &&\text{on }\partial\Regionplus. \\
    \end{aligned}
    \right.
\end{equation}
Here, $C>0$ is a positive constant. Note that $\Regionplus$ is not necessarily a Cartesian product of an open set in $\RR^\dimension$, the spatial domain, and an open set in $\RR$, the time domain.

We consider now two collections of functions which we define based on our boundary function $f$. These are either hypertemperatures bounded from below by $f$ or hypotemperatures bounded from above by $f$.
\begin{align*}
    \hyperspace&(\Regionplus;f) := \\
    &\set{w\in\hyperspace(\Regionplus)\suchthat 
        \begin{aligned}
            \liminf_{(x,t)\to (y,s)}  w(x,t) &\geq f(y,s) 
                & (x,t)\in\Regionplus,\, (y,s)\in\partial_n\Regionplus \\
            \liminf_{(x,t)\to(y,s^+)} w(x,t) &\geq f(y,s) 
                & (x,t)\in\Regionplus,\, (y,s)\in\partial_{ss}\Regionplus
        \end{aligned}
    }, \\
    \hypospace&(\Regionplus;f) := \\
    &\set{w\in\hyperspace(\Regionplus)\suchthat 
        \begin{aligned}
            \limsup_{(x,t)\to (y,s)}  w(x,t) &\leq f(y,s) 
                & (x,t)\in\Regionplus,\, (y,s)\in\partial_n\Regionplus \\
            \limsup_{(x,t)\to(y,s^+)} w(x,t) &\leq f(y,s) 
                & (x,t)\in\Regionplus,\, (y,s)\in\partial_{ss}\Regionplus
        \end{aligned}
    }.
\end{align*}
These spaces of functions are each nonempty since $w\equiv\infty$ is in the first and $w\equiv-\infty$ is in the second.

We next define the upper and lower candidates for the solution to the heat equation. 
\begin{align}
    \label{eqn:PWBcandidates}
    \begin{split}
        \wupper(p)  &:= \inf\set{w(p) \suchthat w\in\hyperspace(\Regionplus;f)}, \\
        \wlower(p)  &:= \sup\set{w(p) \suchthat w\in\hypospace(\Regionplus;f)}.
    \end{split}
\end{align}
These define two functions $\wupper$ and $\wlower$ on $\Regionplus$. If they agree (and are sufficiently regular), then the function $f$ is said to be \textit{resolutive}.

\begin{Definition}[Resolutive Function]
    \label{def:resolutiveFn}
    \index{Resolutive function}
    \index{Heat equation!Resolutive boundary conditions}
    A function $f$ on $\partial_e\Regionplus$, the essential boundary of $\Regionplus\subseteq\RRNplus$, is said to be \textbf{resolutive} if the following hold. Firstly, the functions defined by \eqref{eqn:PWBcandidates} are the same, i.e. $\wupper\equiv\wlower$, and in this case, we define $\wsoln:=\wupper=\wlower$. Secondly, we require that $\wsoln\in C^{2,1}(\Regionplus)$, i.e. $\wsoln$ is twice continuously differentiable in its spatial coordinates and once continuously differentiable in its time coordinate, with $C^{2,1}(\Regionplus)$ as defined in Section~\ref{ssc:temperatures}.
\end{Definition}

When the boundary function $f$ is resolutive, we obtain a \textit{temperature} function $\wsoln$, as it follows that since $\wsoln\in C^{2,1}(\Regionplus)$, $\heatop\wsoln=0$. From the definitions of the function spaces $\hyperspace(\Regionplus,f)$ and $\hypospace(\Regionplus,f)$, note that whenever the limit of $\wsoln$ as $p$ approaches a point $q$ in the essential boundary $\partial_e\Regionplus$ exists, the value of the limit must necessarily equal $f(q)$ since $\wsoln\in\hyperspace(\Regionplus,f)\cap \hypospace(\Regionplus,f)$. Therefore, $\wsoln$ converges to $f$ on the regular essential boundary of $\Regionplus$. This is exactly what it means for $\wsoln$ to be a solution to a Dirichlet problem for the heat equation in the sense of Perron-Wiener-Brelot. 
\begin{Proposition}[Perron-Wiener-Brelot (PWB) Solution]
    \label{def:PWBsolution}
    \index{Heat equation!Perron-Wiener-Brelot solution}
    Let $\Regionplus\subseteq\RRNplus$ and suppose that $f$ is a \textit{resolutive} function in the sense of Definition~\ref{def:resolutiveFn}. Then we say that $\wsoln$ is the \textbf{Perron-Wiener-Brelot solution} to Problem~\ref{prob:arbitraryHeatProb} and we have that $\wsoln$ converges to $f$ on the regular essential boundary of $\Regionplus$ in the sense of Definition~\ref{def:convergenceOnEssenBd}.
\end{Proposition}

\subsection{Parabolic Measure and Resolutive Functions}

The final ingredient in our study of PWB solutions is the last but most important remaining question. Given an open set $\Regionplus\subseteq\RRNplus$, when is a function $f$ on $\partial_e\Regionplus$ resolutive? Per Proposition~\ref{def:PWBsolution}, this condition is exactly what is required for a PWB solution to exist with the Dirichlet boundary conditions specified by $f$. 

Luckily, such functions are plentiful. In fact, any continuous function $f\in C^0(\partial_e\Regionplus)$ on the essential boundary is resolutive; see for instance \cite[Theorem~8.26]{Wat12}. We can give a characterization of the class of resolutive functions in terms of an integrability condition relating to a family of measures called \textit{parabolic measures}. 

Given the open set $\Regionplus\subseteq\RRNplus$ and a point $p\in\Regionplus$, it can be shown (see, for example, \cite[Theorem~8.27]{Wat12}) that there exists a unique nonnegative Borel probability measure $\parameasure$ on $\partial_e\Regionplus$ such that for any continuous function $f\in C^0(\partial_e\Regionplus)$, 
\begin{equation}
    \label{eqn:defPoissonInt}
    \wsoln(p) = \int_{\partial_e\Regionplus} f\,d\parameasure.
\end{equation}
An integral of this form may be called a \textit{Poisson integral} owing to their connection to the classical theory of Poisson kernels and integral representations (in the setting of harmonic functions which can be extended to the heat equation). 

\begin{Definition}[Parabolic Measure and Parabolic Integrability]
    \label{def:paraMeasure}
    \index{Parabolic measure}
    \index{Measure!Parabolic measure}
    \index{Parabolic measure!Parabolical integrability}
    The completion of the measure defined by \eqref{eqn:defPoissonInt}, which we shall also denote by $\parameasure$, is called the \textbf{parabolic measure} relative to $\Regionplus\subseteq\RRNplus$ and $p\in\Regionplus$. 
    \medskip 

    A function $f$ on $\partial_e\Regionplus$ is said to be \textbf{parabolically integrable} if it is $\parameasure$-integrable for any $p\in\Regionplus$. 
\end{Definition}

The set of resolutive functions $f$ for a given set $\Regionplus$ is exactly the class of \textit{parabolically integrable} functions on $\partial_e(\Regionplus)$ (see for instance Corollary~8.34 of \cite{Wat12}). The integral representation given by \eqref{eqn:defPoissonInt} means that the PWB solution $\wsoln$ to Problem~\ref{prob:arbitraryHeatProb} at a point $p$ is exactly the parabolic average of the boundary value $f$ with respect to the associated parabolic measure $\parameasure$.

\section{Mellin Transforms}
\label{app:Mellin}

We first define Mellin transforms and their truncated counterparts, which are our main tool in solving scaling functional equations. The Mellin transform is an integral transform associated with the Haar measure of the positive real line with respect to multiplication in the same sense that the Fourier transform is associated with the Haar measure of the real line with respect to addition (which is simply the Lebesgue measure). The underlying space $(\RR^+,\cdot)$, the positive real line $\RR^+=(0,\infty)$ viewed as a group with respect to multiplication, can be thought of as the space of multiplicative scales. As fractals generally have scale invariance or approximate scale invariance, the Mellin transform associated with this space is well suited to study such shapes. 

\subsection{Mellin Transform}

Let $f:\RR^+\to\RR$ be a locally Lebesgue integrable function. The standard \index{Mellin transform}\textbf{Mellin transform} of $f$ is the integral 
\begin{equation}
    \label{eqn:defMellinTransform}
    \Mm[f](s) := \int_0^\infty x^{s-1}f(x)\,dx,
\end{equation}
defined for all $s\in\CC$ for which the Lebesgue integral converges. Strictly speaking, we identify the Mellin transform with its analytic continuation (which we shall restrict to be defined on a connected open subset of the complex plane). The factor $x^{s-1}$ has an extra factor of $x^{-1}$ since the Haar measure associated to the group $(\RR^+,\cdot)$ is $dx/x$, where $dx$ is the Lebesgue measure on $\RR$. Under a scaling transformation $S_\lambda(x)=\lambda x$, note that $d(\lambda x)/\lambda x=dx/x$ is invariant.

Integrability of the function $f$ and the existence of some polynomial growth conditions are sufficient for the transform to exist on a vertical strip in $\CC$. However, we will often simply impose that $f$ be continuous on $\RR^+$ since the heat contents we consider in Section~\ref{sec:heatAnalysis} are continuous.

\subsection{Truncated Mellin Transforms}

A truncated, or equivalently a restricted, Mellin transform is simply an integral of the same integrand as \eqref{eqn:defMellinTransform}, but over an interval (i.e. connected subset) of $(0,\infty)$. 
\begin{Definition}[Truncated Mellin Transform]
    \label{def:truncatedMellin}
    \index{Mellin transform!Truncated Mellin transform}
    Let $f\in C^0(\RR^+,\RR)$ and fix $\alpha,\beta \geq 0$, with $\alpha<\beta$. The truncated Mellin transform of $f$, denoted by $\Mm_\alpha^\beta[f]$, is given by 
    \[ \Mm_\alpha^\beta[f](s) := \int_\alpha^\beta t^{s-1}f(t)\,dt \]
    for all $s\in\CC$ for which the Lebesgue integral is convergent.
\end{Definition}
Typically, we will identify $\Mm_\alpha^\beta[f](s)$ with its analytic continuation to an open connected domain of $\CC$. (In this work, we shall restrict ourselves to the case when these analytic continuations live on a connected open subset of the complex plane, rather than a more general \textit{Riemann surface}.) If $\alpha=0$, we write $\Mm^\beta$ for the truncated transform. If $\alpha=0$ and we take $\beta=\infty$, then $\Mm=\Mm_0^\infty$ is simply the standard Mellin transform. Note that we may equivalently define $\Mm_\alpha^\beta[f]$ to be the Mellin transform of $f$ times the characteristic function of the interval $(\alpha,\beta)$, viz. 
\[ \Mm_\alpha^\beta[f] = \Mm[f\cdot\1_{[\alpha,\beta]}]. \] 
This allows us to compare the convergence of the two directly, and it shows that the truncated transform inherits the properties of its standard counterpart, for example its linearity. 

\subsection{Convergence of Mellin Transforms}

The convergence properties of the truncated transform follow from standard results on the Mellin transform. In what follows, let 
\begin{align*}
    \HH_{\sigma}    :=& \set{s\in\CC\suchthat \sigma<\Re(s)}, \\
    \HH_{a}^{b}     :=& \set{s\in\CC\suchthat a<\Re(s)<b}
\end{align*}
denote an \index{Half-plane}\textbf{open right half-plane} and an \textbf{open vertical strip} in $\CC$, respectively. If $\sigma=-\infty$, then $\HH_{-\infty}=\CC$ and if $\sigma=+\infty$, then $\HH_{+\infty}=\emptyset$. It is known (see, for example, \cite[Chapter 6]{Gra10}) that $\Mm[f](s)$ is holomorphic in the vertical strip $\HH_{\sigma_-}^{\sigma_+}$, where 
\begin{align}
    \label{eqn:defMellinAbscissa}
    \begin{split}
        \sigma_- :=& \inf\set{\sigma\in\RR : f(x) = O(x^{-\sigma}) \text{ as }x\to0^+},     \\
        \sigma_+ :=& \sup\set{\sigma\in\RR : f(x) = O(x^{-\sigma}) \text{ as }x\to+\infty}.
    \end{split}
\end{align}
When $\alpha<\beta<\infty$, $f\cdot\1_{[\alpha,\beta]}\equiv 0$ as $x\to+\infty$, whence $\sigma_+=+\infty$. In this case, we say that $\sigma_-$ is the \textbf{abscissa of absolute convergence} of $\Mellin_\alpha^\beta[f]$, denoted by $\sigma_{ac}$, and it follows that $\Mellin_\alpha^\beta[f]$ is holomorphic in the open right half-plane $\HH_{\sigma_{ac}}$. Similarly, if $0<\alpha<\beta$ then $f\cdot\1_{[\alpha,\beta]}\equiv 0$ as $x\to0^+$, whence $\sigma_-=-\infty$. So, if $0<\alpha<\beta<\infty$, then $\Mellin_\alpha^\beta[f]$ is automatically entire, i.e. holomorphic in all of $\CC$, in which case $\sigma_{ac}=-\infty$ (formally). We collect these observations into the following lemma, which will be of use later. 
\begin{Lemma}[Holomorphicity of Truncated Mellin Transforms]
    \label{lem:MellinHolo}
    \index{Mellin transform!Convergence}
    Let $f$ be integrable on $(\alpha,\beta)\subset\RR^+$ and let $\Mellin_\alpha^\beta[f]$ be its (truncated) Mellin transform.
    \begin{itemize}
        \item Let $\beta<\infty$. Suppose that $f$ is bounded away from $0$ (i.e. on any interval of the form $[\e,\beta]$ for $\e>0$) and that $f(x)=O(x^{-\sigma_0})$ as $x\to0^+$, for some $\sigma_0\in\RR$. Then $\Mellin^\beta[f]$ is absolutely convergent and holomorphic in the open right half-plane $\HH_{\sigma_0}$.
        \item Let $0<\alpha<\beta<\infty$. If $f$ is bounded on $[\alpha,\beta]$, then $\Mellin_\alpha^\beta[f]$ is entire. 
    \end{itemize}
\end{Lemma}
\begin{proof}
    First, consider the case where $\beta<\infty$ and $\alpha=0$. Since $f(t)=O(t^{-\sigma_0})$ as $t\to0^+$ and is bounded on any interval of the form $[\e,\delta]$, there exists $C>0$ so that $|f(t)|\leq C\,t^{-\sigma_0}$ for all $t\in (0,\delta]$. Thus, we may estimate the Mellin transform to find that 
    \begin{align}
        \label{eqn:MellinEstimate}
        \begin{split}
            |\Mellin_0^\beta[f](s)| 
            &\leq \int_0^\delta t^{\Re(s)-1}|f(t)|\,dt 
            \leq C\int_0^\delta t^{\Re(s)-1-\sigma_0}\,dt 
            = \frac{C}{\Re(s)-\sigma_0}\,\delta^{\Re(s)-\sigma_0},
        \end{split}
    \end{align}
    provided $\Re(s)>\sigma_0$. It follows that $\Mellin_0^\beta[f](s)$ is convergent in $\HH_{\sigma_0}$. Further, this bound also implies that it is holomorphic in $\HH_{\sigma_0}$. This may be seen, for instance, by the application of Lebesgue's dominated convergence theorem to compute its derivative or alternatively by application of Morera's theorem and the Fubini-Tonelli theorem. 

    Now let $0<\alpha<\beta<\infty$. Then the function $f(t)\1_{(\alpha,\beta)}(t)$ is integrable and in fact $\Mellin_\alpha^\beta[f](s)=\Mellin^\beta[f(t)\1_{(\alpha,\beta)}(t)](s)$ for any $s$ for which it is convergent, since $f(t)=f(t)\1_{(\alpha,\beta)(t)}(t)$ for all $t\in(\alpha,\beta)$. Note that $f(t)\1_{(\alpha,\beta)}(t)$ is bounded on $[0,\beta]$ and that for any $n>0$, we have that $f(t)=O(t^n)$ as $t\to0^+$ since $\1_{(\alpha,\beta)}(t)\equiv 0$ whenever $t<\alpha$. By the first part of the proof, we have that $\Mellin^\delta[f(t)\1_{(\alpha,\beta)}(t)]$ is a convergent integral and is holomorphic in any half-plane of the form $\HH_{-n}$, $n\in\NN$. Consequently, $\Mellin_\alpha^\beta[f]$ converges for all $s\in\CC$ and is an entire function. 
\end{proof}

Suppose now that $\zeta_f(s):=\Mellin_\alpha^\beta[f](s)$ is holomorphic in $\HH_{\sigma_{ac}}$ and that $\zeta_f$ admits an analytic continuation to a half-plane of the form $\HH_\sigma$, with $\sigma\leq\sigma_{ac}$.
In this case, $\zeta_f$ is a \textbf{tamed Dirichlet-type integral} (DTI) in the sense of \cite[Definition A.1.2]{LRZ17_FZF}. More generally, $\zeta_f$ may admit holomorphic continuation to an open connected set $U$ in $\CC$ containing $\HH_{\sigma_{ac}}$. Typically, we will assume that $U$ is a connected open neighborhood of the set $W\setminus\Dd_f(W)$, where $\Dd_f(W)$ is a discrete subset of $W$ which contains the possible singularities of $\zeta_f$. If $\zeta_f$ is meromorphic in $U$, then its poles in $W$ are contained in this set $\Dd_f(W)$. In this case, we say that $\zeta_f$ admits a \textbf{meromorphic continuation} to a connected open neighborhood of $W$.

Next, we note that the estimates in \eqref{eqn:MellinEstimate} are \textit{independent of the imaginary part} of the parameter. Thus, if the real part is bounded (as is the case in a horizontally bounded vertical strip), we may obtain uniform estimates for the Mellin transform. If $\alpha=0$, note that we must choose a vertical strip which lies in the half-plane $\HH_{\sigma_0}$, where $f(t)=O(t^{-\sigma_0})$, but the restriction is removed when $\alpha>0$ since we may choose $\sigma_0$ to be arbitrarily small. 

\begin{Corollary}[Uniform Boundedness on Vertical Strips]
    \label{cor:MellinBounds}
    Let $f$ be integrable on $(\alpha,\beta)\subset\RR^+$ and let $\Mellin_\alpha^\beta[f]$ be its (truncated) Mellin transform, with $0\leq\alpha<\beta<\infty$. Suppose that either of the two hypotheses of Lemma~\ref{lem:MellinHolo} hold. In the first case, let $a>\sigma_0$ and in the second let $a\in\RR$ be arbitrary. 
    \medskip

    Then $\Mellin_\alpha^\beta[f]$ is bounded in any vertical strip of the form 
    \[ \HH_a^b=\set{s\in\CC\suchthat a<\Re(s)<b} \]
    where either $a>\sigma_0$ in the first case (when $\alpha=0$ and $f(t)=O(t^{-\sigma_0})$ as $t\to0^+$) or where $a$ is arbitrary when $\alpha>0$. We may also choose any vertical line of the form $\Re(s)\equiv a$ or the closed strip with $a\leq \Re(s)\leq b$. 
\end{Corollary}
\begin{proof}
    It is easiest to prove for closed vertical strips and deduce the other results for open strips and vertical lines as corollaries. Suppose that $\sigma_0<a\leq \Re(s)\leq b$. Then $|\Re(s)-\sigma_0|\leq a-\sigma_0>0$ is bounded from below, whence its reciprocal is bounded. The function $\delta^{t}$ is continuous on $[a,b]$, and thus bounded. It follows from \eqref{eqn:MellinEstimate} that $\Mellin_\alpha^\beta[f]$ is bounded on this closed vertical strip. If $\alpha>0$, for any $a\in\RR$ choose $\sigma_0<a$ and apply the previous argument.  

    For vertical lines, take $\sigma_0<a=b$. For an open strip, note that it is contained in its closure where the function is bounded by the result for closed strips, and by assumption $\sigma_0<a$. 
\end{proof}

Notably, the restricted transform may converge even if the full Mellin transform integral diverges. Indeed, this improvement to the convergence is the main reason to use this modification of the Mellin transform, as later we will see that truncation complicated the scaling property of the transform (viz. Lemma~\ref{lem:MellinScaling}) slightly. Perhaps the most important class of functions for which this occurs is that of polynomials. Without loss of generality, let $f(t)=t^k$ be a monomial. If $\Re(s)>-k$, then $\Mm^\beta[f](s)$ converges; however, $\Mm[f](s)$ is divergent for all $s\in\CC$. Thus, for a general polynomial $p(t)=\sum_{k=0}^n a_kt^k$, we have that 
\[ \Mm^\beta[p](s) = \sum_{k=0}^n \frac{a_k}{s+k}\beta^{s+k},   \]
which is valid when $\Re(s)>\max\set{-k: a_k\neq 0}$. Negative powers become admissible if the lower bound $\alpha$ of the truncation is positive. 

\subsection{Scaling Properties of Mellin Transforms}

Owing to the role of the scale invariant Haar measure $dx/x$ on $(\RR^+,\cdot)$ defining the Mellin transform, it possesses a very simple formula for scaling. Let $\lambda>0$ and define the scaling function $S_\lambda(x):=\lambda x$. Then for the standard Mellin transform,
\begin{equation}
    \label{eqn:MellinScaling}
    \Mellin[f\circ S_\lambda](s) = \lambda^{-s} \Mellin[f](s). 
\end{equation}
This property can be seen as a consequence of the change of variables formula, the scale invariance of the Haar measure $dx/x$, and the scale invariance of the domain of integration, $\RR^+$. 

For the truncated Mellin transforms, a property similar to \eqref{eqn:MellinScaling} holds. However, the domain of integration, $(\alpha,\beta)$, is \textit{not} scale invariant. Thus, the cutoffs of the truncated transform change. In what follows, we assume that the function is defined (and continuous, for simplicity) on all of $\RR^+$ so that the change of domain does not present any new issues.
\begin{Lemma}[Scaling Property of Truncated Mellin Transforms]
    \label{lem:MellinScaling}
    \index{Mellin transform!Scaling properties}
    Let $f\in C^0(\RR^+)$ , let $\alpha,\beta\in[0,\infty)$ with $\alpha<\beta$, let $\lambda>0$, and define $S_\lambda(x):=\lambda x$. 

    Then, provided that the transforms are convergent, we have that 
    \begin{equation}
        \label{eqn:truncMellinScaling}
        \Mellin_\alpha^\beta[f\circ S_\lambda](s) = \lambda^{-s}\Mellin_{\lambda\alpha}^{\lambda\beta}[f](s).
    \end{equation}
    If $\alpha=0$, then \eqref{eqn:truncMellinScaling} becomes
    \begin{equation}
        \label{eqn:MellinScalingFunctional}
        \Mellin^\beta[f\circ S_\lambda](s)=\lambda^{-s}\Mellin^\beta[f](s) 
            + \lambda^{-s}\Mellin_{\beta}^{\lambda\beta}[f](s).
    \end{equation}
\end{Lemma}
The purpose of \eqref{eqn:MellinScalingFunctional} is clear in our usage of this property. Essentially, it establishes a functional relation akin to the standard property of Mellin transforms (viz. \eqref{eqn:MellinScaling}) up to the addition of an entire function, $\lambda^{-s}\Mellin_{\beta}^{\lambda\beta}[f](s)$. Crucially, this means that $\Mellin^\beta[f\circ S_\lambda]$ and $\lambda^{-s}\Mellin^{\beta}[f]$ have the exact same set of singularities.

\section*{Acknowledgements}
Parts of this work have been submitted by William E. Hoffer in partial fulfillment of the requirements for Doctor of Philosophy at the University of California, Riverside.

\section*{Funding} 

The research of M.L. Lapidus was partially supported by grants from the (French) Agence Nationale de la Recherche (ANR FRACTALS; ANR-24-CE45-3362), by the Centre National de la Recherche Scientifique (CNRS) through the MITI interdisciplinary program (MITI CNRS Conditions Extr\^emes), and by the Burton Jones Endowed Chair in Pure Mathematics. The research of W.E. Hoffer was partially supported by the Burton Jones Fellowship, the John C. Fay Fellowship, and the Dissertation Completion Fellowship (University of California, Riverside).

\printbibliography

\medskip 
\noindent
\textbf{William E. Hoffer}

\noindent
Department of Mathematics, University of California, Riverside, 900 University Avenue, Riverside, CA 92521-0135, USA. 

\noindent
\textit{Email address:} \href{mailto:whoff003@ucr.edu}{whoff003@ucr.edu}

\noindent
\textit{Website:} \url{https://willhoffer.com}
\\

\noindent
\textbf{Michel L. Lapidus}

\noindent
Department of Mathematics, University of California, Riverside, 900 University Avenue, Riverside, CA 92521-0135, USA. 

\noindent
\textit{Email address:} \href{mailto:lapidus@ucr.edu}{lapidus@ucr.edu}

\noindent
\textit{Website:} \url{https://math.ucr.edu/~lapidus/}

\end{document}